# Asynchronous Correspondences Between Hybrid Trajectory Semantics

Patrick Cousot[0000−0003−0101−9953]

CS, CIMS, New York University, USA
pcousot@cims.nyu.edu https://cs.nyu.edu/~pcousot/

*Dedicated to Thomas Henzinger for his 60th birthday*

**Abstract.** We formalize the semantics of hybrid systems as sets of hybrid trajectories, including those generated by an hybrid transition system. We study the abstraction of hybrid trajectory semantics for verification, static analysis, and refinement. We mainly consider abstractions of hybrid semantics which establish a correspondence between trajectories derived from a correspondence between states such as homomorphisms, simulations, bisimulations, and preservations with progress. We also consider abstractions that cannot be defined stepwise like discretization. All these abstractions are Galois connections between concrete and abstract hybrid trajectory or discrete trace semantics. In contrast to semantic based abstractions, we investigate the problematic trace-based composition of abstractions.

## 1 Introduction

State and transition-based abstractions such as homomorphisms, simulations, bisimulations [26], and preservations with progress (as used in type theory [38]) formalize a correspondence between concrete and abstract discrete semantics. They have been successfully applied to the verification, analysis, and refinement of programs. In program refinement, such state and transition-based abstractions are used to transform specifications into implementations. In program verification and analysis they are used to simplify the reasoning on properties of program executions.

All these abstractions have two fundamental properties. The first is that a reasoning on computation steps (via a transition system) is sufficient to establish a correspondence between program semantics (which is the set of all their possible maximal executions). The second is that they compose. For example the composition of simulations is a simulation. This allows, for example, for stepwise refinement in program construction or composing successive sound abstractions in program verification.

Our objective is to extend and study these state and transition-based abstractions for dynamical systems that exhibits both continuous and discrete dynamic behavior as found in cyber-physical systems. We consider concrete and abstract hybrid semantics (that is sets of sequences of configurations specifying continuous behaviors between discrete changes of modes) that allow for arbitrary timings, arbitrary continuous dynamic mode changes, and arbitrary evolutions of the states over time. We also consider hybrid semantics generated by hybrid transition systems hoping that, as in the discrete case, the abstraction of transition systems will induce the abstraction of the hybrid semantics. But contrary to the discrete case, this is problematic.

Such hybrid trajectory semantics can be understood as specifications, implementations, or abstractions of hybrid dynamical systems. They are more general than particular abstract models of hybrid systems such as synchronous systems [7], timed automata [2], switched systems (for which the sequence of modes and mutation times are known in advance) [22], hybrid automata [1], including restrictions for decidability subclasses [4,18], Simulink [24], and so on. Hybrid trajectory semantics can also be used to specify the semantics of these abstract models, that is, the set of possible behaviors that they describe.

We study homomorphisms, simulations, (bisimulations, preservations with progress in the appendix) between concrete and abstract hybrid semantics as well as discretization of hybrid semantics to establish a correspondence between an hybrid system and a discrete system (such as a computer). Considered as semantic transformers they all form Galois connections and so do compose. However, when considering individual concrete and abstract trajectories, the problem is that in full generality, these abstraction may not compose well. For examples the discretization of two trajectories of (bi)simular hybrid systems may not be (bi)simular discrete traces. We investigate sufficient conditions to solve this compositionally problem when reasoning on individual trajectories.

The paper organized as follows. In section 2 we recall the definitions of Galois connections, Galois relations (ordered logical relations), and tensor products. In section 3, we introduce hybrid trajectory semantics to define the arbitrary evolution of hybrid systems over time. In section 4, we introduce hybrid transition systems that can be used to generate hybrid trajectory semantics (the same way that discrete transition systems generate a discrete trace-based operational semantics for discrete systems). In section 5, we consider the abstraction of hybrid trajectory semantics by reasoning on trajectories, that is executions of the hybrid system as defined by its semantics. It is often considered that reasoning on states, or consecutive states, is simpler that reasoning on full trajectories (although less general). This is the objective of section 6, where an abstraction of states is shown to induce an abstraction of hybrid trajectories, hence of hybrid semantics (which are sets of hybrid trajectories). In case the hybrid semantics is defined by a transition system, we consider in section 7 the abstraction of transition systems by homomorphisms, simulations, (bisimulation in section A.1, and preservation with progress in section A.2 of the appendix) and study which abstraction of trajectories and hybrid semantics they induce. The main difficulty is that concrete and abstract trajectories may have different, not necessarily comparable timelines, that is timings for mode changes. A difficulty, in particular for discretization, is that the abstraction of transition systems may not be an abstraction of their hybrid semantics (which is never the case for discrete systems). We solve the problem under sufficient conditions. We conclude in section 8.

## 2 Galois connections and relations

### 2.1 Galois connections

A Galois connection $\langle C,\, \sqsubseteq\rangle \xleftrightarrows[\alpha]{\gamma} \langle A,\, \preccurlyeq\rangle$[1] between posets $\langle C,\, \sqsubseteq\rangle$ and $\langle A,\, \preccurlyeq\rangle$ is a pair of an abstraction function $\alpha$ and a concretization function $\gamma$ such that

$$\langle C,\, \sqsubseteq\rangle \xleftrightarrows[\alpha]{\gamma} \langle A,\, \preccurlyeq\rangle \triangleq \begin{cases} \alpha \in C \xrightarrow{\nearrow} A & \text{is increasing} & (1.\text{a}) \\ \gamma \in A \xrightarrow{\nearrow} C & \text{is increasing} & (1.\text{b}) \\ \gamma \circ \alpha & \text{is an upper closure} & (1.\text{c}) \\ \alpha \circ \gamma & \text{is a lower closure} & (1.\text{d}) \end{cases} \qquad (1)$$

where an upper closure is increasing, idempotent, and extensive ($\forall c \in C \,.\, x \sqsubseteq \gamma \circ \alpha(x)$) while a lower closure is increasing, idempotent, and reductive ($\forall y \in A \,.\, \alpha \circ \gamma(y) \preccurlyeq y$). An equivalent definition of a Galois connection is a pair of increasing functions satisfying

$$\forall x \in C \,.\, \forall y \in A \,.\, \alpha(x) \preccurlyeq y \Longrightarrow x \sqsubseteq \gamma(y) \quad \wedge \qquad (2)$$

$$\alpha(x) \preccurlyeq y \Longleftarrow x \sqsubseteq \gamma(y) \qquad (3)$$

*Example 1 (Classic examples of Galois connections).* Set transformers form Galois connections

$$\langle \wp(\mathsf{S}),\, \subseteq\rangle \xleftrightarrows[\mathsf{pre}[r]]{\widetilde{\mathsf{post}}[r]} \langle \wp(\overline{\mathsf{S}}),\, \subseteq\rangle \text{ and } \langle \wp(\mathsf{S}),\, \subseteq\rangle \xleftrightarrows[\mathsf{post}[r]]{\widetilde{\mathsf{pre}}[r]} \langle \wp(\overline{\mathsf{S}}),\, \subseteq\rangle \qquad (4)$$

where $r \in \wp(\mathsf{S} \times \overline{\mathsf{S}})$, $\mathsf{post}[r]P \triangleq \{y \mid \exists x \in P \,.\, \langle x,\, y\rangle \in r\}$, $\mathsf{pre}[r] = \mathsf{post}[r^{-1}]$, $r^{-1} \triangleq \{\langle y,\, x\rangle \mid \langle x,\, y\rangle \in r\}$, $\tilde{f} \triangleq \neg \circ f \circ \neg$, and $(f \circ g)(x) = f(g(x))$ is function composition.

Another classic example is an homomorphic abstraction, where given $h \in \mathsf{S} \to \overline{\mathsf{S}}$, $\alpha_h(X) \triangleq \{h(x) \mid x \in X\}$, and $\gamma_h(Y) \triangleq \{x \in \mathsf{S} \mid h(x) \in Y\}$, we have

$$\langle \wp(\mathsf{S}),\, \subseteq\rangle \xleftrightarrows[\alpha_h]{\gamma_h} \langle \wp(\overline{\mathsf{S}}),\, \subseteq\rangle \qquad (5) \qquad \square$$

Interpreting $C$ in (1) as a concrete semantics (e.g. a set of execution discrete traces or hybrid trajectories) and $A$ as an abstract semantics, the concretization

[1] see an introduction in [11, Ch. 11]

$\gamma(y)$ is the concrete semantics corresponding to the abstract semantics $y \in A$, that is its concrete meaning. Conversely, $\alpha(x)$ is the abstraction of the concrete semantics $x \in C$.

The conditions (1.a) and (1.b) of order preservation express that the notions of over approximation in the concrete and the abstract are the same.

Condition (1.c) implies $x \sqsubseteq \gamma(\alpha(x))$. This expresses that $\alpha(x)$ is an abstract sound over approximation of $x$.

Condition (1.c) with $y = \alpha(x)$ implies (3) which expresses that $\alpha(x)$ is the best abstraction of $x$ (since given any other abstraction of $x$ which is sound, that is $x \sqsubseteq \gamma(y)$, $\alpha(x)$ is more precise since $\alpha(x) \preccurlyeq y$).

### 2.2 Galois relations

Any Galois connection $\langle \mathcal{C},\ \sqsubseteq\rangle \xleftrightarrows[\alpha]{\gamma} \langle \mathcal{A},\ \preccurlyeq\rangle$ can be encoded by a *Galois relation* $R_\alpha \in \wp(\mathcal{C} \times \mathcal{A})$ (also called ordered logical relations) defined as

$$R_\alpha \triangleq \{\langle x,\ y\rangle \in C \times A \mid \alpha(x) \preccurlyeq y\} = \{\langle x,\ y\rangle \in C \times A \mid x \sqsubseteq \gamma(y)\} \qquad (6)$$

If $\langle \mathcal{C},\ \sqsubseteq,\ \bigsqcup\rangle$ and $\langle \mathcal{A},\ \preccurlyeq,\ \curlywedge\rangle$ are complete lattices such relations $R_\alpha$ satisfy the following characteristic properties of Galois relations $R$.

$$\begin{aligned}
&(x \sqsubseteq x' \wedge \langle x',\ y'\rangle \in R \wedge y' \preccurlyeq y) \Longrightarrow (\langle x,\ y\rangle \in R) && \text{(a)}\\
&(\forall i \in \Delta\ .\ \langle x_i,\ y\rangle \in R) \Longrightarrow \langle \bigsqcup_{i\in\Delta} x_i,\ y\rangle \in R && \text{(b)}\\
&(\forall i \in \Delta\ .\ \langle x,\ y_i\rangle \in R) \Longrightarrow \langle x,\ \curlywedge_{i\in\Delta} y_i\rangle \in R && \text{(c)}
\end{aligned} \qquad (7)$$

The tensor product $\langle \mathcal{C},\ \sqsubseteq\rangle \otimes \langle \mathcal{A},\ \preccurlyeq\rangle$ of two complete lattices $\langle \mathcal{C},\ \sqsubseteq\rangle$ and $\langle \mathcal{A},\ \preccurlyeq\rangle$ is [34]

$$\langle \mathcal{C},\ \sqsubseteq\rangle \otimes \langle \mathcal{A},\ \preccurlyeq\rangle \triangleq \{R \in \wp(\mathcal{C} \times \mathcal{A}) \mid R \text{ is a relation satisfying } (7)\} \qquad (8)$$

Galois connections and relations are mathematically equivalent. If $\langle \mathcal{C}, \sqsubseteq\rangle$ and $\langle \mathcal{A}, \preccurlyeq\rangle$ be complete lattices then $\langle \mathcal{C}, \sqsubseteq\rangle \xleftrightarrows[\alpha]{\gamma} \langle \mathcal{A}, \preccurlyeq\rangle$ if and only if $R_\alpha \in \langle \mathcal{C}, \sqsubseteq\rangle \otimes \langle \mathcal{A}, \preccurlyeq\rangle$ where $R_\alpha$ is defined in (6) and, conversely, $\alpha(x) \triangleq \curlywedge\{y \mid \langle x,\ y\rangle \in R_\alpha\}$ and $\gamma(y) \triangleq \bigsqcup\{x \mid \langle x,\ y\rangle \in R_\alpha\}$.

Dual definitions of Galois connections and relations can be used to cope with under approximation.

## 3 Hybrid trajectory semantics

**Time.** We let the time $t$ run over the set $\mathbb{R}_{\geqslant 0}$ of all positive reals.

**States and flows.** We let $\mathsf{S}$ be a set of states. In our pictures, we use Cartesian coordinates where the horizontal axis is time and the vertical axis is the set of states (which we take to be $\mathsf{S} = \mathbb{R}$).

**Flows.** A flow $f \in \mathsf{F} \triangleq \mathbb{R}_{\geqslant 0} \nrightarrow \mathsf{S}$ is a partial map from time to states representing the evolution of the state over time. Flows can be specified e.g. by ODEs over a period of time (with appropriate hypothesis, see e.g. [21, Ch. XIX], [19, ch. 8 & 9], [20], and [30]).

**Time intervals.** If $t_1 \in \mathbb{R}_{\geqslant 0}$, $t_2 \in \mathbb{R}_{\geqslant 0} \cup \{\infty\}$, and $t_1 < t_2$ then $[t_1, t_2[ \triangleq \{t \in \mathbb{R}_{\geqslant 0} \mid t_1 \leqslant t < t_2\}$ is the interval of time between $t_1$ and $t_2$, the lower bound $\mathsf{b}([t_1, t_2[) \triangleq t_1$ being included while the upper bound $\mathsf{e}([t_1, t_2[) \triangleq t_2$ is excluded. The set of all such time intervals is

$$i \in \mathsf{I} \triangleq \{[t_1, t_2[ \mid t_1 \in \mathbb{R}_{\geqslant 0} \wedge t_2 \in \mathbb{R}_{\geqslant 0} \cup \{\infty\} \wedge t_1 + \zeta \leqslant t_2\} \tag{9}$$

where $\zeta > 0$ is any arbitrarily chosen infinitesimal defining the minimal duration $\mathsf{d}(i) \triangleq \mathsf{e}(i) - \mathsf{b}(i)$ of a time interval $i$. This implies that the duration of successive configurations cannot tend to 0 so we exclude *zeno* systems (with infinitely many successive configurations in a finite interval of time [39]).

The closure of an interval $\mathsf{cl}([t_1, t_2[) \triangleq [t_1, t_2]$ if $t_2 \neq \infty$ and $\mathsf{cl}([t_1, \infty[) = [t_1, \infty[$ includes the upper bound unless it is infinite. By convention, $[t_1, \infty] = [t_1, \infty[ = \{t \in \mathbb{R}_{\geqslant 0} \mid t_1 \leqslant t\}$. We let $\mathsf{cl}(\mathsf{I}) \triangleq \{\mathsf{cl}(i) \mid i \in \mathsf{I}\}$.

**Configurations.** A configuration is a pair of a flow and a time interval

$$c \in \mathsf{C} \triangleq \{\langle f,\ i\rangle \in \mathsf{F} \times \mathsf{I} \mid \forall t \in i\ .\ f(t) \in \mathsf{S}\} \tag{10}$$

while final configurations include the upper bound

$$c \in \mathsf{cl}(\mathsf{C}) \triangleq \{\langle f,\ i\rangle \in \mathsf{F} \times \mathsf{cl}(\mathsf{I}) \mid \forall t \in i\ .\ f(t) \in \mathsf{S}\} \tag{11}$$

such that the flow is well-defined in the set of states $\mathsf{S}$ on the time interval, i.e. $i \subseteq \mathsf{dom}(f)$. A configuration $c = \langle f,\ i\rangle$ starts a time $\mathsf{b}(c) = \mathsf{b}(i)$ and ends at time $\mathsf{e}(c) = \mathsf{e}(i)$, excluded in (10) and included in (11). We call $\mathsf{dom}(c) = i$ the time interval of configuration $c$. Notice that by the choice of the infinitesimal $\zeta > 0$ and the definition (9) of $\mathsf{I}$, the intervals $i \in \mathsf{I}$ in (10) and (11) cannot be empty.

Configurations $c$ record the evolution of the state as specified by the flow during the period of time $\mathsf{dom}(\mathrm{c})$. During that time interval the definition of the flow $f$, which is the law of continuous evolution of the system as a function of the time, is fixed. It may be different in the next configuration of the system. In that case, it is common to say that the mode of the hybrid system has changed. The duration $\mathsf{d}(c) = \mathsf{e}(c) - \mathsf{b}(c) \geqslant \zeta$ of the configuration is lower-bounded by $\zeta > 0$ so that infinite sequences of configurations are always nonzeno. Additional hypotheses might be necessary on the flow $f$ of configurations $\langle f,\ i\rangle$ such as continuity, uniform continuity, Lipschitz continuity, etc. However, discontinuities are always allowed (but not mandatory) when changing mode between consecutive configurations.

By convention the state of a configuration $c$ at time $t \in \mathbb{R}_{\geqslant 0}$ is

$$\begin{aligned} c(t) &\triangleq f(t) && \text{if } c = \langle f,\ i\rangle \text{ and } t \in i \\ &\triangleq \text{undefined} && \text{otherwise} \end{aligned} \tag{12}$$

Let us define the concatenation of two consecutive configurations $\langle f,\ i\rangle \in \mathsf{C}$ and $\langle f',\ i'\rangle \in \mathsf{C} \cup \mathsf{cl}(\mathsf{C})$ where $\mathsf{e}(i) = \mathsf{b}(i')$ (i.e. the concatenation is undefined for non-consecutive intervals).

$$\langle f,\ i\rangle \mathbin{\text{\textfrac{o}{9}}} \langle f',\ i'\rangle \triangleq \langle f'',\ i \cup i'\rangle \text{ where } \begin{cases} f''(t) = f(t) \text{ when } t \in i \\ f''(t) = f'(t) \text{ when } t \in i' \end{cases} \tag{13}$$

Since the state at the beginning of a configuration may be different from the state at the end of the previous configuration at the same time, definitions (13), (15), and (23) favor states at the beginning of configurations (because intervals are left closed and open right).

To simplify notations, the empty configuration is, by convention, $\varepsilon \triangleq \langle\emptyset,\ \emptyset\rangle$ where $\emptyset$ is the empty set, that is, the everywhere undefined function. By convention, $\mathsf{b}(\varepsilon) \triangleq +\infty$ and $\mathsf{e}(\varepsilon) = -\infty$ so that $\min(t, \mathsf{b}(\varepsilon)) = \max(t, \mathsf{e}(\varepsilon)) = t$ when $t \in \mathbb{R}_{\geqslant 0}$. Observe that although $\varepsilon \notin \mathsf{C} \cup \mathsf{cl}(\mathsf{C})$ since the time interval is empty, we nevertheless have $\langle f,\ i\rangle \mathbin{⨾} \varepsilon \triangleq \varepsilon \mathbin{⨾} \langle f,\ i\rangle \triangleq \langle f,\ i\rangle$, for ease of writing.

The selection of a time slice during the configuration time interval.

$$\begin{array}{lll} \langle f,\ i\rangle(\!|t_1, t_2]\!) \triangleq \langle f,\ i \cap [t_1, t_2]\rangle & \text{where} & \mathsf{b}(i \cap [t_1, t_2]) + \zeta \leqslant \mathsf{e}(i \cap [t_1, t_2]) \\ \langle f,\ i\rangle(\!|t_1, t_2(\!| \triangleq \langle f,\ i \cap [t_1, t_2[\rangle & & \mathsf{b}(i \cap [t_1, t_2[) + \zeta \leqslant \mathsf{e}(i \cap [t_1, t_2[) \end{array} \tag{14}$$

In particular, we define $\varepsilon(\!|t_1, t_2]\!) \triangleq \varepsilon(\!|t_1, t_2(\!| \triangleq \varepsilon$.

**Trajectories** The trajectories over configurations $\mathsf{C}$ are nonempty finite or infinite sequences of contiguous configurations.

$$\begin{array}{ll} \mathsf{T}^n_{\mathsf{C}} \triangleq \{\sigma \in [0, n] \to \mathsf{cl}(\mathsf{C}) \mid \mathsf{b}(\sigma_0) = 0 \wedge \forall i \in [0, n[\ .\ \sigma_i \in \mathsf{C} \wedge \\ \qquad\qquad \mathsf{e}(\sigma_i) = \mathsf{b}(\sigma_{i+1}) \wedge \sigma_n \in \mathsf{cl}(\mathsf{C})\} \\ & \text{finite trajectories } \sigma \in \mathsf{T}^n_{\mathsf{C}} \text{ of length } |\sigma| = n+1,\ n \in \mathbb{N} \\ \mathsf{T}^+_{\mathsf{C}} \triangleq \bigcup_{n\in\mathbb{N}} \mathsf{T}^n_{\mathsf{C}} & \text{finite nonempty trajectories} \\ \mathsf{T}^\infty_{\mathsf{C}} \triangleq \{\sigma \in \mathbb{N} \to \mathsf{C} \mid \mathsf{b}(\sigma_0) = 0 \wedge \forall i \in \mathbb{N}\ .\ \mathsf{e}(\sigma_i) = \mathsf{b}(\sigma_{i+1})\} \\ & \text{infinite trajectories } \sigma \in \mathsf{T}^\infty_{\mathsf{C}} \text{ of length } |\sigma| = \infty \\ \mathsf{T}^{+\infty}_{\mathsf{C}} \triangleq \mathsf{T}^+_{\mathsf{C}} \cup \mathsf{T}^\infty_{\mathsf{C}} & \text{nonempty trajectories} \end{array} \tag{15}$$

A finite or infinite trajectory $\sigma \in [0, |\sigma|[ \to \mathsf{C}$ is a sequence of configurations that will be denoted $\sigma = \langle\sigma_i,\ i \in [0, |\sigma|[\rangle$. Such a trajectory $\sigma$ records the evolution of the state along discrete changes of the flows encoded by configurations. The state at the end of a configuration is that of the next state, if any. Therefore, the configuration intervals are open right and consecutive except for the last one in finite trajectories which is closed. No configuration in a trajectory can be empty.

We let $\sigma[i, j]$ denote the subsequence of configurations in $\sigma$ of ranks $i$ to $j$, $i, j \in [0, |\sigma|[$. $\sigma[i, j[$ excludes $j$ (usually $\infty$).

**Traces.** We let traces $\varsigma \in \mathsf{T}_{\mathsf{S}}^{+\infty}$ be discrete finite or infinite untimed sequences of states in $\mathsf{S}$ and use the same notations for continuous trajectories and discrete traces. The homomorphic timeline abstraction ((_?_ : _) is the conditional)

$$\alpha_{tl}(\sigma) \triangleq \boldsymbol{\lambda}\, i \in [0, |\sigma|] \bullet (i = 0 \;?\; 0 \;:\; (i = \infty \;?\; \infty \;:\; \mathsf{e}(\sigma_{i-1})))$$
$$\alpha_{tl}(T) \triangleq \{\alpha_{tl}(\sigma) \mid \sigma \in T\}$$

such that, by (5), $\langle \mathsf{T}_{\mathsf{C}}^{+\infty}, \subseteq\rangle \xleftarrow[\alpha_{tl}]{\gamma_{tl}} \langle \mathsf{T}_{\mathbb{R}_{\geqslant 0}\cup\{\infty\}}^{+\infty}, \subseteq\rangle$ is an example of abstraction of trajectories into traces (by projection of the mode change timings).

**Hybrid trajectory semantics and properties.** Given a set $\mathsf{S}$ of states and the corresponding configurations $\mathsf{C}$ in (10), a hybrid trajectory semantics $\mathcal{S}_{\mathsf{C}} \in \wp(\mathsf{T}_{\mathsf{C}}^{+\infty})$ is a subset of all possible trajectories (15). Properties of hybrid trajectory semantics belong to $\wp(\wp(\mathsf{T}_{\mathsf{C}}^{+\infty}))$ (sometimes called hyper properties) while there abstraction $\alpha_{\cup}(P) = \bigcup P$ into trajectory properties belong to $\wp(\mathsf{T}_{\mathsf{C}}^{+\infty})$.

Similarly a trace semantics $\mathcal{S}_{\mathsf{S}} \in \wp(\mathsf{T}_{\mathsf{S}}^{+\infty})$ is a subset of all possible traces.

**Trajectory states.** The duration $[\![\sigma]\!]$ of a trajectory $\sigma$ is

$$[\![\sigma]\!] \triangleq \sum_{k=0}^{n} \mathsf{e}(\sigma_i) - \mathsf{b}(\sigma_i) = \mathsf{e}(\sigma_n) \qquad \text{when} \quad \sigma \in \mathsf{T}_{\mathsf{C}}^{n} \tag{16}$$
$$\triangleq \sum_{k=0}^{\infty} \mathsf{e}(\sigma_i) - \mathsf{b}(\sigma_i) = \infty \qquad \text{when} \quad \sigma \in \mathsf{T}_{\mathsf{C}}^{\infty} \qquad \text{(nonzeno hypothesis)}$$

as indicated by the time at which the last configuration in the trajectory ends or $\infty$ for infinite trajectories.

**Time-evolution law abstraction.** A trajectory $\sigma$ can be abstracted into a function $\alpha_{tr}(\sigma) \in \mathbb{R}_{\geqslant 0} \to \mathsf{S}$ mapping time to a state such that

$$\mathsf{dom}(\alpha_{tr}(\sigma)) \triangleq [0, [\![\sigma]\!]] \qquad \text{(by convention, excluding } \infty \text{ if } [\![\sigma]\!] = \infty)$$
$$\alpha_{tr}(\sigma)(t) \triangleq f(t) \text{ such that } \exists k \in [0, |\sigma|[\;.\; \sigma_k = \langle f,\, i\rangle \wedge t \in i \tag{17}$$
$$\sigma_t \triangleq \alpha_{tr}(\sigma)(t) \qquad \text{(abbreviated notation)}$$

So we have two different representations of trajectories, $\sigma$ in (15) and $\alpha_{tr}(\sigma)$ in (17), this second representation being closer to the time-evolution law of the theory of dynamical systems [20]. Notice that $\alpha_{tr}(\sigma)$ is a function defined by parts on the timeline abstraction $\alpha_{tl}(\sigma)$ of the trajectory $\sigma$ so the time-evolution law $\alpha_{tr}(\sigma)$ is not simpler that the trajectory $\sigma$ to reason upon, in particular because the timeline information is abstracted away.

We leave this $\alpha_{tr}$ abstraction implicit and use the same notation for both cases. Therefore a trajectory $\sigma$ is either a discrete sequence of configurations $\sigma = \langle\sigma_i,\ i \in [0, |\sigma|[\rangle$ or a state function of the time $\sigma = \langle\sigma_t,\ t \in [0, [\![\sigma]\!][\rangle$ where $\sigma_t \triangleq \alpha_{tr}(\sigma)(t)$. By homomorphic abstraction (5), this extends to hybrid trajectory semantics $T$ with $\alpha_{tr}(T) \triangleq \{\alpha_{tr}(\sigma) \mid \sigma \in T\}$

$$\langle\wp(\mathsf{T}_{\mathsf{C}}^{+\infty}),\ \subseteq\rangle \underset{\alpha_{tr}}{\overset{\gamma_t}{\leftrightarrows}} \langle\wp(\mathbb{R}_{\geqslant 0} \to \mathsf{S}),\ \subseteq\rangle \tag{18}$$

**Maximal trajectory semantics.** A trajectory semantics $T \in \wp(\mathsf{T}_{\mathsf{C}}^{+\infty})$ on configurations $\mathsf{C}$ is a set of finite or infinite trajectories. Let us define the maximal trajectories of $T$ as those without strict prefixes

$$\max(T) \triangleq \{\langle\sigma_i,\ i \in [0, |\sigma|[\rangle \in T \mid \forall n < |\sigma| \ .\ \langle\sigma_i,\ i \in [0, n[\rangle \notin T\}$$

A maximal trajectory semantics has no strict prefixes, that is $\max(T) = T$.

*Example 2 (Specification of a water tank [17]).* A water tank (or water dam) runs for ever with a continuous inflow and a valve (or spillway floodgate) than can be opened or shut to control the outflow. The objective is to design a controller to maintain the water level $y$ between 0 and 3 (for some length unit). When the valve is opened, the water level $y$ decreases while, when the valve is shut down, the water level $y$ increases. The tank should never remain empty more than $\zeta$ units of time.

Define states

$$s \in \mathsf{S} \triangleq \mathbb{R} \times \{\mathit{open}, \mathit{shut}\} \tag{19}$$

such that $s.y \in \mathbb{R}$ and $s.v \in \{\mathit{open}, \mathit{shut}\}$. Let $\mathsf{C}$ be the corresponding set (10) of configurations. The above informal specification can be formalized by the following abstract hybrid semantics of the water tank.

$$\begin{aligned}
P(\sigma) \triangleq \forall t \in \mathbb{R}_{\geqslant 0} \ .\ 0 \leqslant \sigma(t).y \leqslant 3 \wedge \forall t_2 > t_1 \geqslant 0 \ . & \quad \text{(a)} \\
\forall t \in [t_1, t_2] \ .\ \sigma(t).v = \mathit{open} \Longrightarrow \sigma(t_1).y > \sigma(t_2).y \wedge & \quad \text{(b)} \\
\forall t \in [t_1, t_2] \ .\ \sigma(t).v = \mathit{shut} \Longrightarrow \sigma(t_1).y < \sigma(t_2).y \wedge & \quad \text{(c)} \\
\forall t \in \mathbb{R}_{\geqslant 0} \ .\ \sigma(t).y = 0 \Longrightarrow \sigma(t + \zeta).y > 0 & \quad \text{(d)}
\end{aligned} \tag{20}$$

The hybrid semantics specification the water tank is then

$$\mathcal{S}^{\text{\tiny 2}} \triangleq \{\sigma \in \{0\} \to \mathsf{C} \mid \mathsf{b}(\sigma_0) = 0 \wedge \mathsf{e}(\sigma_0) = \infty \wedge P(\sigma_0)\} \tag{21}$$

with only one configuration, or using the homomorphic abstraction (17),

$$\mathcal{S}^{\text{\tiny 2}} \triangleq \{\sigma \in \mathbb{R}_{\geqslant 0} \to \mathsf{S} \mid P(\sigma)\}$$

□

## 4 Transition-based hybrid trajectory semantics

As in the discrete case, a simple way to define a hybrid trajectory semantics, is to first define a hybrid transition system and then to consider the hybrid semantic defined as the set of all possible trajectories generated by the hybrid transition system. As is the case for discrete trace semantics, not all hybrid semantics can be generated by a hybrid transition system on the same set of configurations (which cannot e.g. express fairness without adding a scheduler to the transition system or adding conditions on the generated traces).

**Hybrid transition system.** A hybrid transition system is defined by a triple $\langle \mathsf{C},\ \mathsf{C}^0,\ \tau\rangle$ of a set of configurations $\mathsf{C}$, initial configurations $\mathsf{C}^0$ and a transition relation $\tau \in \wp(\mathsf{C} \times (\mathsf{C} \cup \mathsf{cl}(\mathsf{C})))$ such that

$$\begin{array}{ll} \text{initial configurations} & \mathsf{C}^0 \subseteq \{c \in \mathsf{C} \mid \mathsf{b}(c) = 0\} \\ \text{consecutiveness} & \forall \langle c,\ c'\rangle \in \tau \,.\, c \in \mathsf{C} \wedge \mathsf{e}(c) = \mathsf{b}(c') \\ \text{closeness of final configurations} & \forall c \,.\, (\forall c' \,.\, \langle c,\ c'\rangle \notin \tau) \iff c \in \mathsf{cl}(\mathsf{C}) \end{array} \tag{22}$$

**Maximal trajectory semantics of a transition system.** A transition semantics $\langle \mathsf{C},\ \mathsf{C}^0,\ \tau\rangle$ is usually used to define a hybrid trajectory semantics $[\![\langle \mathsf{C},\ \mathsf{C}^0,\ \tau\rangle]\!]$ abbreviated $[\![\tau]\!]$, for example the maximal one.

$$\begin{aligned} [\![\tau]\!]^n &\triangleq \{\sigma \in \mathsf{T}^n_{\mathsf{C}} \mid \sigma_0 \in \mathsf{C}^0 \wedge \forall i \in [0, n[\,.\, \langle \sigma_i,\ \sigma_{i+1}\rangle \in \tau \wedge \forall c \,.\, \langle \sigma_n,\ c\rangle \notin \tau\} \\ [\![\tau]\!]^+ &\triangleq \bigcup_{n \in \mathbb{N}} [\![\tau]\!]^n \\ [\![\tau]\!]^\infty &\triangleq \{\sigma \in \mathsf{T}^\infty_{\mathsf{C}} \mid \sigma_0 \in \mathsf{C}^0 \wedge \forall i \in \mathbb{N} \,.\, \langle \sigma_i,\ \sigma_{i+1}\rangle \in \tau\} \\ [\![\tau]\!] &\triangleq [\![\tau]\!]^+ \cup [\![\tau]\!]^\infty \end{aligned} \tag{23}$$

The trajectories of $[\![\tau]\!]$ are maximal, that is,

$$\max([\![\tau]\!]) = [\![\tau]\!] \tag{24}$$

*Example 3 (Water tank automaton [17]).* Continuing example 2, the water tank specification can be implemented as described by the following hybrid automaton.

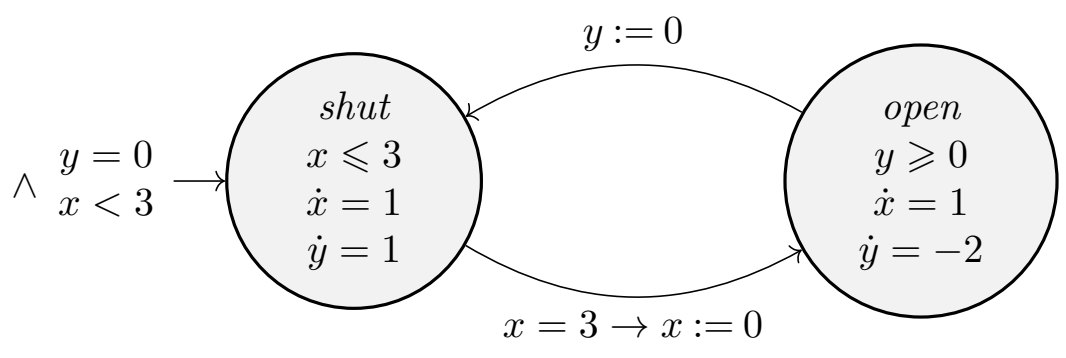

As soon as the tank is empty, the valve is shut down. The valve is reopened after 3 units of time.

The states, configurations, initial configurations, and transitions are (we write $\dot{x}$ for the derivative $\frac{dx}{dt}$ of the everywhere differentiable (hence continuous) real-valued function $x(t)$ of the time $t$).

$$\begin{aligned} \mathsf{S} &\triangleq \{\mathit{open}, \mathit{shut}\} \times \mathbb{R} \times \mathbb{R} \\ \mathsf{C}^{\mathit{shut}} &\triangleq \{\langle f,\ [t_1, t_2[\rangle \mid \exists x, y \,.\, \forall t \in [t_1, t_2] \,.\, f(t) = \langle \mathit{shut},\ x(t),\ y(t)\rangle \wedge \\ &\qquad (t = t_1 \Longrightarrow y(t) = 0) \wedge x(t) \leqslant 3 \wedge (x(t) = 3 \Longrightarrow t = t_2) \\ &\qquad \wedge \dot{x}(t) = 1 \wedge \dot{y}(t) = 1\} \\ \mathsf{C}^{\mathit{open}} &\triangleq \{\langle f,\ [t_1, t_2[\rangle \mid \exists x, y \,.\, \forall t \in [t_1, t_2] \,.\, f(t) = \langle \mathit{open},\ x(t),\ y(t)\rangle \wedge \\ &\qquad (t = t_1 \Longrightarrow x(t) = 0) \wedge y(t) \geqslant 0 \wedge (y(t) = 0 \Longrightarrow t = t_2) \\ &\qquad \wedge \dot{x}(t) = 1 \wedge \dot{y}(t) = -2\} \\ \mathsf{C} &\triangleq \mathsf{C}^{\mathit{shut}} \cup \mathsf{C}^{\mathit{open}} \end{aligned}$$

$$\mathsf{C}^0 \triangleq \{\langle f,\ [0,t[\rangle \in \mathsf{C}^{shut} \mid t > 0 \wedge \exists x < 3\ .\ f(0) = \langle shut,\ x,\ 0\rangle\} \in \mathsf{C}^{shut}$$
$$\tau^3 \triangleq (\mathsf{C}^{shut} \times \mathsf{C}^{open}) \cup (\mathsf{C}^{open} \times \mathsf{C}^{shut}) \text{ as restricted by (22)} \tag{25}$$

Notice that the final time $t_2$ is not part of the time interval of configurations in $\mathsf{C}^{shut}$ but, by (22), the starting time of the next configuration in $\mathsf{C}^{open}$. Therefore, at that time the value of $x$ is 0, not 3. So in this example, $f$ is continuous on $[t_1, t_2[$ that is continuous on $]t_1, t_2[$ and right continuous at $t_1$. Same for $y$ in $\mathsf{C}^{open}$. An example of execution is given in figure (5.b). The hybrid semantics $[\![\tau^3]\!]$ of the water tank automaton is given by (23). □

**Lemma 1.** [2]
*If* $\tau \subseteq \tau'$ *and the* blocking condition *holds, i.e.*

$$\forall c\ .\ (\forall c'\ .\ \langle c,\ c'\rangle \notin \tau) \Longrightarrow (\forall c'\ .\ \langle c,\ c'\rangle \notin \tau') \tag{26}$$

*then* $[\![\tau]\!] \subseteq [\![\tau']\!]$.

(so if $[\![\tau']\!]$ has trajectory property $P \in \wp(\mathsf{T}_{\mathsf{C}}^{+\infty})$ (i.e. $[\![\tau']\!] \subseteq P$) then lemma 1 implies that $[\![\tau]\!]$ has the same property $P$.)

Observe that the transition of one configuration to the next in (22) requires the specification of the time at which the next configuration will terminate. As shown by the water tank automaton example 3 of [17], this is not a problem when the duration of the configuration is specified by a condition on the flow.

## 5 Trajectory-based hybrid trajectory semantics abstraction

In many program verification and refinement methods, the hybrid semantics is abstracted or concretized to simplify soundness and completeness proofs. One way of simplifying the proofs is to reason on an abstraction of trajectories, by applying an homomorphic abstraction (5) to these trajectories.

A classic example is sampling in signal processing, to reduce a continuous-time signal to a discrete-time signal. For an hybrid semantics, this is defined as follows.

Let $\delta > 0$ be a sampling interval (see [29, Ch. 9] for an adequate choice of the sampling rate). Define

$$h_\delta(\sigma) \triangleq \langle \sigma_{n\delta},\ n \in \mathbb{N} \wedge n\delta \leqslant [\![\sigma[\!]\rangle \tag{27}$$
$$\alpha_\delta(T) \triangleq \{h_\delta(\sigma) \mid \sigma \in T\}$$

which, by (5), is an homomorphic Galois connection

$$\langle \wp(\mathsf{T}_{\mathsf{C}}^{+\infty}),\ \subseteq\rangle \xleftarrow[\alpha_\delta]{\gamma_\delta} \langle \wp(\mathsf{T}_{\mathsf{S}}^{+\infty}),\ \subseteq\rangle \tag{28}$$

[2] Underlined equation or theorem numbers link to proofs given in the appendix .

where $\gamma_\delta(\Theta) \triangleq \{\sigma \in \mathsf{T}_{\mathsf{C}}^{+\infty} \mid h_\delta(\sigma) \in \Theta\}$.

(In general a trajectory $\sigma$ cannot be regained from its discretization $h_\delta(\sigma)$. This might be possible under specific hypotheses. For example, the Nyquist–Shannon sampling theorem [28,33] establishes a sufficient condition for a sample rate that permits a discrete sequence of samples to capture all the information from a continuous-time signal of finite bandwidth.)

Trajectory based abstractions are useful to prove trajectory properties of hybrid systems by considering one possible trajectory at a time (but inadequate to prove (hyper) properties relating two of more trajectories). But reasoning on a complete trajectory is often complicated, in which case local reasonings relating states or transitions locally are preferred.

## 6 State-based hybrid trajectory semantics abstraction

Since reasoning on discrete execution traces (hence on hybrid trajectories) is difficult, a number of proof techniques have been developed to reduce the reasoning on trajectories to reasonings on states (or pairs of states, that is transitions). Examples are discrete simulations that we extend to hybrid trajectories (and bisimulation [26] as well as preservation with progress [38] considered in the appendix). Sampling in (28) is a counter example since, in general, sampling must be defined by reasoning on trajectories, not states and transitions.

Our objective is to show that a relation between states can be extended to configurations, then to trajectories, and then to hybrid semantics (independently of whether trajectories are generated by transition systems or not).

### 6.1 Relation between states

For timed trajectories, the relation $r$ between concrete states $\mathsf{S}$ to abstract states $\overline{\mathsf{S}}$ is a function of the time.

$$r \in \mathbb{R}_{\geqslant 0} \to \wp(\mathsf{S} \times \overline{\mathsf{S}}) \tag{29}$$

For simplicity, we assume $r$ to be a total function of the time. If necessary, a partial function could be encoded using an undefined element (like $\perp$ in denotational semantics).

### 6.2 Relation between configurations

Let us define a partial relation between configurations with related states

$$\gamma(r) \triangleq \{\langle\langle f,\, i\rangle,\, \langle\overline{f},\, \overline{i}\rangle\rangle \mid i \cap \overline{i} \neq \emptyset \wedge \forall t \in i \cap \overline{i}\;.\; \langle f(t),\, \overline{f}(t)\rangle \in r(t)\} \tag{30}$$

(which is said to be total when $i = \overline{i}$ e.g. for homomorphic abstractions or well-nested when $i \subseteq \overline{i}$). Define

$$\alpha(R) \triangleq \boldsymbol{\lambda}\, t \bullet \{\langle f(t),\, \overline{f}(t)\rangle \mid \exists i, \overline{i}\;.\; t \in i \cap \overline{i} \wedge \langle\langle f,\, i\rangle,\, \langle\overline{f},\, \overline{i}\rangle\rangle \in R\} \tag{31}$$

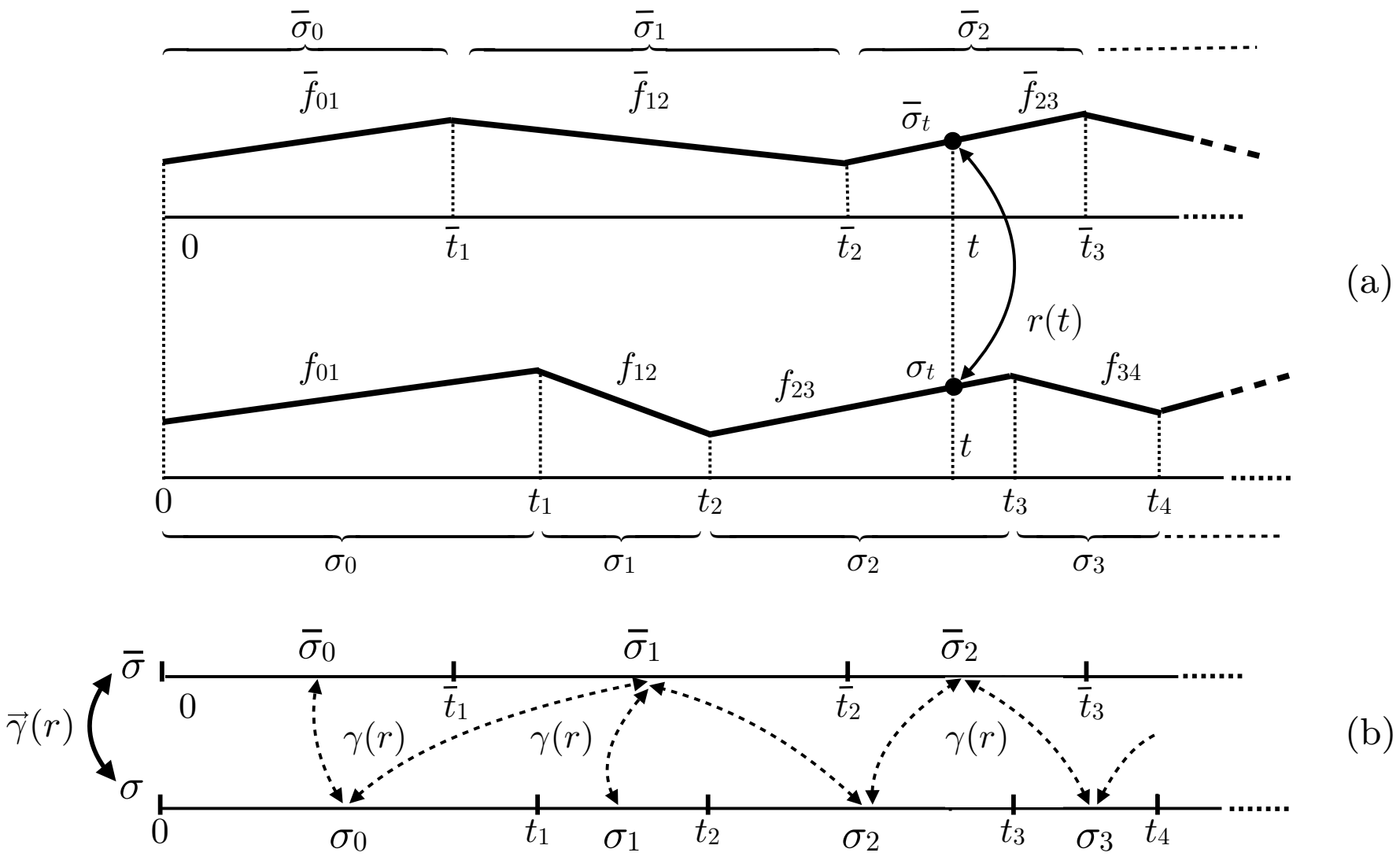

**Fig. 1.** Relations $r$ between states in (a), and relations $\gamma(r)$ between configurations and $\vec{\gamma}(r)$ between trajectories in (b)

Define the set of all relations between overlapping configurations as

$$\mathsf{R_C} \triangleq \{R \in \wp(\mathsf{C} \times (\mathsf{C} \cup \mathsf{cl}(\mathsf{C}))) \mid \forall \langle\langle f,\, i\rangle,\, \langle \bar{f},\, \bar{i}\rangle\rangle \in R \,.\, i \cap \bar{i} \neq \emptyset\} \tag{32}$$

We have a Galois isomorphism

$$\langle \mathsf{R_C},\, \subseteq\rangle \underset{\alpha}{\overset{\gamma}{\leftrightarrows}} \langle \mathbb{R}_{\geqslant 0} \to \wp(\mathsf{S} \times \mathsf{S}),\, \dot{\subseteq}\rangle \tag{33}$$

where $\dot{\subseteq}$ is the pointwise extension of set inclusion $\subseteq$. So when abstracting trajectories by abstraction of their configurations, we can equivalently start from a relation $r$ between states and use the relation $\gamma(r) \in \mathsf{R_C}$ or start from a relation between configurations $R \in \mathsf{R_C}$ which induces a relation $\alpha(R)$ between states. In discrete systems, the two notions of state and configuration coincide.

### 6.3 Relation between trajectories

Let us also define a relation between trajectories so as to relate states of trajectories

$$\vec{\gamma}(r) \triangleq \{\langle \sigma,\, \bar{\sigma}\rangle \mid \forall t \in [0, \min(\llbracket \sigma \rrbracket, \llbracket \bar{\sigma} \rrbracket)[\ .\ \langle \sigma_t,\, \bar{\sigma}_t\rangle \in r(t)\} \tag{34}$$

as illustrated in figure (1.a) Notice that in the definition (34) of related trajectories, we do not use the relation $\gamma(r)$ in (30) between configurations since the states of the configurations with the same ranks in the concrete and abstract in trajectories may be unrelated while the timings are (i.e. the concrete and abstract configurations of same rank $k$ may not even overlap in time). However, we have the following equivalent definition using ranks of configurations in trajectories, as illustrated in figures (1.b) and (2).

$$\vec{\gamma}(r) \triangleq \vec{\gamma}_c(r) \cap \vec{\gamma}_a(r) \quad (35)$$
$$\vec{\gamma}_c(r) \triangleq \{\langle\sigma,\, \overline{\sigma}\rangle \mid \forall j < |\sigma| \,.\, (\mathsf{e}(\sigma_j) \leqslant [\![\overline{\sigma}]\!]) \Longrightarrow (\exists k < |\overline{\sigma}| \,.\, \langle\sigma_j,\, \overline{\sigma}_k\rangle \in \gamma(r))\} \quad \text{(a)}$$
$$\vec{\gamma}_a(r) \triangleq \{\langle\sigma,\, \overline{\sigma}\rangle \mid \forall k < |\overline{\sigma}| \,.\, (\mathsf{e}(\overline{\sigma}_k) \leqslant [\![\sigma]\!]) \Longrightarrow (\exists j < |\sigma| \,.\, \langle\sigma_j,\, \overline{\sigma}_k\rangle \in \gamma(r))\} \quad \text{(b)}$$

(By the isomorphism (33), there is a definition equivalent to (35) using $R \in \mathsf{R_C}$ instead of $\gamma(r)$.)

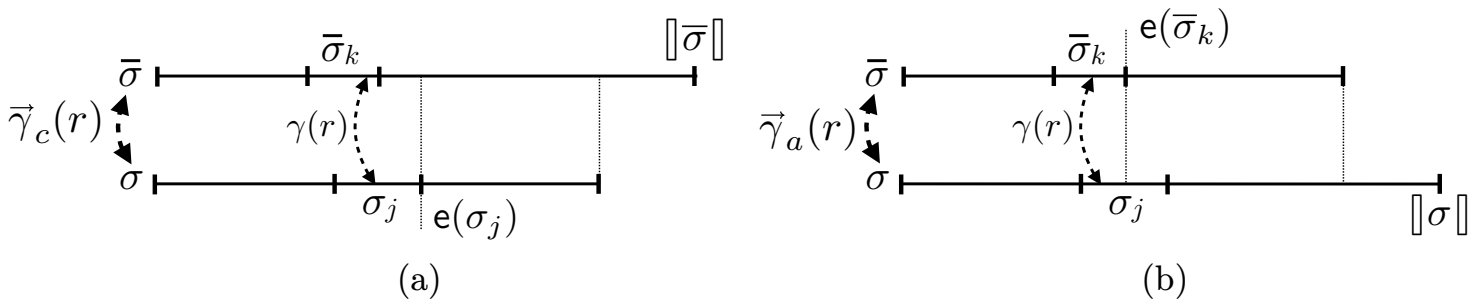

**Fig. 2.** Relations $\vec{\gamma}_c(r)$ and $\vec{\gamma}_a(r)$ between traces

Defining $\vec{\alpha}(\vec{R}) \triangleq \boldsymbol{\lambda}\, t \bullet \{\langle\sigma_t,\, \overline{\sigma}_t\rangle \mid \langle\sigma,\, \overline{\sigma}\rangle \in \vec{R} \wedge t \in [0, \min([\![\sigma]\!], [\![\overline{\sigma}]\!])[\}$, this is a Galois connection

$$\langle\wp(\mathsf{T}_{\mathsf{C}}^{+\infty} \times \mathsf{T}_{\overline{\mathsf{C}}}^{+\infty}),\ \subseteq\rangle \xleftrightarrow[\vec{\alpha}]{\vec{\gamma}} \langle\mathbb{R}_{\geqslant 0} \rightarrow \wp(\mathsf{S} \times \overline{\mathsf{S}})),\ \dot{\subseteq}\rangle \quad (36)$$

### 6.4 Relation between hybrid trajectory semantics

The abstraction (36) is then extended to hybrid trajectory semantics through a preorder, a common one being the overapproximation in verification ($\overline{T}$ is an abstraction of $T$ since $\overline{T}$ has more possible behaviors than $T$) and underapproximation in refinement ($T$ is a refinement of $\overline{T}$ since $T$ has less behaviors that the specification $\overline{T}$), that is

$$\begin{aligned}\vec{\gamma}(R) &\triangleq \{\langle T,\, \overline{T}\rangle \mid T \subseteq \mathsf{pre}[R]\overline{T}\} \quad (37)\\ &= \{\langle T,\, \overline{T}\rangle \mid \forall\sigma \in T \,.\, \exists\overline{\sigma} \in \overline{T} \,.\, \langle\sigma,\, \overline{\sigma}\rangle \in R\}\end{aligned}$$

Defining $\vec{\alpha}(P) \triangleq \{\langle\sigma,\, \overline{\sigma}\rangle \mid \exists\overline{T} \,.\, \langle\{\sigma\},\, \overline{T}\rangle \in P \wedge \overline{\sigma} \in \overline{T}\}$, we have the Galois connection

$$\langle\{\langle T,\, \overline{T}\rangle \in \wp(\mathsf{T}_{\mathsf{C}}^{+\infty}) \otimes \wp(\mathsf{T}_{\overline{\mathsf{C}}}^{+\infty}) \mid \overline{T} = \emptyset \Longrightarrow T = \emptyset\},\ \supseteq\rangle \xleftrightarrow[\vec{\alpha}]{\vec{\gamma}} \langle\wp(\mathsf{T}_{\mathsf{C}}^{+\infty} \times \mathsf{T}_{\overline{\mathsf{C}}}^{+\infty}),\ \supseteq\rangle \quad (38)$$

where by (37), (4), and (8), the concrete domain is a tensor product.

*Example 4 (The water tank automaton is a state-based refinement of the specification).* Le us define the state-based relation

$$r^{(39)}(t) \triangleq \{\langle\langle v,\ x,\ y\rangle,\ \langle v,\ y\rangle\rangle \mid v \in \{\mathit{shut}, \mathit{open}\} \wedge x, y \in \mathbb{R}\} \quad (39)$$

between states (19) and (25) of the water tank specification and automaton.

This induces a relation (30) between configurations, as follows.

$$\gamma(r^{(39)}) = \{\langle\langle \boldsymbol{\lambda}\, t \bullet \langle v(t),\ x(t),\ y(t)\rangle,\ i\rangle,\ \langle \boldsymbol{\lambda}\, t \bullet \langle \overline{v}(t),\ \overline{y}(t)\rangle,\ \overline{i}\rangle\rangle \mid i \cap \overline{i} \neq \emptyset \wedge \\ \forall t \in i \cap \overline{i}\ .\ v(t) = \overline{v}(t) \in \{\mathit{shut}, \mathit{open}\} \wedge y(t) = \overline{y}(t)\} \tag{40}$$

i.e. at any time in overlapping configurations, the water height and the state of the valve coincide. This induces a relation (35) between trajectories, as follows.

$$\vec{\gamma}(r^{(39)}) = \mathsf{let}\ \rho(c, \overline{c}) \triangleq \exists v, x, y, i, \overline{v}, \overline{y}, \overline{i}\ .\ c = \langle \boldsymbol{\lambda}\, t \bullet \langle v(t),\ x(t),\ y(t)\rangle,\ i\rangle \wedge \\ \overline{c} = \langle \boldsymbol{\lambda}\, t \bullet \langle \overline{v}(t),\ \overline{y}(t)\rangle,\ i\rangle \wedge i \cap \overline{i} \neq \emptyset \wedge \forall t \in i\ \cap i\ .\ v(t) = \overline{v}(t) \wedge \\ y(t) = \overline{y}(t)\ \mathsf{in} \\ \{\langle \sigma,\ \overline{\sigma}\rangle \mid \forall j < |\sigma|\ .\ (\mathsf{e}(\sigma_j) \leqslant [\![\overline{\sigma}]\!]) \Longrightarrow (\exists k < |\overline{\sigma}|\ .\ \rho(\sigma_j, \overline{\sigma}_k) \wedge \\ \forall k < |\overline{\sigma}|\ .\ (\mathsf{e}(\overline{\sigma}_k) \leqslant ]\!]\sigma]\!]) \Longrightarrow (\exists j < |\sigma|\ .\ \rho(\sigma_j, \overline{\sigma}_k)\} \tag{41}$$

Let us prove that the hybrid semantics $[\![\tau^3]\!]$ (25) of the water tank automaton of example 3 is a state based refinement of the water tank specification $\mathcal{S}^2$ of example 2 for $r^{(39)}$ in (39) (denoted $r^{(39)}$ to avoid confusions), meaning that

$$\langle [\![\tau^3]\!],\ \mathcal{S}^2\rangle \in \vec{\gamma}(\overline{\gamma}(r^{(39)}))$$

or equivalently

$$\forall \sigma \in [\![\tau^3]\!]\ .\ \exists \overline{\sigma}\ .\ P(\overline{\sigma}) \wedge \forall t \geqslant 0\ .\ \sigma(t).y = \overline{\sigma}(t).y \wedge \sigma(t).v = \overline{\sigma}(t).v \tag{42}$$

By definition (20) of $P$, we have to show that $\forall t \in \mathbb{R}_{\geqslant 0}\ .\ 0 \leqslant \sigma(t).y \leqslant 3$. In a shut configuration of $\mathsf{C}^{shut}$, $y(t) = 0$ at the beginning, $y$ evolves as the same rate as $x$, and $x(t)$ is bounded by 3 so that that $y(t)$ is also bounded by 3. By definition of initial configurations $\mathsf{C}^0$, any trajectory of $[\![\tau^3]\!]$ starts with a *shut* configuration, and so, by definition (25) of the transition relation $\tau^3$, any open configuration of $\mathsf{C}^{open}$ follows a *shut* configuration. At the end $t$ of this *shut* configuration, and so at the beginning $t$ of the following *open* configuration, we have shown that $\sigma(t).y \leqslant 3$. In the *open* configuration, $y$ decreases by $\dot{y} = -2$ and remains positive, so the invariant holds.

Moreover, we must show that if the valve remains opened, then $y$ decreases. If $\forall t \in [t_1, t_2]\ .\ \sigma(t).v = \mathit{open}$ then $t$ is within an open configuration, so $\dot{y} = -2$ implies that $y$ decreases between $t_1$ and $t_2$. Similarly, if $\forall t \in [t_1, t_2]\ .\ \sigma(t).v = \mathit{shut}$ then $t$ is within a shut configuration, so $\dot{y} = 1$ implies that $y$ increases.

Finally, if at some point $t$ of time, $y(t) = 0$ then if we are in an *open* configuration, the system instantaneously moves to a *shut* configuration which last at least $\zeta$ by the nonzeno hypothesis, and so, by $\dot{y} = 1$, we have $\sigma(t+\zeta).y > 0$. □

## 7 Transition-based hybrid trajectory semantics abstraction

Reasonings on trajectories is often considered difficult and reasonings involving only one computation step at a time are preferred. An example is Turing/Naur/Floyd/Hoare invariance proof method where verification conditions involve only one computation step at a time.

So we assume that the concrete and abstract semantics are generated by transitions systems $\langle \mathsf{C},\ \mathsf{C}^0,\ \tau\rangle$ and $\langle \overline{\mathsf{C}},\ \overline{\mathsf{C}}^0,\ \overline{\tau}\rangle$, that is $T = [\![\tau]\!]$ and $\overline{T} = [\![\overline{\tau}]\!]$, and, given a relation (29) between states, we study relations between transition relations which enable us to define relations (34) between trajectories hence relations (38) between trajectory semantics. In the literature of abstraction of discrete transition systems, basic state and transition-based abstractions are homomorphisms, simulations, bisimulations, and preservations with progress, which we extend to hybrid transition systems, adding discretization.

### 7.1 Homomorphisms

Homomorphisms are the case when relation $r$ in (29) is given by a function $h(t) \in \mathsf{S} \to \overline{\mathsf{S}}$ at time $t$. Following (30), the homomorphism is extended to configurations as

$$\alpha_h(\langle f,\ i\rangle) \triangleq \langle h \circ f,\ i\rangle \tag{43}$$

The function $h$ is composed with the flow and the timings remain the same. The extension to trajectories is

$$\alpha_h(\langle \sigma_i,\ i \in [0, |\sigma|[\rangle) \triangleq \langle \alpha_h(\sigma_i),\ i \in [0, |\sigma|[\rangle \tag{44}$$

and to trajectory semantics

$$\alpha_h(T) \triangleq \{\alpha_h(\sigma) \mid \sigma \in T\} \tag{45}$$

which, by (5), is a Galois connection

$$\langle \wp(\mathsf{T}_{\mathsf{C}}^{+\infty}),\ \subseteq\rangle \underset{\alpha_h}{\overset{\gamma_h}{\leftrightarrows}} \langle \wp(\mathsf{T}_{\overline{\mathsf{C}}}^{+\infty}),\ \subseteq\rangle \tag{46}$$

The homomorphic abstraction of a transition system is

$$\alpha_h(\langle \mathsf{C},\ \mathsf{C}^0,\ \tau\rangle) \triangleq \langle \{h(c) \mid c \in \mathsf{C}\},\ \{h(c) \mid c \in \mathsf{C}^0\},\ \{\langle h(c),\ h(c')\rangle \mid \langle c,\ c'\rangle \in \tau\}\rangle \tag{47}$$

For brevity, we write $\alpha_h(\tau)$ for $\alpha_h(\langle \mathsf{C},\ \mathsf{C}^0,\ \tau\rangle)$. The homomorphic abstraction of the trajectory semantics generated by the concrete transition system is the abstract trajectory semantics generated by the homomorphic abstraction of the concrete transition system

**Theorem 1.**

$$\alpha_h([\![\tau]\!]) = [\![\alpha_h(\tau)]\!] \tag{48}$$

The *verification* of a property of an hybrid system $[\![\tau]\!]$ defined by a transition relation $\tau$ can be done in the abstract, as follows.

**Theorem 2.** *For any abstract hybrid trajectory property $\overline{P} \in \wp(\mathsf{T}_{\overline{\mathsf{C}}}^{+\infty})$,*

$$\frac{\alpha_h(\tau) \subseteq \overline{\tau}, \quad (26), \quad [\![\overline{\tau}]\!] \subseteq \overline{P}}{[\![\tau]\!] \subseteq \gamma_h(\overline{P})} \tag{49}$$

i.e. a sound abstract small-step semantics $\overline{\tau}$ overapproximating the concrete semantics $\tau$ is designed so that the concretization $\gamma_h(\overline{P})$ of its trace properties $\overline{P}$ holds for the concrete semantics $[\![\tau]\!]$.

Given a specification in the form of an abstract transition system $\overline{\tau}$, *refinement* consists in designing a concrete transition system $\tau$ such that $[\![\tau]\!] \subseteq \gamma_h(\{[\![\overline{\tau}]\!]\})$. By induction principle (49) where $\overline{P} = \{[\![\overline{\tau}]\!]\}$, it is sufficient to ensure that $\alpha_h(\tau) \subseteq \overline{\tau}$ and the blocking condition (26).

Finally, the homomorphic abstraction is preserved by discretization (27).

**Theorem 3.**

$$\alpha_\delta(\alpha_h(T)) = \alpha_h(\alpha_\delta(T)) \tag{50}$$

In conclusion of this section 7.1, homomorphic abstractions are very simple since they compose (because $h(t) \in \mathsf{S} \to \overline{\mathsf{S}}$ and $\overline{h}(t) \in \overline{\mathsf{S}} \to \overline{\overline{\mathsf{S}}}$ implies $\overline{h}(t) \circ h(t) \in \mathsf{S} \to \overline{\overline{\mathsf{S}}}$), there is a unique best abstract homomorphic abstract hybrid semantics (by (46)), they extend from hybrid transition systems to hybrid semantics (by theorem 1), allow proofs of trajectory properties by abstraction (by theorem 2), and are preserved by discretization (by theorem 3). Homomorphic abstractions seem to be almost the only ones considered in model-checking [5, pp. 499–504].

However, homomorphic abstractions are very restrictive in that the relation between flows is deterministic and the concrete and abstract timelines must be exactly the same[3].

### 7.2 Simulations

Simulations were introduced by Robin Milner [26] to relate discrete transitions systems hence, implicitly, their trace semantics or abstractions of these trace semantics. They have been used for program verification and refinement. Notice that Robin Milner originally used (bi)simulation relations to abstract reachability/invariance properties for which reasoning on transitions and their reflexive closure is sound and complete. So there was no need to consider (bi)similar traces.

Various extensions to continuous and hybrid systems have been proposed such as [23,3,31,13,14,25,15,12,15,16,35,6,36,9] among others. In contrast with this previous work, our definition of (bi)simulation takes into account the fact that concrete and abstract trajectories may have different durations and not necessarily comparable timelines for mode changes.

[3] One could argue that the time-evolution low abstraction of (17) applied to the concrete and abstract trajectories would solve the problem of having the same timeline by merging the trajectories into a single configuration, but then the original timelines are hidden in the flow functions, which does not make time-dependent reasonings simpler.

**Definition of asynchronous hybrid simulations.** A relation $R \in \wp(\mathsf{C} \times \overline{\mathsf{C}})$ between concrete and abstract configurations (which can be the extension (30) $R = \gamma(r)$ of the timed relation $r$ between states in (29)) is a *hybrid simulation* between the transition relations $\tau$ and $\overline{\tau}$ if and only if

$$\begin{aligned}&\forall c, \overline{c}, c' \,.\, \exists \overline{c}' \,.\, (\langle c,\ \overline{c}\rangle \in R \wedge (\langle c,\ c'\rangle \in \tau \vee c' = \varepsilon)) \Longrightarrow \\ &\quad ((\langle \overline{c},\ \overline{c}'\rangle \in \overline{\tau} \vee \overline{c}' = \varepsilon) \wedge \langle c \mathbin{\text{;}} c'(\!|\min(\mathsf{b}(c'), \mathsf{b}(\overline{c}')), \min(\mathsf{e}(c'), \mathsf{e}(\overline{c}'))|\!), \\ &\qquad\qquad \overline{c} \mathbin{\text{;}} \overline{c}'(\!|\min(\mathsf{b}(c'), \mathsf{b}(\overline{c}')), \min(\mathsf{e}(c'), \mathsf{e}(\overline{c}'))|\!)\rangle \in R)\end{aligned} \tag{51}$$

To simplify notations, we write $c' = \varepsilon$ for $\mathsf{e}(\overline{c}') \leqslant \mathsf{e}(c) \wedge c' = \varepsilon$ and similarly $\overline{c}' = \varepsilon$ stands for $\mathsf{e}(c') \leqslant \mathsf{e}(\overline{c}) \wedge \overline{c}' = \varepsilon$, see figure 3.

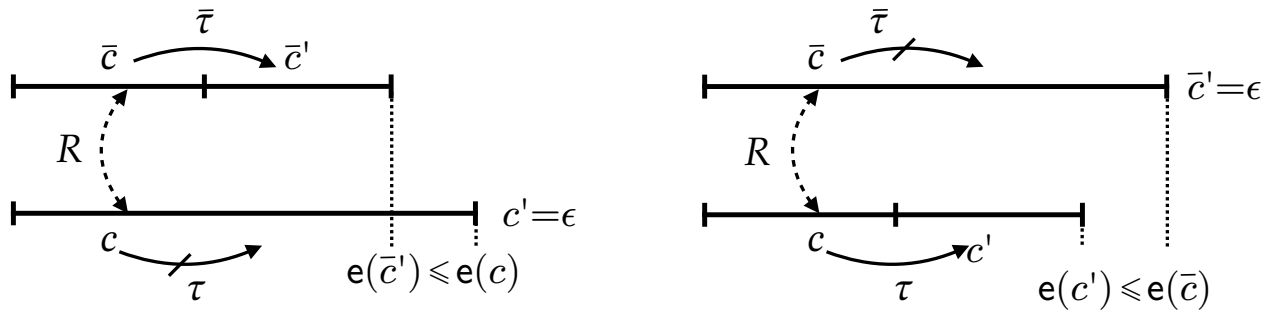

**Fig. 3.** Empty successor configurations

As shown in figure 4, this definition of an *asynchronous simulation* takes into account the fact that the concrete and abstract configurations may correspond to different timelines. Concrete and abstract configurations $\langle c, \overline{c}\rangle \in R$ are related and there is a concrete transition $\langle c,\ c'\rangle \in \tau$ from $c$ to $c'$ so there must exists an abstract transition $\langle \overline{c},\ \overline{c}'\rangle \in \overline{\tau}$ such that $c'$ and $\overline{c}'$ are related. But since $c'$ and $\overline{c}'$ may have different timings, one of them is extended in the past by the previous configuration ($\overline{c}'$ extended to $t_2 = \min(t_2, \overline{t}_2) = \min(\mathsf{b}(c'), \mathsf{b}(\overline{c}'))$ using the previous $\overline{c}$ in figure 4) while one of them, maybe the same, is truncated in the future to the first terminating configuration ($c'$ truncated to $\overline{t}_3 = \min(t_3, \overline{t}_3) = \min(\mathsf{e}(c'), \mathsf{e}(\overline{c}'))$ in figure 4).

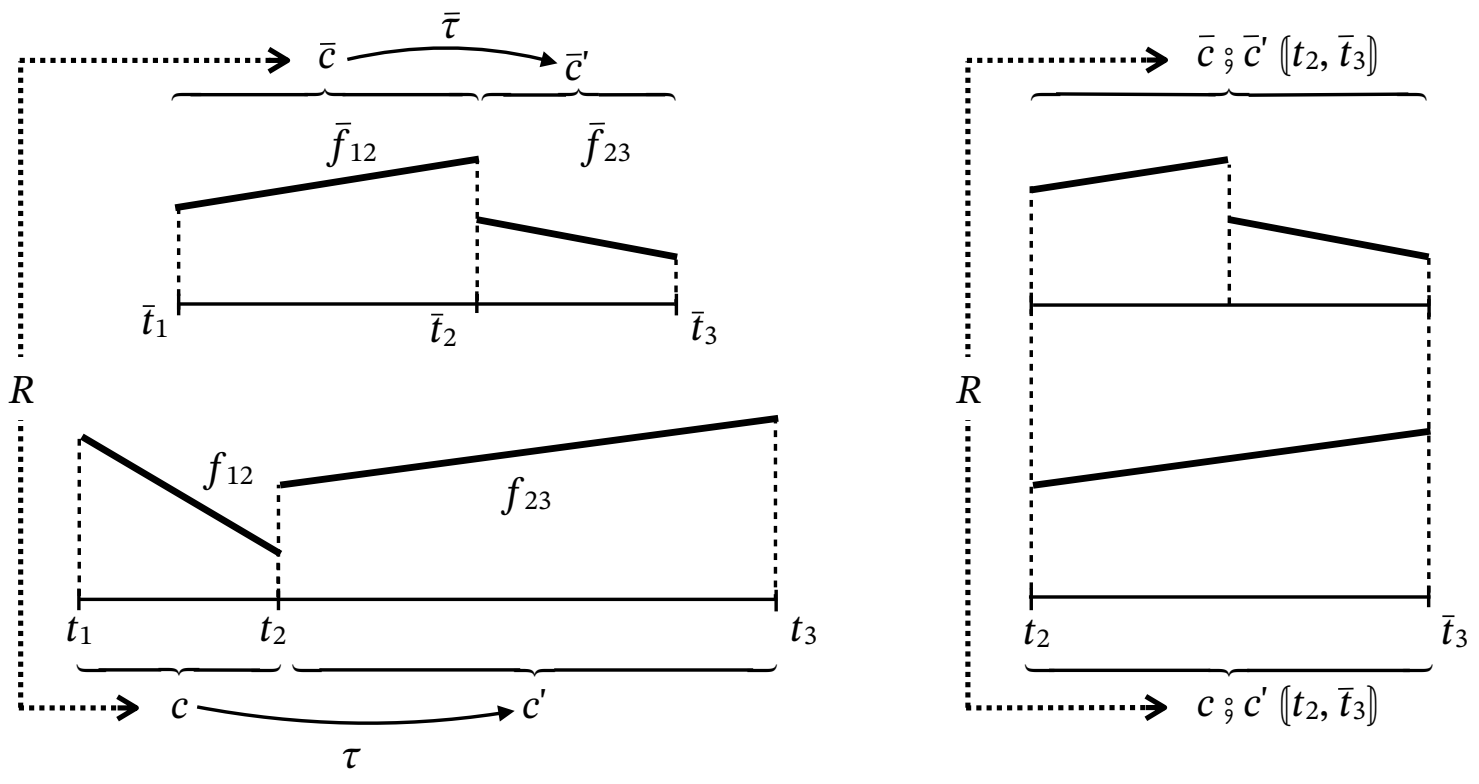

**Fig. 4.** Asynchronous hybrid simulation

The simulation $R = \{\langle c, \alpha_h(c)\rangle \mid c \in \mathsf{C}\}$ may be the homomorphic abstraction (43) which would enforce the concrete and abstract timings to be the same. More generally, the simulation $R$ may be many-to-many. For example the concrete states can be equipped with a distance and $R$ would ensure that, at each time instant, the concrete state is in a ball around the abstract state [9,16], the size of the ball evolving over time (this would, for example, account for cumulated rounding errors when the abstract states are reals and the refined concrete states are floats).

Another particular case is that of a *synchronous simulation* of well-nested configurations when concrete timelines are subdivisions of the abstract timelines (that is, if $\langle f,\ i\rangle \in \mathsf{C}$, $\langle \overline{f},\ \overline{i}\rangle \in \overline{\mathsf{C}}$, and $i \cap \overline{i} \neq \emptyset$ then $\mathsf{b}(\overline{i}) \leqslant \mathsf{b}(i) < \mathsf{e}(i) \leqslant \mathsf{e}(\overline{i})$). If moreover all configurations have at least one successor, there are no blocking configurations so that (51) becomes

$$\forall c, \overline{c}, c' \;.\; \exists \overline{c}' \;.\; (\langle c,\ \overline{c}\rangle \in R \wedge (\langle c,\ c'\rangle \in \tau)) \Longrightarrow \\ ((\langle \overline{c},\ \overline{c}'\rangle \in \overline{\tau} \vee \overline{c}' = \varepsilon) \wedge \langle c',\ \overline{c}'(\!|\mathsf{b}(c'), \mathsf{e}(c')|\!)\rangle \in R) \tag{52}$$

*Example 5 (Change of variables).* Let $\langle c_i,\ i \in [0, |c|[\rangle$ be a concrete semantics with concrete configurations $c_i = \langle \boldsymbol{\lambda}\, t \bullet f_i(t - t_i^\ell),\ [t_i^\ell, t_i^h[\rangle$ where $f_i(t)$ is given by the Cauchy-Euler implicit ordinary differential (ODE) equation $t^2 f_i''(t) + a_i t f_i'(t) + b_i f_i(t) = 0$. Under appropriate continuity hypotheses, a classic resolution method [30, ch.19, p. 170] consists in applying the change of variable $t = \ln(\overline{t})$, that is $\overline{t} = e^t$ to get $\varphi_i(\overline{t})$ solution of $\varphi_i''(\overline{t}) + (a_i - 1)\varphi_i'(\overline{t}) + b_i \varphi_i(\overline{t}) = 0$ which is a linear ODE solved via its characteristic polynomial. Let the abstract hybrid semantics be $\langle \overline{c}_i,\ i \in [0, |c|[\rangle$ with abstract configurations $\overline{c}_i = \langle \boldsymbol{\lambda}\, \overline{t} \bullet \varphi_i(\overline{t} - e^{t_i^\ell}),\ [e^{t_i^\ell}, e^{t_i^h}[\rangle$. This is a hybrid simulation (indeed a bisimulation) $\gamma(r)$ for $r(t) = \{\langle f_i(t - t_i^\ell),\ \varphi_i(e^t - e^{t_i^\ell})\rangle \mid i \in [0, |c|[ \wedge t \in [t_i^\ell, t_i^h[\}$. □

*Example 6.* Continuing the water tank automaton example 3, we refine the tank specification by taking some time $\epsilon$ to close (in *off* configuration) and open (in *on* configuration) the valve while in *shut* mode. We assume $\epsilon > \zeta$ to ensure that the duration of the valve opening and closing is not infinitesimal. The water inflow $\dot{y}$ is increased to compensate for this delay. We assume that $\epsilon$ is large enough for the valve opening and closing to be mechanically feasible in this period of time. We assume that $\epsilon$ is small enough so that the duration of the *shut* configuration in (25) is much larger than $2\epsilon$. In particular it must be chosen so that the increase of the inflow is physically possible.

The *open* mode is unchanged, as shown in figure (5.a). A formal definition of example 6 is given in the appendix.

The relation $R^{(53)}$ between configurations of concrete trajectories in (5.a) and the configurations of abstract trajectories in (5.b) is the following.

$$R^{(53)} \triangleq \big\{\langle\langle \boldsymbol{\lambda}\, t \bullet \langle m_t,\ x_t,\ y_t\rangle,\ [t_1, t_2[\rangle,\ \langle \boldsymbol{\lambda}\, t \bullet \langle \overline{m}_t,\ \overline{x}_t,\ \overline{y}_t\rangle,\ [\overline{t}_1, \overline{t}_2[\rangle\rangle \mid \\ P^{(53)}(m, x, y, t_1, t_2, \overline{m}, \overline{x}, \overline{y}, \overline{t}_1, \overline{t}_2)\big\} \tag{53}$$

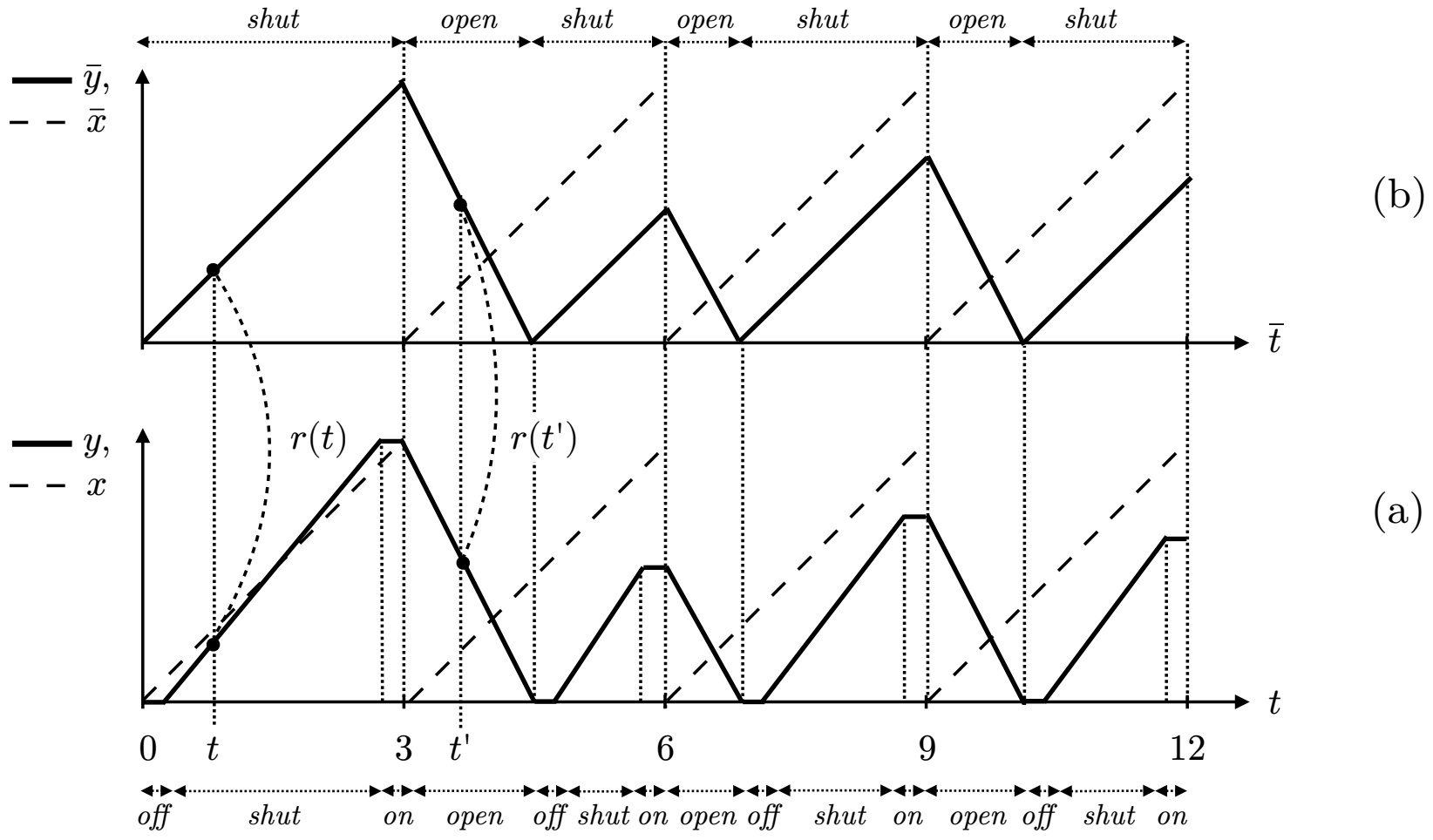

**Fig. 5.** Concrete (a) and abstract (b) tank trajectories

$$
\begin{aligned}
&P^{(53)}(m,x,y,t_1,t_2,\overline{m},\overline{x},\overline{y},\overline{t}_1,\overline{t}_2) \triangleq \forall t \in [\overline{t}_1,\overline{t}_2[\ .\ x_t = \overline{x}_t \wedge \qquad (54)\\
&\big((\overline{m}_t = \mathit{shut} \wedge\\
&\quad ((m_t = \mathit{off} \wedge y = 0 \wedge \overline{y}_t = t - t_1 \wedge t_1 = \overline{t}_1 \wedge t_2 = \overline{t}_1 + \epsilon)\\
&\quad \vee (m_t = \mathit{shut} \wedge t_1 = \overline{t}_1 + \epsilon \wedge t_2 = \overline{t}_2 - \epsilon \wedge \overline{y}_t = y_t + \epsilon\Big(1 - \frac{2(t_2 - t)}{t_2 - t_1}\Big))\\
&\quad \vee (m_t = \mathit{on} \wedge t_1 = \overline{t}_2 - \epsilon \wedge t_2 = \overline{t}_2 \wedge y_t = \overline{y}_t + \epsilon\Big(\frac{\overline{t}_2 - t}{\overline{t}_2 - t_2}\Big))))\\
&\vee (\overline{m}_t = m_t = \mathit{open} \wedge y_t = \overline{y_t} \wedge t_1 = \overline{t}_1 \wedge t_2 = \overline{t}_2)\big)
\end{aligned}
$$

By the Galois isomorphism (33), the relation $R^{(53)}$ between configurations defines the relation $r^{(53)}$ between states as a function of the time.

$$
\begin{aligned}
r^{(53)} \triangleq \big\{\langle\langle m_t,\ x_t,\ y_t\rangle,\ \langle\overline{m}_t,\ \overline{x}_t,\ \overline{y}_t\rangle\rangle \ \big|\ \exists[t_1,t_2[\ \subseteq [\overline{t}_1,\overline{t}_2[\ .\ t \in [t_1,t_2[ \qquad (55)\\
\wedge P^{(53)}(m,x,y,t_1,t_2,\overline{m},\overline{x},\overline{y},\overline{t}_1,\overline{t}_2)\big\}
\end{aligned}
$$

$R^{(53)}$, that is $\gamma(r^{(53)})$, is a synchronous simulation (52), where in *shut* mode the concrete level is within $\epsilon$ of the abstract water level while in *open* mode they are the same. □

**Trace abstraction by asynchronous hybrid simulations.** Our objective is now to generalize the results of section 7.1 on homomorphisms to (weaker ones for) simulations. Similar to (48) for homomorphic abstractions, simulations induce related hybrid trajectory semantics.

**Theorem 4.** *If the timed relation $r$ between states in* (29) *is such that its extension $\gamma(r)$ to configurations in* (30) *is a simulation* (51) *between* $\langle \mathsf{C},\ \mathsf{C}^0,\ \tau\rangle$ *and* $\langle \overline{\mathsf{C}},\ \overline{\mathsf{C}}^0,\ \overline{\tau}\rangle$ *satisfying the* initialization hypothesis

$$\forall c \in \mathsf{C}^0 \,.\, \exists \overline{c} \in \overline{\mathsf{C}}^0 \,.\, \langle c,\, \overline{c}\rangle \in \gamma(r) \tag{56}$$

*then* $\langle [\![\tau]\!],\, [\![\overline{\tau}]\!]\rangle \in \vec{\gamma}(\vec{\gamma}_c(r))$. *If moreover, the* blocking hypothesis

$$\forall c, \overline{c} \,.\, (\langle c,\, \overline{c}\rangle \in \gamma(r) \wedge \forall c' \,.\, \langle c,\, c'\rangle \notin \tau) \Longrightarrow (\forall \overline{c}' \,.\, \langle \overline{c},\, \overline{c}'\rangle \notin \overline{\tau}) \tag{57}$$

*holds then*

$$\langle [\![\tau]\!],\, [\![\overline{\tau}]\!]\rangle \in \vec{\gamma}(\vec{\gamma}(r)) \tag{58}$$

(*that is, by* (37)*,* $\forall \sigma \in [\![\tau]\!] \,.\, \exists \overline{\sigma} \in [\![\overline{\tau}]\!] \,.\, \langle \sigma,\, \overline{\sigma}\rangle \in \vec{\gamma}(r)$ *and so, by* (34)*,* $\forall t \in [0, \min([\![\sigma]\!], [\![\overline{\sigma}]\!])[ \cap \mathsf{dom}(r) \,.\, \langle \sigma_t,\, \overline{\sigma}_t\rangle \in r(t))$.

*Example 7.* Continuing the water tank automaton in examples 3 and 6, $R^{(53)}$ is a synchronous simulation (52). So the hybrid semantics of example 6 is a simulation of (58) of the hybrid semantics $[\![\tau^3]\!]$ of example 3. □

**Simulations are abstractions.** Observe that theorem 4 implies that hybrid simulations are Galois connection-based abstractions (38).

**Compositionality of simulations.** The composition of simulations may not, in general, correspond to the composition of their timed relations between states (defined as $(r_1 \circ r_2)(t) = r_1(t) \circ r_2(t)$). This is because the intermediate trajectories may be shorter that those in the composition, as shown in figure 6.

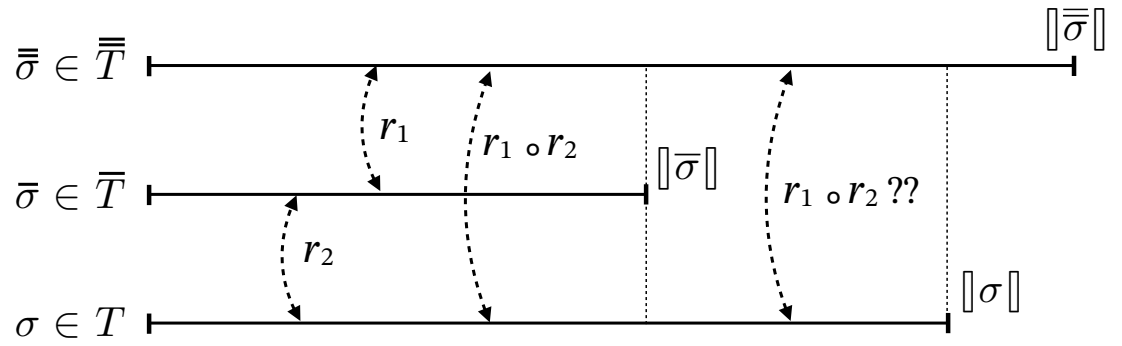

**Fig. 6.** Non-composition due to short intermediate trajectory duration

Another problem is that of interval mismatches where the intervals along trajectories thus leaving some states time-unrelate in the composition, see figure 7.

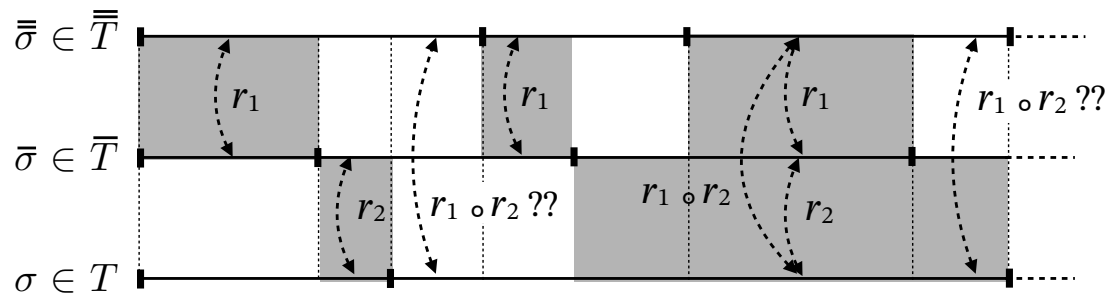

**Fig. 7.** Non-nested intervals

A sufficient condition for compositionally is that the involved trajectories be all infinite with well nested interval, meaning

$$\forall \langle \langle f_j,\, i_j\rangle,\, j \in \mathbb{N}\rangle \in T \,.\, \forall \langle \langle \overline{f}_k,\, \overline{i}_k\rangle,\, k \in \mathbb{N}\rangle \in \overline{T} \,. \tag{59}$$
$$\forall j, k \in \mathbb{N} \,.\, (i_j \cap \overline{i}_k \neq \emptyset) \Longrightarrow (i_j \subseteq \overline{i}_k)$$

**Theorem 5.** *If* $T \in \mathsf{T}^{\infty}_{\mathsf{C}}$, $\overline{T} \in \mathsf{T}^{\infty}_{\overline{\mathsf{C}}}$, $\overline{\overline{T}} \in \mathsf{T}^{\infty}_{\overline{\overline{\mathsf{C}}}}$ *are well-nested,* $\langle T,\ \overline{T}\rangle \in \vec{\gamma}(\vec{\gamma}_c(r_1))$ *and* $\langle \overline{T},\ \overline{\overline{T}}\rangle \in \vec{\gamma}(\vec{\gamma}_c(r_2))$ *then* $\langle T,\ \overline{\overline{T}}\rangle \in \vec{\gamma}(\vec{\gamma}_c(r_1 \circ r_2))$.

*Example 8 (Composition of the water tank simulations).* Continuing the water tank specification in example 2, automaton in example 3, and implementation in example 6, the hybrid trajectory semantics are well-nested according to (59). The hybrid semantics $[\![\tau^6]\!]$ of example 6 is a simulation of the hybrid semantics $[\![\tau^3]\!]$ of example 3 by $r^{(53)}$, which itself is a simulation of the specification $\mathcal{S}^2$ by $r^{(39)}$. So, by theorem 5, their composition $r^{(53)} \circ r^{(39)}$ holds at any time between the implementation $[\![\tau^6]\!]$ and the specification $\mathcal{S}^2$.

This may look paradoxical because if $\epsilon > \zeta$ then water in the implementation will remain at the zero level longer than prescribed by the specification (20.d).

However, this is not an anomaly since the composition is

$$r^{(53)} \circ r^{(39)} \triangleq \{\langle\langle m_t,\ x_t,\ y_t\rangle,\ \langle \overline{m}_t,\ \overline{y}_t\rangle\rangle \mid \exists [t_1, t_2[\ \subseteq [\overline{t}_1, \overline{t}_2[\ .\ t \in [t_1, t_2[\ \wedge \\ P^{(53)}(m_t, x_t, y_t, t_1, t_2, \overline{m}_t, x_t, \overline{y}_t, \overline{t}_1, \overline{t}_2)\} \tag{60}$$

By definition (53), this expresses that the height $\overline{y}_t$ of the water in the specification when the valve is *off* for $\epsilon$ units of time is equal to $t - t_1 = t - \overline{t}_1$, not to the level of water $y_t = 0$ in the implementation. So, although each simulation $r^{(39)}$ and $r^{(53)}$ is a satisfactory specification, their composition is an incomplete refinement of the expected water tank behavior. □

**Greatest simulation.** (51) can be rewritten as

$$R \subseteq F^s_{\tau,\overline{\tau}}(R) \qquad \text{with} \tag{61}$$

$$F^s_{\tau,\overline{\tau}}(R) \triangleq \{\langle c,\ \overline{c}\rangle \mid \forall c'\ .\ (\langle c,\ c'\rangle \in \tau) \Longrightarrow \\ (\exists \overline{c}'\ .\ \langle \overline{c},\ \overline{c}'\rangle \in \overline{\tau} \wedge \langle c \mathbin{⨾} c'(\!|\min(\mathsf{b}(c'), \mathsf{b}(\overline{c}')), \min(\mathsf{e}(c'), \mathsf{e}(\overline{c}'))|\!), \\ \overline{c} \mathbin{⨾} \overline{c}'(\!|\min(\mathsf{b}(c'), \mathsf{b}(\overline{c}')), \min(\mathsf{e}(c'), \mathsf{e}(\overline{c}'))|\!)\rangle \in R)\}$$

where $F^s_{\tau,\overline{\tau}}$ is increasing on the complete lattice $\langle \wp(\mathsf{C} \times \overline{\mathsf{C}}), \subseteq\rangle$ so that by Tarski's fixpoint theorem [37] there exists a greatest simulation between $\tau$ and $\overline{\tau}$, thus extending Robin Milner's classic result [27, Proposition 16, section 4.6] to hybrid simulations (and the least fixpoint is $\emptyset$).

**Verification of trace properties by simulation.** The homomorphic induction principle (49) can be generalized to hybrid asynchronous simulations $\vec{\gamma}(r)$ as follows

$$\frac{\vec{\gamma}(r) \subseteq F^s_{\tau,\overline{\tau}}(\vec{\gamma}(r)), \quad (56), \quad (57), \quad [\![\overline{\tau}]\!] \subseteq \overline{P}}{\langle [\![\tau]\!],\ \overline{P}\rangle \in \vec{\gamma}(\vec{\gamma}(r))} \tag{62}$$

**Discretization by sampling.**

*Discretization of a hybrid transition system.* We have defined the discretization (27) of a hybrid trajectory semantics. In general the discretization of a hybrid transition system (22) and that of the generated trajectories (23) do not coincide, as shown by the following counterexample (for which the discretization of the trajectory and that of the configurations $c_1, c_2, c_3, \ldots$ in the transition relation $\tau$ do not coincide).

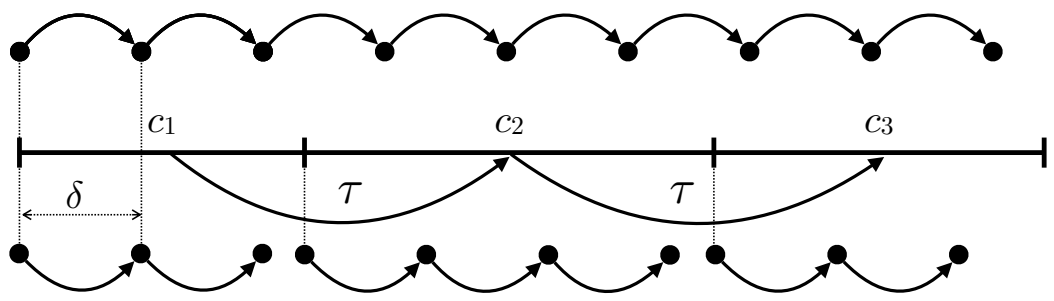

To solve this dependency, it is generally assumed that the start time and duration of configurations is a multiple of the discretization step

$$\forall c \in \mathsf{C} \,.\, \exists k, k' \in \mathbb{N} \,.\, \mathsf{b}(c) = k\delta < k'\delta = \mathsf{e}(c). \tag{63}$$

The relation (29) between states is time-dependent. Simply ignoring the discrete time $\langle n\delta,\ n \in \mathbb{N}\rangle$ might create circularities (see example 10 thereafter).

One solution is to incorporate the time (or at least the rank $n \in \mathbb{N}$) into states to make the relation time-independent as in classic simulations. So (27) becomes

$$\alpha_\delta(\sigma) \triangleq \langle\langle\sigma_{n\delta},\ n\rangle,\ n \in \mathbb{N} \wedge n\delta \in [0, \llbracket\sigma\rrbracket]\rangle \tag{64}$$

and (29) becomes

$$\alpha_\delta(r) \triangleq \{\langle\langle s,\ n\rangle,\ \langle\overline{s},\ n\rangle\rangle \mid n \in \mathbb{N} \wedge n\delta \in \mathsf{dom}(r) \wedge \langle s,\ \overline{s}\rangle \in r(n\delta)\}. \tag{65}$$

The timeful discretization of the hybrid transition system is

$$\begin{aligned}
\alpha_\delta(\tau) \triangleq \{\langle\langle s,\ n\rangle,\ \langle s',\ n+1\rangle\rangle \mid & & (66)\\
& (\exists c \,.\, (c \in \mathsf{C}^0 \vee \exists c' \,.\, \langle c',\ c\rangle \in \tau) \wedge & \text{(a)}\\
& \quad \mathsf{b}(c) \leqslant n\delta < (n+1)\delta < \mathsf{e}(c) \wedge s = c_{n\delta} \wedge s' = c_{(n+1)\delta}) & \\
& \vee (\exists\langle c,\ c'\rangle \in \tau \,.\, (n+1)\delta = \mathsf{e}(c) \wedge s = c_{n\delta} \wedge s' = c'_{(n+1)\delta}) & \text{(b)}\\
& \vee (\exists c \in \mathsf{C} \,.\, \forall c' \,.\, \langle c,\ c'\rangle \notin \tau \wedge (n+1)\delta = \mathsf{e}(c) \wedge & \text{(c)}\\
& \quad s = c_{n\delta} \wedge s' = c_{(n+1)\delta})\} & \\
\alpha_\delta(\mathsf{C}^0) \triangleq \{\langle c_0,\ 0\rangle \mid c \in \mathsf{C}^0\} & & \text{(d)}
\end{aligned}$$

which is well-defined by (22) and since the durations of the configurations are assumed to be multiples of $\delta$. (66.a) covers discrete transitions within a configuration but the last one. The last transition is either to the first state of the next configurations (66.b), or in absence of any successor configuration, to the last state of the current configuration (66.c). This condition (66.c) solves the problem of having open right time intervals in configurations by defining the last state of the last configuration of finite trajectories. This discretization applies to both concrete and abstract transition systems.

*Example 9 (hybrid transition discretization).* The various cases in (66) are illustrated below.

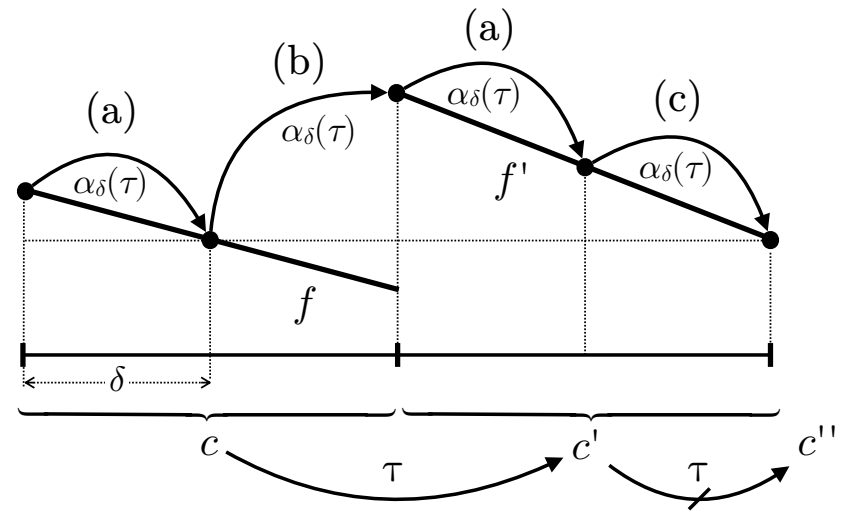

- By (66.a), the initial configuration $c = \langle f, [0, 2\delta]\rangle$ starting at time 0 of duration $2\delta$ has an internal discrete transition $\langle\langle f(0),\ 0\rangle,\ \langle f(\delta),\ 1\rangle\rangle \in \alpha_\delta(\tau)$ between its states at times 0 and $\delta$;
- Similarly, by (66.a), the successor configuration $c' = \langle f',\ [2\delta, 4\delta]\rangle$ starting at time $2\delta$ of duration $2\delta$ has an internal discrete transition $\langle\langle f'(2\delta), 2\rangle, \langle f'(3\delta), 3\rangle\rangle \in \alpha_\delta(\tau)$;
- By (66.b), the last discrete transition of configuration $c$ is toward the beginning state $\langle f'(2\delta),\ 2\rangle$ of its successor configuration(s) $c'$ (and not toward its final state $\langle f(2\delta),\ 2\rangle$);
- In contrast, by (66.c), the last discrete transition for configuration $c'$ which has no possible successor by $\tau$ is toward its final state $\langle f'(4\delta),\ 4\rangle$.

Observe that $f(\delta) = f'(4\delta)$ but they are distinguished by incorporating the rank $n$ of discrete times $n\delta$. □

*Example 10 (timeful and timeless abstraction).* Consider $\mathsf{S} = \{s\}$, $\mathsf{C} = \mathsf{C}^0 = \{c\}$, $\tau = \emptyset$ where $c = \langle f,\ [0, 2]\rangle$ with $\forall t \in [0, 2]\ .\ f(t) = s$, and $\delta = 1$. We have $[\![\tau]\!] = \{c\}$ and $\alpha_\delta([\![\tau]\!]) = \{\langle s,\ 0\rangle\langle s,\ 1\rangle\}$ as well as $\alpha_\delta(\tau) = \emptyset$ and $[\![\alpha_\delta(\tau)]\!] = \{\langle s, 0\rangle\langle s,\ 1\rangle\}$, that is, (67).

Ignoring time, we would have $\widetilde{\alpha}_\delta([\![\tau]\!]) = \{ss\}$ while the transition abstraction $\widetilde{\alpha}_\delta(\tau) = \{\langle s,\ s\rangle\}$ yields a circularity so that $[\![\widetilde{\alpha}_\delta(\tau)]\!] = s^+|s^\infty$ and in general $\widetilde{\alpha}_\delta([\![\tau]\!]) \subseteq [\![\widetilde{\alpha}_\delta(\tau)]\!]$ which would be a rather imprecise overapproximation. □

By definition of $\alpha_\delta$, it follows that

**Theorem 6.** *The timed discretization of the semantics is the semantics of the timeful discretized transition system, formally*

$$\alpha_\delta([\![\tau]\!]) = [\![\alpha_\delta(\tau)]\!] \tag{67}$$

**Is the discretization of a hybrid simulation a discrete simulation?** If the simulation of a hybrid transition system is a generalization of Robin Milner's simulation of discrete transition systems [26] there should be an abstraction of time mapping the hybrid simulation of the hybrid transition system into a discrete simulation for the discretized transition system. This is our next objective.

Sampling in (28) is a discretization. But this discretization of a hybrid simulation may not be a discrete simulation, even when configuration durations are a multiples of a base duration $\delta$, as assumed in (63).

For a counter example, on figure (8.a), we have a hybrid simulation since states are related by $r$ on the common interval of time of $c$ and $\overline{c}$. But in the discretization, concrete state $s$ has a successor while the related state $\overline{s}$ has none, so this is not a discrete simulation.

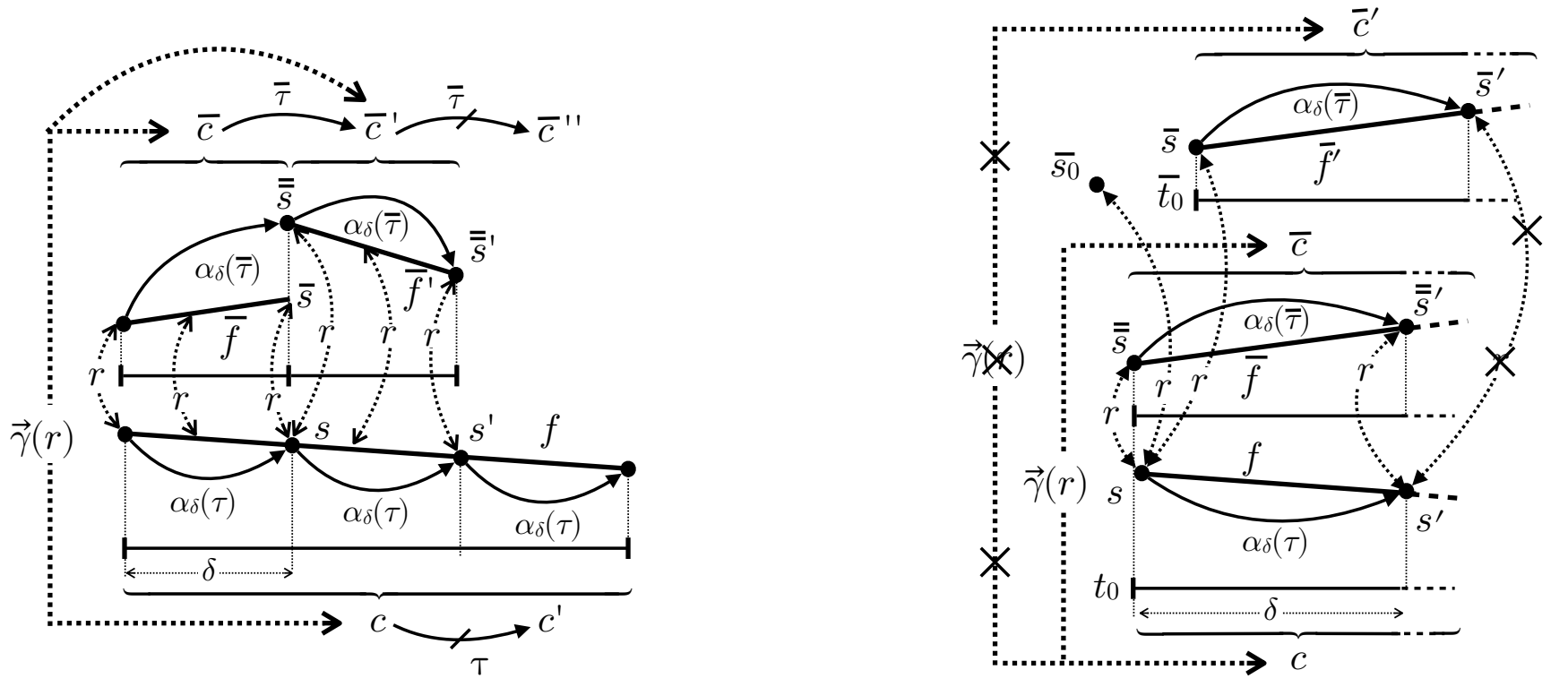

(a) blocking states $\overline{s}$ and $\overline{\overline{s}}'$

(b) related states $s$ and isolated state $\overline{s}_0$ or $s$ and $\overline{s}$ of unrelated configurations $c$ and $\overline{c}'$

**Fig. 8.** Effects of asynchronous discretization

A second counterexample is given in figure (8.b) when $t_0 = \overline{t}_0$. We have $\langle s, \overline{s}_0\rangle \in r(\overline{t}_0)$ but $\overline{s}_0$ does not belong to any configuration and so has no successor by $\alpha_\delta(\overline{\tau})$. So the discretization of the hybrid simulation is not a discrete simulation.

A third counterexample is also given on figure (8.b) Configurations $c$ and $\overline{c}$ are related because relation $r$ between their states during the first period of time (end included in successor) while $c$ and $\overline{c}'$ are not since $r$ holds between $s$ and $\overline{s}$ whereas it does not hold between $s'$ and $\overline{s}'$. After discretization, the configuration $\overline{c}'$ generates a transition $\alpha_\delta(\overline{\tau})$ from $\overline{s}$ to $\overline{s}'$. Now $\langle s, \overline{s}\rangle \in r^{-1} \wedge \langle s, s'\rangle \in \alpha_\delta(\tau)$ but the only successor $\overline{s}'$ of $\overline{s}$ by $\alpha_\delta(\overline{\tau})$ is not related to $s'$. So the discretization of the hybrid simulation is not a discrete simulation.

Moreover, the relation $r$ in (29) is the partial function of the time, whereas its abstraction $\alpha_\delta(r)^{-1}$ in Robin Milner's simulation $\alpha_\delta(r)^{-1} \circ \alpha_\delta(\tau) \dot{\subseteq} \alpha_\delta(\overline{\tau}) \circ \alpha_\delta(r)^{-1}$ is a well-defined relation between states. So, in case $r$ in (29) is not total, and to be compatible with Robin Milner's definition, we must assume that $r$ is well-defined at the discretization points

$$\forall n \in \mathbb{N} \,.\, (\exists c \in \mathsf{C} \,.\, n\delta \in \mathsf{dom}(c)) \Longrightarrow (n\delta \in \mathsf{dom}(r)). \tag{68}$$

To prevent the case of isolated state $\overline{s}_0$ in (8.b), we assume that related states must come from related configurations (either initial or successor ones).

$$\forall c \in \mathsf{C}, \overline{s} \in \overline{\mathsf{S}}, n \in \mathbb{N} \,.\, (n\delta \in \mathsf{dom}(c) \cap \mathsf{dom}(r) \wedge \langle c_{n\delta},\, \overline{s}\rangle \in r(n\delta)) \Longrightarrow \\ (\exists \overline{c} \in \overline{\mathsf{C}} \,.\, (\overline{c} \in \overline{\mathsf{C}}^0 \vee \exists \overline{c}' \,.\, \langle \overline{c}',\, \overline{c}\rangle \in \overline{\tau}) \wedge n\delta \in \mathsf{dom}(\overline{c}) \wedge \overline{c}_{n\delta} = \overline{s}) \tag{69}$$

Beyond an initialization hypothesis (similar to (56)), a common hypothesis for discrete simulations is the *non-blocking condition*, which, for hybrid simulations, translates into

$$\forall c \in \mathsf{C}, \overline{c} \in \overline{\mathsf{C}} \,.\, (\exists t \in \mathsf{dom}(c) \cap \mathsf{dom}(\overline{c}) \cap \mathsf{dom}(r) \,.\, \langle c_t,\, \overline{c}_t\rangle \in r(t) \wedge \\ \forall \overline{c}' \,.\, \langle \overline{c},\, \overline{c}'\rangle \notin \overline{\tau}) \Longrightarrow (\mathsf{e}(\overline{c}) = \mathsf{e}(c)) \tag{70}$$

The non-blocking condition (70) will avoid the blocking state $\overline{s}$ in figure 8, on the left, since concrete blocking configurations can only be related to abstract configurations with the same ending time.

Moreover, we request the relation $r$ between states to be compatible with the discretization (66) of transition relations. If a concrete configuration $c$ and an abstract one $\overline{c}$ have related states at some time $t$ then their states must be related at any time $n\delta$ in their common time intervals, except maybe at the end of these time intervals (71.a).

$$\begin{array}{lr}
\forall c \in \mathsf{C}, \overline{c} \in \overline{\mathsf{C}} \,.\, (\exists t \in \mathsf{dom}(c) \cap \mathsf{dom}(\overline{c}) \cap \mathsf{dom}(r) \,.\, \langle c_t,\, \overline{c}_t\rangle \in r(t)) \Longrightarrow & (71)\\
\quad (\forall n\delta \in (\mathsf{dom}(c) \cap \mathsf{dom}(\overline{c})) \setminus \{\mathsf{e}(c), \mathsf{e}(\overline{c})\} \,.\, \langle c_{n\delta},\, \overline{c}_{n\delta}\rangle \in r(n\delta)) \wedge & \text{(a)}\\
\quad (\forall n \,.\, n\delta = \mathsf{e}(c) \in \mathsf{dom}(\overline{c})) \Longrightarrow & \text{(b)}\\
\qquad (\forall c' \,.\, (\langle c,\, c'\rangle \in \tau) \Longrightarrow (\langle c'_{n\delta},\, \overline{c}_{n\delta}\rangle \in r(n\delta))) \wedge & \text{(b.1)}\\
\qquad ((\forall c' \,.\, \langle c,\, c'\rangle \in \tau \Longrightarrow c_{n\delta} \neq c'_{n\delta}) \Longrightarrow (\langle c_{n\delta},\, \overline{c}_{n\delta}\rangle \notin r(n\delta)))) \wedge & \text{(b.2)}\\
\qquad ((\forall c' \,.\, \langle c,\, c'\rangle \notin \tau) \Longrightarrow (\langle c_{n\delta},\, \overline{c}_{n\delta}\rangle \in r(n\delta))) \wedge & \text{(b.3)}\\
\quad (\forall n \,.\, (n\delta = \mathsf{e}(\overline{c}) \in \mathsf{dom}(c)) \Longrightarrow & \text{(c)}\\
\qquad (\forall \overline{c}' \,.\, (\langle \overline{c},\, \overline{c}'\rangle \in \overline{\tau}) \Longrightarrow (\langle c_{n\delta},\, \overline{c}'_{n\delta}\rangle \in r(n\delta))) \wedge & \text{(c.1)}\\
\qquad ((\forall \overline{c}' \,.\, \langle \overline{c},\, \overline{c}'\rangle \in \overline{\tau} \Longrightarrow \overline{c}_{n\delta} \neq \overline{c}'_{n\delta}) \Longrightarrow (\langle c_{n\delta},\, \overline{c}_{n\delta}\rangle \notin r(n\delta))) \wedge & \text{(c.2)}\\
\qquad ((\forall \overline{c}' \,.\, \langle \overline{c},\, \overline{c}'\rangle \notin \overline{\tau}) \Longrightarrow (\langle c_{n\delta},\, \overline{c}_{n\delta}\rangle \in r(n\delta)))) & \text{(c.3)}
\end{array}$$

The relations between states at the end of a concrete configuration $c$ are illustrated in figure 9. In case (71.b.1), the state $c'_{n\delta}$ at the beginning of the next concrete configuration $c'$ is related to the abstract state $\overline{c}_{n\delta}$ at the end of this concrete configuration $c$.

Case (71.b.2) states that if there is no concrete configuration $c'$ which initial state $c'_{n\delta}$ is equal to the last state $c_{n\delta}$ of the previous configuration $c$ then $c_{n\delta}$ should *not* be related to the abstract state $\overline{c}_{n\delta}$ at the end of this concrete configuration $c$.

Case (71.b.3) states that if the concrete configuration $c$ ending at time $n\delta$ has no successor then its last state should be related to the abstract state $\overline{c}_{n\delta}$ at the end of this concrete configuration $c$.

Cases (71.c) in figure 10 are symmetrical.

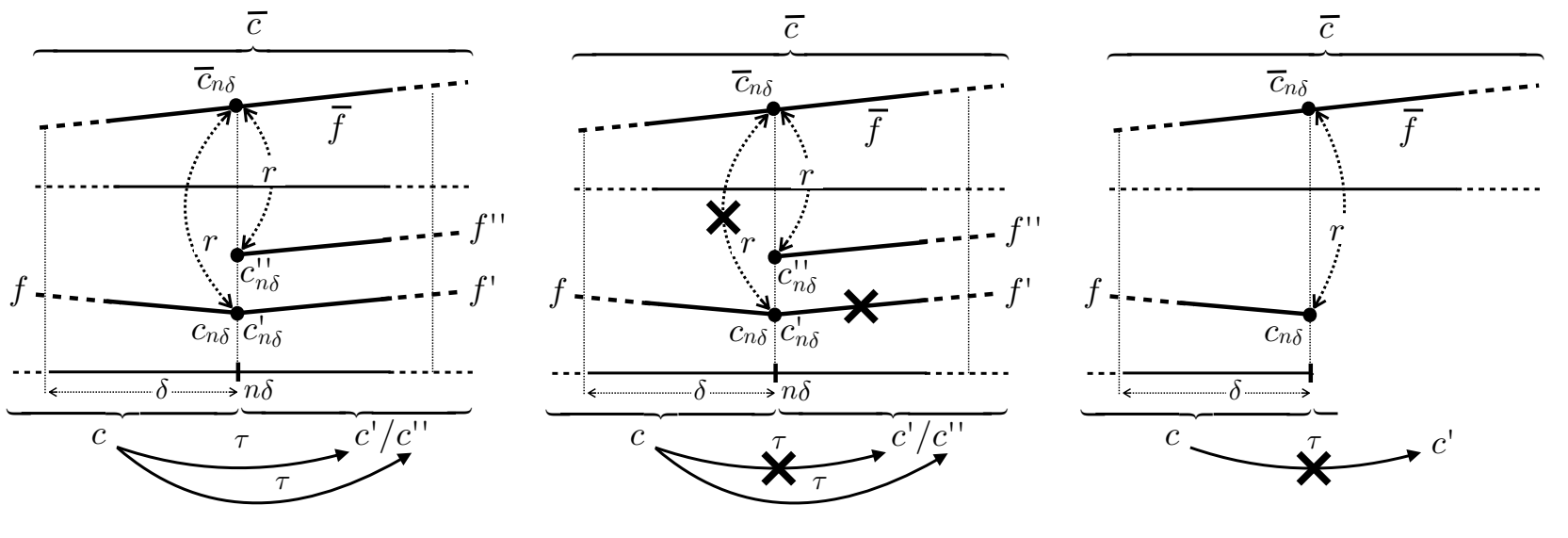

(71.b.1) (71.b.2) (71.b.3)

**Fig. 9.** Relation between states after discretization of concrete configuration transitions

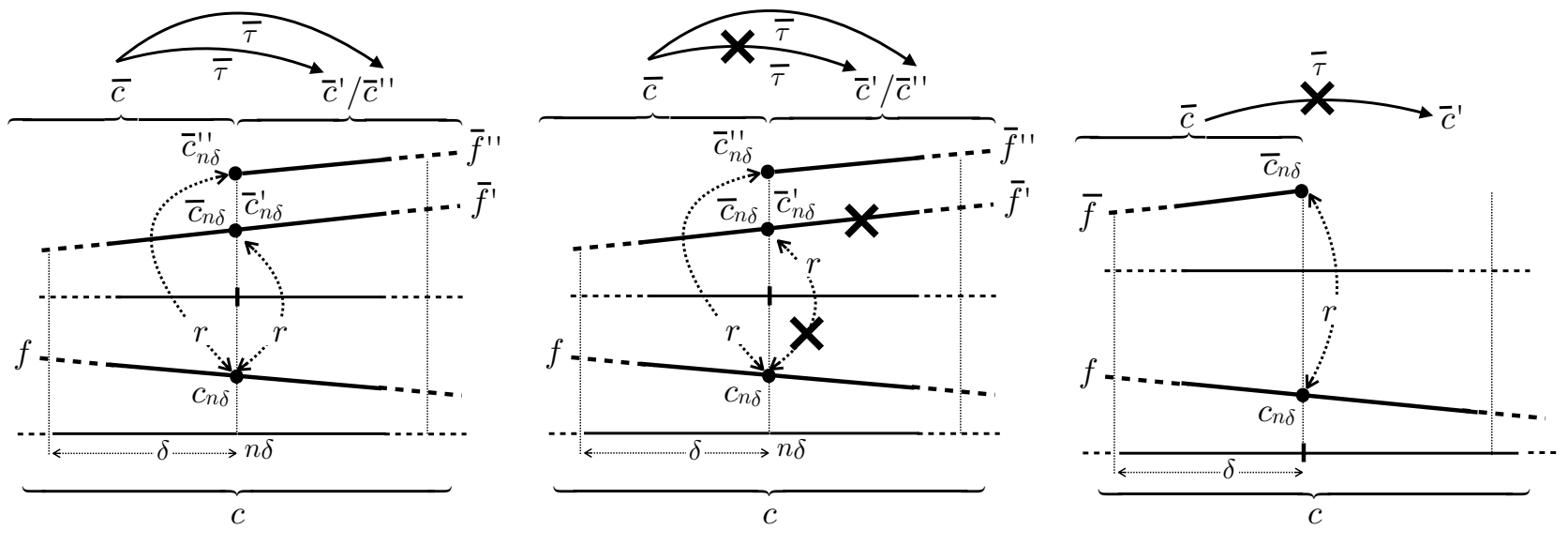

(71.c.1) (71.c.2) (71.c.3)

**Fig. 10.** Relation between states after discretization of abstract configuration transitions

Then, we have the following result (72) that supports the intuition that state-based hybrid simulations $\vec{\gamma}(r)$ satisfying (51) (or equivalently (61)) are a meaningful generalization of Robin Milner discrete simulations (i.e. $R \circ t \mathrel{\dot{\subseteq}} \overline{t} \circ R$).

**Theorem 7.**

$$(\vec{\gamma}(r) \subseteq F^{s}_{\tau,\overline{\tau}}(\vec{\gamma}(r)) \wedge (68) \wedge (69) \wedge (70) \wedge (71)) \Longrightarrow \alpha_{\delta}(r)^{-1} \circ \alpha_{\delta}(\tau) \mathrel{\dot{\subseteq}} \alpha_{\delta}(\overline{\tau}) \circ \alpha_{\delta}(r)^{-1} \quad (72)$$

## 8 Conclusion

We have studied correspondences between trajectory semantics of hybrid systems with possibly different durations and timelines (that is different timing for mode changes), including the case where the hybrid semantics is generated by an hybrid transition system.

The abstraction relation between semantics can be derived from relations between trajectories, possibly themselves derived from relations between configurations, possibly themselves derived from timed relations between states. Such correspondences include the particular cases of homomorphisms, simulations, and discretization (as well as bisimulations, preservation, progress considered in the appendix). They induce abstractions of the hybrid semantics that are Galois connections. So the abstractions between hybrid semantics defined by the correspondences between trajectories do compose.

However, contrary to the discrete case [26,32] and with the exception of homomorphic trajectory abstraction in section 7.1, the correspondences between trajectories or configurations do not necessarily compose with the correspondence between states. For example, the discretization of similar hybrid trajectories may not be similar discrete traces. The problem does not appear in Milner's original discrete definition [26] because the notion of state and configuration as well as timings do coincide. We have studied sufficient conditions for composability of trajectory correspondences to hold.

Like in the case of discrete systems [10], further abstractions of the hybrid trajectory semantics will lead to a hierarchy of semantics, verification, and static analysis methods. The most common abstraction is the reachability abstraction $\alpha(\{\langle\sigma^j_{i^j},\ i^j \in [0, |\sigma^j|[\rangle \mid j \in \Delta\}) \triangleq \{\sigma^j_{i^j}(t) \mid j \in \Delta \wedge i^j \in [0, |\sigma^j|[ \wedge t \in [0, |\sigma^j|[\}$, see [8] for a documented and comprehensive survey.

**Acknowledgements**

I thank Dominique Méry for his suggestions and encouragements, "la discrétisation, c'est coton". I thank the two anonymous referees for their constructive criticisms.

## References

1. Rajeev Alur, Costas Courcoubetis, Thomas A. Henzinger, and Pei-Hsin Ho. Hybrid automata: An algorithmic approach to the specification and verification of hybrid systems. In *Hybrid Systems*, volume 736 of *Lecture Notes in Computer Science*, pages 209–229. Springer, 1992.
2. Rajeev Alur and David L. Dill. A theory of timed automata. *Theor. Comput. Sci.*, 126(2):183–235, 1994.
3. Rajeev Alur, Thomas A. Henzinger, Gerardo Lafferriere, and George J. Pappas. Discrete abstractions of hybrid systems. *Proceedings of the IEEE*, 88(7):971–984, July 2000.
4. Rajeev Alur and P. Madhusudan. Decision problems for timed automata: A survey. In *SFM*, volume 3185 of *Lecture Notes in Computer Science*, pages 1–24. Springer, 2004.
5. Christel Baier and Joost-Pieter Katoen. *Principles of model checking.* MIT Press, 2008.
6. Richard Banach, Michael J. Butler, Shengchao Qin, Nitika Verma, and Huibiao Zhu. Core hybrid Event-B I: single hybrid Event-B machines. *Sci. Comput. Program.*, 105:92–123, 2015.

7. Paul Caspi and Nicolas Halbwachs. An approach to real time systems modeling. In *ICDCS*, pages 710–716. IEEE Computer Society, 1982.
8. Xin Chen and Sriram Sankaranarayanan. Reachability analysis for cyber-physical systems: Are we there yet? In *NFM*, volume 13260 of *Lecture Notes in Computer Science*, pages 109–130. Springer, 2022.
9. Zheng Cheng and Dominique Méry. A refinement strategy for hybrid system design with safety constraints. In *MEDI*, volume 12732 of *Lecture Notes in Computer Science*, pages 3–17. Springer, 2021.
10. Patrick Cousot. Constructive design of a hierarchy of semantics of a transition system by abstract interpretation. *Theor. Comput. Sci.*, 277(1-2):47–103, 2002.
11. Patrick Cousot. *Principles of Abstract Interpretation*. MIT Press, September 2021.
12. Alessandro D'Innocenzo, A. Agung Julius, George J. Pappas, Maria Domenica Di Benedetto, and Stefano Di Gennaro. Verification of temporal properties on hybrid automata by simulation relations. In *CDC*, pages 4039–4044. IEEE, 2007.
13. Laurent Doyen, Thomas A. Henzinger, and Jean-François Raskin. Automatic rectangular refinement of affine hybrid systems. In *FORMATS*, volume 3829 of *Lecture Notes in Computer Science*, pages 144–161. Springer, 2005.
14. Goran Frehse. On timed simulation relations for hybrid systems and compositionality. In *FORMATS*, volume 4202 of *Lecture Notes in Computer Science*, pages 200–214. Springer, 2006.
15. Antoine Girard, A. Agung Julius, and George J. Pappas. Approximate simulation relations for hybrid systems. *Discret. Event Dyn. Syst.*, 18(2):163–179, 2008.
16. Antoine Girard and George J. Pappas. Approximate bisimulation: A bridge between computer science and control theory. *Eur. J. Control*, 17(5-6):568–578, 2011.
17. Thomas A. Henzinger and Pei-Hsin Ho. A note on abstract interpretation strategies for hybrid automata. In *Hybrid Systems*, volume 999 of *Lecture Notes in Computer Science*, pages 252–264. Springer, 1994.
18. Thomas A. Henzinger, Peter W. Kopke, Anuj Puri, and Pravin Varaiya. What's decidable about hybrid automata? In *STOC*, pages 373–382. ACM, 1995.
19. Eugene Isaacson and Herbert Bishop Keller. *Analysis of Numerical Methods*. Dover, 1994.
20. Anatole Katok and Biros Hasselblatt. *Introduction to the Theory of Dynamical Systems*. Cambridge University Press, 1999.
21. Serge Lang. *Undergraduate Analysis*. Springer, 2 edition, 1997.
22. Daniel Liberzon. *Switching in Systems and Control*. Birkhäuser, 2003.
23. Nancy A. Lynch. Simulation techniques for proving properties of real-time systems. In *REX School/Symposium*, volume 803 of *Lecture Notes in Computer Science*, pages 375–424. Springer, 1993.
24. MathWorks. Simulation and model-based design. `https://www.mathworks.com/products/simulink.html`, 2022.
25. Larissa Meinicke and Ian J. Hayes. Continuous action system refinement. In *MPC*, volume 4014 of *Lecture Notes in Computer Science*, pages 316–337. Springer, 2006.
26. Robin Milner. An algebraic definition of simulation between programs. In *Proceedings IJCAI 1971*, pages 481–489, 1971.
27. Robin Milner. *Communication and concurrency*. PHI Series in computer science. Prentice Hall, 1989.
28. Harry Nyquist. Certain topics in telegraph transmission theory. *Proceedings of the IEEE*, 47(2):617–644, April 1928.
29. John G. Proakis and Dimitris G Manolakis. *Digital Signal Processing*. Pearson, 4 edition, 2006.

30. James C. Robinson. *An Introduction to Ordinary Differential Equations.* Cambridge University Press, 2004.
31. Mauno Rökkö, Anders P. Ravn, and Kaisa Sere. Hybrid action systems. *Theor. Comput. Sci.*, 290(1):937–973, 2003.
32. Davide Sangiorgi. *Introduction to Bisimulation and Coinduction.* Cambridge University Press, 2011.
33. Claude E. Shannon. Communication in the presence of noise. *Proceedings of the I.R.E.*, pages 10–21, January 1949.
34. Zahava Shmuely. The structure of Galois connections. *Pacific Journal of Mathematics*, 54(2):209—225, 1974.
35. Wen Su, Jean-Raymond Abrial, and Huibiao Zhu. Formalizing hybrid systems with Event-B and the Rodin platform. *Sci. Comput. Program.*, 94:164–202, 2014.
36. Yong Kiam Tan and André Platzer. An axiomatic approach to liveness for differential equations. In *FM*, volume 11800 of *Lecture Notes in Computer Science*, pages 371–388. Springer, 2019.
37. Alfred Tarski. A lattice theoretical fixpoint theorem and its applications. *Pacific J. of Math.*, 5:285–310, 1955.
38. Andrew K. Wright and Matthias Felleisen. A syntactic approach to type soundness. *Inf. Comput.*, 115(1):38–94, 1994.
39. Jun Zhang, Karl Henrik Johansson, John Lygeros, and Shankar Sastry. Dynamical systems revisited: Hybrid systems with zeno executions. In *HSCC*, volume 1790 of *Lecture Notes in Computer Science*, pages 451–464. Springer, 2000.

# A Transition-based hybrid trajectory semantics abstraction (continued)

## A.1 Bisimulations

Bisimulations were introduced by Robin Milner [26] under the name "mutual simulations". They can be generalized to hybrid transition systems by

$$R \subseteq F^{s}_{\tau,\overline{\tau}}(R) \wedge R^{-1} \subseteq F^{s}_{\overline{\tau},\tau}(R^{-1}) \tag{73}$$

Results similar to those of simulations can be proved for bisimulations, in particular, hybrid bisimulations are Galois connection-based abstractions (38).

## A.2 Preservations with Progress

**Definition of asynchronous hybrid preservations and progress.** Preservations and progress where introduced by Andrew Wright and Matthias Felleisen [38] to prove the soundness of type systems by subject reduction. A relation $R \in \wp(\mathsf{C} \times \overline{\mathsf{C}})$ between concrete and abstract configurations is a preservation between the transition relations $\tau$ and $\overline{\tau}$ if and only if [4]

[4] Often, in typing, abstract states in $\overline{\mathsf{C}}$ are types, $\tau$ is the reduction of an expression, and $\overline{\tau}$ is the identity.

$$\forall c, \overline{c}, c', \overline{c}' \;.\; (\langle c,\ \overline{c}\rangle \in R \wedge (\langle c,\ c'\rangle \in \tau \vee c' = \varepsilon) \wedge (\langle \overline{c},\ \overline{c}'\rangle \in \overline{\tau} \vee \overline{c}' = \varepsilon) \wedge) \Longrightarrow (\ \langle c \mathbin{⨟} c'(\!|\min(\mathsf{b}(c'), \mathsf{b}(\overline{c}')), \min(\mathsf{e}(c'), \mathsf{e}(\overline{c}'))|\!), \overline{c} \mathbin{⨟} \overline{c}'(\!|\min(\mathsf{b}(c'), \mathsf{b}(\overline{c}')), \min(\mathsf{e}(c'), \mathsf{e}(\overline{c}'))|\!)\rangle \in R)) \tag{74}$$

Simulations $S$ (with an $\exists\overline{c}'$ in (51)) and preservations $P$ (with a $\forall\overline{c}'$ in (74)) below are different, as shown in figure 11.

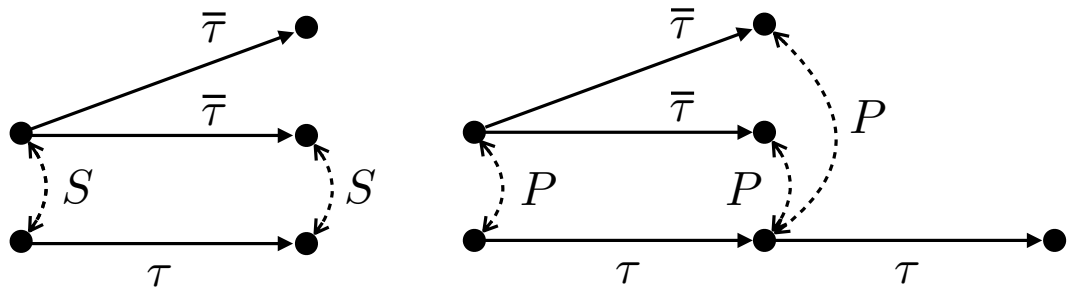

(a) simulation not preservation (b) preservation not simulation (c) preservation and non-blocking simulation

**Fig. 11.** Simulation versus preservation

**Greatest preservation.** (74) can be rewritten as

$$R \subseteq F^{p}_{\tau,\overline{\tau}}(R) \qquad \text{with} \tag{75}$$

$$F^{p}_{\tau,\overline{\tau}}(R) \triangleq \{\langle c,\ \overline{c}\rangle \mid \forall c', \overline{c}' \;.\; (\langle c,\ c'\rangle \in \tau \wedge \langle \overline{c},\ \overline{c}'\rangle \in \overline{\tau}) \Longrightarrow (\ \langle c \mathbin{⨟} c'(\!|\min(\mathsf{b}(c'), \mathsf{b}(\overline{c}')), \min(\mathsf{e}(c'), \mathsf{e}(\overline{c}'))|\!), \overline{c} \mathbin{⨟} \overline{c}'(\!|\min(\mathsf{b}(c'), \mathsf{b}(\overline{c}')), \min(\mathsf{e}(c'), \mathsf{e}(\overline{c}'))|\!)\rangle \in R))\}$$

so that by Tarski's fixpoint theorem [37] there exists a greatest preservation between $\tau$ and $\overline{\tau}$ (and $\emptyset$ is the smallest one).

**Progress.** Preservation often goes with a notion of progress (which holds for simulations)

$$\forall c, \overline{c}, c' \;.\; (\langle c,\ \overline{c}\rangle \in R \wedge \langle c,\ c'\rangle \in \tau) \Longrightarrow (\exists \overline{c}' \;.\; \langle \overline{c},\ \overline{c}'\rangle \in \overline{\tau}) \tag{76}$$

so that preservation and progress is a simulation but not conversely (except for deterministic transition systems).

**Trace abstraction by asynchronous hybrid preservations.** Similar to theorem 4, we have

**Theorem 8.** *If the timed relation $r$ between states in (29) is such that its extension $\gamma(r)$ to configurations in (30) is a preservation (74) between $\langle \mathsf{C},\ \mathsf{C}^0,\ \tau\rangle$ and $\langle \overline{\mathsf{C}},\ \overline{\mathsf{C}}^0,\ \overline{\tau}\rangle$ satisfying the* initialization hypothesis (56) *and the* progress hypothesis (76) *then $\langle [\![\tau]\!],\ [\![\overline{\tau}]\!]\rangle \in \vec{\gamma}(\vec{\gamma}_c(r))$. If moreover, the* blocking hypothesis (57) *holds then $\langle [\![\tau]\!],\ [\![\overline{\tau}]\!]\rangle \in \vec{\gamma}(\vec{\gamma}(r))$.*

**Preservations are abstractions.** Observe that theorem 8 implies that hybrid preservations are Galois connection-based abstractions (38).

The other results for simulations in section 7.2 extend equally well to preservations with progress.

# B Proofs for Section 4 (Transition-based hybrid trajectory semantics)

*Proof of lemma 1.* Let $\sigma = \langle \sigma_i,\ i \in [0, |\sigma|[\rangle \in [\![\tau]\!]$. If $|\sigma| = \infty$, then $\tau \subseteq \tau'$ implies that $\sigma \in [\![\tau']\!]$. Otherwise, $\sigma$ is finite and being maximal it's last state has no successor by $\tau$, hence by the blocking condition no successor by $\tau'$. Since $\tau \subseteq \tau'$ all non-terminal states in $\sigma$ have a successor by $\tau$ hence by $\tau'$ so that $\sigma \in [\![\tau']\!]$. □

# C Proofs for Section 6 (State-based hybrid trajectory semantics abstraction)

*Proof of* (33).

– $\alpha(R) \mathrel{\dot{\subseteq}} r$

$\iff \forall t \in \mathbb{R}_{\geqslant 0} \,.\, \alpha(R)(t) \subseteq r(t)$ ⟅pointwise definition of $\dot{\subseteq}$⟆

$\iff \forall t \in \mathbb{R}_{\geqslant 0} \,.\, \{\langle f(t),\ \overline{f}(t)\rangle \mid \exists i, \overline{i} \,.\, t \in i \cap \overline{i} \wedge \langle\langle f,\ i\rangle,\ \langle \overline{f},\ \overline{i}\rangle\rangle \in R\} \subseteq r(t)$
⟅definition (31) of $\alpha(R)$⟆

$\iff \forall t \in \mathbb{R}_{\geqslant 0} \,.\, \{\langle f(t), \overline{f}(t)\rangle \mid \exists i, \overline{i} \,.\, i\cap\overline{i} \neq \emptyset \wedge t \in i\cap\overline{i} \wedge \langle\langle f, i\rangle, \langle \overline{f}, \overline{i}\rangle\rangle \in R\} \subseteq r(t)$
⟅since $t \in i \cap \overline{i}$, we have $i \cap \overline{i} \neq \emptyset$⟆

$\iff \forall t \in \mathbb{R}_{\geqslant 0} \,.\, \forall \langle f,\ i\rangle, \langle \overline{f},\ \overline{i}\rangle \,.\, \wedge i \cap \overline{i} \neq \emptyset((t \in i \cap \overline{i} \wedge \langle\langle f,\ i\rangle,\ \langle \overline{f},\ \overline{i}\rangle\rangle \in R) \Longrightarrow$
$\langle f(t),\ \overline{f}(t)\rangle \in r(t))$ ⟅definition of $\subseteq$⟆

$\iff \forall \langle f,\ i\rangle, \langle \overline{f},\ \overline{i}\rangle \,.\, \langle\langle f,\ i\rangle,\ \langle \overline{f},\ \overline{i}\rangle\rangle \in R \Longrightarrow i \cap \overline{i} \neq \emptyset \wedge \forall t \in \mathbb{R}_{\geqslant 0} \,.\, t \in i \cap \overline{i} \Longrightarrow$
$\langle f(t),\ \overline{f}(t)\rangle \in r(t)$ ⟅definition of $\Longrightarrow$⟆

$\iff R \subseteq \{\langle\langle f,\ i\rangle,\ \langle \overline{f},\ \overline{i}\rangle\rangle \mid i\cap\overline{i} \neq \emptyset \wedge \forall t \in \mathbb{R}_{\geqslant 0} \,.\, t \in i\cap\overline{i} \Longrightarrow \langle f(t),\ \overline{f}(t)\rangle \in r(t)\}$
⟅definition of $\subseteq$⟆

$\iff R \subseteq \{\langle\langle f,\ i\rangle,\ \langle \overline{f},\ \overline{i}\rangle\rangle \mid i \cap \overline{i} \neq \emptyset \wedge \forall t \in i \cap \overline{i} \,.\, \langle f(t),\ \overline{f}(t)\rangle \in r(t)\}$
⟅since $i \cap \overline{i} \subseteq \mathbb{R}_{\geqslant 0}$⟆

$\iff R \subseteq \gamma(r)$ ⟅definition (30) of $\gamma$⟆

– Given any $r \in \mathbb{R}_{\geqslant 0} \to \wp(\mathsf{S} \times \mathsf{S})$, define $R = \langle\langle s,\ [0, \infty[\rangle,\ \langle \overline{s},\ [0, \infty[\rangle\rangle \mid \langle s,\ \overline{s}\rangle \in r$. Then $\alpha(R) = r$, probing that $\alpha$ is surjective, and so, by the Galois connection, $\gamma$ is injective.

– $\gamma \circ \alpha(R)$

$= \{\langle\langle f,\ i\rangle,\ \langle \overline{f},\ \overline{i}\rangle\rangle \mid i \cap \overline{i} \neq \emptyset \wedge \forall t \in i \cap \overline{i} \,.\, \langle f(t),\ \overline{f}(t)\rangle \in \alpha(R)(t)\}$
⟅definition of function composition $\circ$ and (30) of $\gamma$⟆

$= \{\langle\langle f,\, i\rangle,\, \langle \overline{f},\, \overline{i}\rangle\rangle \mid i \cap \overline{i} \neq \emptyset \wedge \forall t \in i \cap \overline{i} \mathrel{.} \langle f(t),\, \overline{f}(t)\rangle \in \{\langle f(t),\, \overline{f}(t)\rangle \mid \exists i, \overline{i} \mathrel{.} t \in i \cap \overline{i} \wedge \langle\langle f,\, i\rangle,\, \langle \overline{f},\, \overline{i}\rangle\rangle \in R\}\}$ ⟅definition (31) of $\alpha$⟆

$= \{\langle\langle f,\, i\rangle,\, \langle \overline{f},\, \overline{i}\rangle\rangle \mid i \cap \overline{i} \neq \emptyset \wedge \forall t \in i \cap \overline{i} \mathrel{.} \exists i', \overline{i}' \mathrel{.} t \in i' \cap \overline{i}' \wedge \langle\langle f,\, i'\rangle,\, \langle \overline{f},\, \overline{i}'\rangle\rangle \in R\}$

⟅definition of $\in$⟆

$\supseteq \{\langle\langle f,\, i\rangle,\, \langle \overline{f},\, \overline{i}\rangle\rangle \mid i \cap \overline{i} \neq \emptyset \wedge \langle\langle f,\, i\rangle,\, \langle \overline{f},\, \overline{i}\rangle\rangle \in R\}$

⟅choosing $i' = i$ and $\overline{i}' = \overline{i}$⟆

$= R$ ⟅by $R \in \mathsf{R_C}$ and definition (32) of $\mathsf{R_C}$⟆

By the Galois connection we also have the converse $R \subseteq \gamma \circ \alpha(R)$ proving by antisymmetry that $\gamma$ is surjective so $\alpha$ is injective and therefore have an isomorphism. □

*Proof of* (35).

– ((34) $\Longrightarrow$ (35))

Let $\langle\sigma,\, \overline{\sigma}\rangle \in \vec{\gamma}(r)$ such that, by (34), $\forall t \in [0, \min(⟦\sigma⟧, ⟦\overline{\sigma}⟧)[ \mathrel{.} \langle\sigma_t,\, \overline{\sigma}_t\rangle \in r(t)$. Assume that $j < |\sigma|$ and $\mathsf{e}(\sigma_j) < ⟦\overline{\sigma}⟧$ by hypothesis in (35). We have $\mathsf{e}(\sigma_j) \leqslant ⟦\sigma⟧$ and so $\mathsf{e}(\sigma_j) \in [0, \min(⟦\sigma⟧, ⟦\overline{\sigma}⟧)]$ which implies $\mathsf{b}(\sigma_j) \in [0, \min(⟦\sigma⟧, ⟦\overline{\sigma}⟧)[$ by the nonzeno hypothesis. It follows, by (34), that $\langle\sigma_{\mathsf{b}(\sigma_j)},\, \overline{\sigma}_{\mathsf{b}(\sigma_j)}\rangle \in r(t)$. Therefore, by (17), $\exists k \in [0, |\overline{\sigma}|[ \mathrel{.} \overline{\sigma}_k = \langle\overline{f},\, \overline{i}\rangle \wedge \mathsf{b}(\sigma_j) \in \overline{i}$. Let, by (10), $\sigma_j = \langle f,\, i\rangle$ so that $\mathsf{b}(\sigma_j) \in i$. It follows that $\mathsf{b}(\sigma_j) \in i \cap \overline{i} \neq \emptyset$. By $i \cap \overline{i} \subseteq [0, \min(⟦\sigma⟧, ⟦\overline{\sigma}⟧)[$, (34) implies that for all $t \in i \cap \overline{i}$ we have $\langle\sigma_t,\, \overline{\sigma}_t\rangle \in r(t)$ that is, by (17), $\langle f(t),\, \overline{f}(t)\rangle \in r(t)$. By (30), that implies that $\langle\langle f,\, i\rangle,\, \langle\overline{f},\, \overline{i}\rangle\rangle = \langle\sigma_j,\, \overline{\sigma}_k\rangle \in \gamma(r)$, proving the first condition (35.a). The proof of the second condition (35.b) is similar (since it is the first one (35.a) for $(\vec{\gamma}(r^{-1}))^{-1} = \vec{\gamma}(r)$).

– ((35.a) $\Longrightarrow$ (34))

Assume that $\langle\sigma,\, \overline{\sigma}\rangle \in \vec{\gamma}(r)$ so that (35.a) and (35.b) do hold and $t \in [0, \min(⟦\sigma⟧, ⟦\overline{\sigma}⟧)[$. By (17), $\exists j \in [0, |\sigma|[ \mathrel{.} \sigma_j = \langle f,\, i\rangle \wedge t \in i \wedge \sigma_t = f(t)$. By (35.a), $\exists k < |\overline{\sigma}| \mathrel{.} \langle\sigma_j,\, \overline{\sigma}_k\rangle \in \gamma(r)$. Let $\overline{\sigma}_k = \langle\overline{f},\, \overline{i}\rangle$.

We claim that there exists one of these $k < |\overline{\sigma}|$ such that $t \in \overline{i}$ since, otherwise, we would have $t < ⟦\overline{\sigma}⟧ \wedge \forall k < |\overline{\sigma}| \mathrel{.} \overline{\sigma}_k = \langle\overline{f}_k,\, \overline{i}_k\rangle \wedge t \notin \overline{i}_k$, in contradiction with the def.(15) of trajectories, nonzenoness, and the def.(16) of the duration of trajectories.

Therefore we have $t \in i \cap \overline{i}$ and by $\langle\langle f,\, i\rangle,\, \langle\overline{f},\, \overline{i}\rangle\rangle = \langle\sigma_j,\, \overline{\sigma}_k\rangle \in \gamma(r)$, and def.(30) of $\gamma(r)$, we have $\langle f(t),\, \overline{f}(t)\rangle \in r(t)$, that is by (17), $\langle\sigma_t,\, \overline{\sigma}_t\rangle \in r(t)$ proving that $\langle\sigma,\, \overline{\sigma}\rangle \in \vec{\gamma}(r)$ in (34). □

*Proof of* (36).

$\vec{R} \subseteq \vec{\gamma}(r)$

$\Longleftrightarrow \vec{R} \subseteq \{\langle\sigma,\, \overline{\sigma}\rangle \mid \forall t \in [0, \min(⟦\sigma⟧, ⟦\overline{\sigma}⟧)[ \cap \mathsf{dom}(r) \mathrel{.} \langle\sigma_t,\, \overline{\sigma}_t\rangle \in r(t)\}$

⟅def.(34) of $\vec{\gamma}$⟆

$\iff \forall\langle\sigma,\,\overline{\sigma}\rangle \in \overrightarrow{R}\,.\,\forall t \in [0, \min(\llbracket\sigma\rrbracket, \llbracket\overline{\sigma}\rrbracket)[\,\cap\,\mathsf{dom}(r)\,.\,\langle\sigma_t,\,\overline{\sigma}_t\rangle \in r(t)$ ⟅def.$\subseteq$⟆

$\iff \forall t \in \mathsf{dom}(r)\,.\,\{\langle\sigma_t,\,\overline{\sigma}_t\rangle \mid \langle\sigma,\,\overline{\sigma}\rangle \in \overrightarrow{R} \wedge t \in [0, \min(\llbracket\sigma\rrbracket, \llbracket\overline{\sigma}\rrbracket)[\} \subseteq r(t)$

⟅def.$\subseteq$⟆

$\iff \forall t \in \mathbb{R}_{\geqslant 0}\,.\,\{\langle\sigma_t,\,\overline{\sigma}_t\rangle \mid \langle\sigma,\,\overline{\sigma}\rangle \in \overrightarrow{R} \wedge t \in [0, \min(\llbracket\sigma\rrbracket, \llbracket\overline{\sigma}\rrbracket)[\} \subseteq r(t)$

⟅since $r(t) \triangleq \mathsf{S} \times \overline{\mathsf{S}}$ when $t \notin \mathsf{dom}(r)$⟆

$\iff \forall t \in \mathbb{R}_{\geqslant 0}\,.\,\vec{\alpha}(\overrightarrow{R})(t) \subseteq r(t)$ ⟅def.$\vec{\alpha}$⟆

$\iff \vec{\alpha}(\overrightarrow{R}) \,\dot{\subseteq}\, r$ ⟅pointwise def.$\dot{\subseteq}$⟆ □

*Proof of* (38).

— $\vec{\alpha}$ and $\vec{\gamma}$ are increasing.

— $\vec{\alpha}(\vec{\gamma}(R))$

$= \{\langle\sigma,\,\overline{\sigma}\rangle \mid \exists\overline{T}\,.\,\langle\{\sigma\},\,\overline{T}\rangle \in \vec{\gamma}(R) \wedge \overline{\sigma} \in \overline{T}\}$ ⟅def. $\vec{\alpha}$⟆

$= \{\langle\sigma,\,\overline{\sigma}\rangle \mid \exists\overline{T}\,.\,\{\sigma\} \subseteq \mathsf{pre}[R]\overline{T} \wedge \overline{\sigma} \in \overline{T}\}$ ⟅def. $\vec{\gamma}$⟆

$= \{\langle\sigma,\,\overline{\sigma}\rangle \mid \exists\overline{T}\,.\,\{\sigma\} \subseteq \{\sigma' \mid \exists\overline{\sigma}' \in \overline{T}\,.\,\langle\sigma',\,\overline{\sigma}'\rangle \in R\} \wedge \overline{\sigma} \in \overline{T}\}$ ⟅def. $\mathsf{pre}[R]$⟆

$= \{\langle\sigma,\,\overline{\sigma}\rangle \mid \exists\overline{T}\,.\,\exists\overline{\sigma}' \in \overline{T}\,.\,\langle\sigma,\,\overline{\sigma}'\rangle \in R \wedge \overline{\sigma} \in \overline{T}\}$ ⟅def. $\subseteq$⟆

$\supseteq$ $R$ ⟅since if $\langle\sigma,\,\overline{\sigma}\rangle \in R$ then choosing $\overline{T} = \{\overline{\sigma}\}$ and $\overline{\sigma}' = \overline{\sigma}$, we have $\langle\sigma,\,\overline{\sigma}'\rangle \in R \wedge \overline{\sigma} \in \overline{T}$⟆

— $\vec{\gamma}(\vec{\alpha}(P))$

$= \{\langle T,\,\overline{T}\rangle \mid \forall\sigma \in T\,.\,\exists\overline{\sigma} \in \overline{T}\,.\,\langle\sigma,\,\overline{\sigma}\rangle \in \vec{\alpha}(P)\}$ ⟅def.$\vec{\gamma}$⟆

$= \{\langle T,\,\overline{T}\rangle \mid \forall\sigma \in T\,.\,\exists\overline{\sigma} \in \overline{T}\,.\,\langle\sigma,\,\overline{\sigma}\rangle \in \{\langle\sigma',\,\overline{\sigma}'\rangle \mid \exists\overline{T}'\,.\,\langle\{\sigma'\},\,\overline{T}'\rangle \in P \wedge \overline{\sigma}' \in \overline{T}'\}\}$ ⟅def.$\vec{\alpha}$⟆

$= \{\langle T,\,\overline{T}\rangle \mid \forall\sigma \in T\,.\,\exists\overline{\sigma} \in \overline{T}\,.\,\exists\overline{T}'\,.\,\langle\{\sigma\},\,\overline{T}'\rangle \in P \wedge \overline{\sigma} \in \overline{T}'\}\}$ ⟅def.$\in$⟆

$\supseteq \{\langle T,\,\overline{T}\rangle \mid \forall\sigma \in T\,.\,\exists\overline{\sigma} \in \overline{T}\,.\,\langle\{\sigma\},\,\overline{T}\rangle \in P \wedge \overline{\sigma} \in \overline{T}\}\}$

⟅since if $\forall\sigma \in T\,.\,\exists\overline{\sigma} \in \overline{T}\,.\,\langle\{\sigma\},\,\overline{T}\rangle \in P \wedge \overline{\sigma} \in \overline{T}$ then choosing $\overline{T}' = \overline{T}$, we have $\forall\sigma \in T\,.\,\exists\overline{\sigma} \in \overline{T}\,.\,\exists\overline{T}'\,.\,\langle\{\sigma\},\,\overline{T}'\rangle \in P \wedge \overline{\sigma} \in \overline{T}'$⟆

$\supseteq$ $P$ ⟅since if $\langle T,\,\overline{T}\rangle \in P$ and $\sigma \in T$ then $\{\sigma\} \subseteq T$ so that, by def.of the tensor product, $\langle\{\sigma\},\,\overline{T}\rangle \in P$ so that $\forall\sigma \in T\,.\,\exists\overline{\sigma} \in \overline{T}\,.\,\langle\{\sigma\},\,\overline{T}\rangle \in P \wedge \overline{\sigma} \in \overline{T}$ if $\overline{T} \neq \emptyset$. Otherwise $\overline{T} = \emptyset$ and so $T = \emptyset$. Therefore $\sigma \notin T$ so $P \subseteq \wp(\mathsf{T}_{\mathsf{C}}^{+\infty}) \times \wp(\mathsf{T}_{\overline{\mathsf{C}}}^{+\infty})$ by def.$\otimes$. ⟆ □

*Proof of* (40).

$\gamma(r^{(39)})$

$= \{\langle\langle f,\,i\rangle,\,\langle\overline{f},\,\overline{i}\rangle\rangle \mid i \cap \overline{i} \neq \emptyset \wedge \forall t \in i \cap \overline{i}\,.\,\langle f(t),\,\overline{f}(t)\rangle \in r^{(39)}(t)\}$

⟅definition (30) of $\gamma$⟆

$= \{\langle\langle \boldsymbol{\lambda}\, t \bullet \langle v(t),\,x(t),\,y(t)\rangle,\,i\rangle,\,\langle\boldsymbol{\lambda}\, t \bullet \langle\overline{v}(t),\,\overline{y}(t)\rangle,\,\overline{i}\rangle\rangle \mid i \cap \overline{i} \neq \emptyset \wedge \forall t \in i \cap \overline{i}\,.\,\langle\langle v(t),\,x(t),\,y(t)\rangle,\,\langle\overline{v}(t),\,\overline{y}(t)\rangle\rangle \in r^{(39)}(t)\}$

⟅definitions (19) and (25) of the states, and (10) of configurations⟆

$= \{\langle\langle \boldsymbol{\lambda}\, t \bullet \langle v(t),\, x(t),\, y(t)\rangle,\, i\rangle,\, \langle \boldsymbol{\lambda}\, t \bullet \langle \overline{v}(t),\, \overline{y}(t)\rangle,\, \overline{i}\rangle\rangle \mid i \cap \overline{i} \neq \emptyset \wedge \forall t \in i \cap \overline{i} \,.\, v(t) = \overline{v}(t) \in \{\mathit{shut}, \mathit{open}\} \wedge y(t) = \overline{y}(t)\}$

$\wr$definition (39) of $r^{(39)}\wr$ □

*Proof of* (41).

$$\vec{\gamma}(r^{(39)})$$

$= \vec{\gamma}_c(r^{(39)}) \cap \vec{\gamma}_a(r^{(39)})$ $\wr$definition (35) of $\vec{\gamma}\wr$

$= \{\langle\sigma,\, \overline{\sigma}\rangle \mid \forall j < |\sigma| \,.\, (\mathsf{e}(\sigma_j) \leqslant \rrbracket\overline{\sigma}\rrbracket) \Longrightarrow (\exists k < |\overline{\sigma}| \,.\, \langle\sigma_j,\, \overline{\sigma}_k\rangle \in \gamma(r^{(39)}))\} \cap \{\langle\sigma,\, \overline{\sigma}\rangle \mid \forall k < |\overline{\sigma}| \,.\, (\mathsf{e}(\overline{\sigma}_k) \leqslant \rrbracket\sigma\rrbracket) \Longrightarrow (\exists j < |\sigma| \,.\, \langle\sigma_j,\, \overline{\sigma}_k\rangle \in \gamma(r^{(39)}))\}$

$\wr$definition (35) of $\vec{\gamma}_c$ and $\vec{\gamma}_a\wr$

$= \mathsf{let}\ \rho(c, \overline{c}) \triangleq \exists v, x, y, i, \overline{v}, \overline{y}, \overline{i} \,.\, c = \langle \boldsymbol{\lambda}\, t \bullet \langle v(t),\, x(t),\, y(t)\rangle,\, i\rangle \wedge \overline{c} = \langle \boldsymbol{\lambda}\, t \bullet \langle \overline{v}(t),\, \overline{y}(t)\rangle,\, i\rangle \wedge i \cap \overline{i} \neq \emptyset \wedge \forall t \in i\ \cap i \,.\, v(t) = \overline{v}(t) \wedge y(t) = \overline{y}(t)\ \mathsf{in}$
$\{\langle\sigma,\, \overline{\sigma}\rangle \mid \forall j < |\sigma| \,.\, (\mathsf{e}(\sigma_j) \leqslant \rrbracket\overline{\sigma}\rrbracket) \Longrightarrow (\exists k < |\overline{\sigma}| \,.\, \rho(\sigma_j, \overline{\sigma}_k) \wedge$
$\forall k < |\overline{\sigma}| \,.\, (\mathsf{e}(\overline{\sigma}_k) \leqslant \rrbracket\sigma\rrbracket) \Longrightarrow (\exists j < |\sigma| \,.\, \rho(\sigma_j, \overline{\sigma}_k)\}$

$\wr$definition (40) of $\gamma(r^{(39)})$ and definition of $\cap\wr$ □

*Proof of* (42).

$$\langle\llbracket\tau^{3}\rrbracket,\, \mathcal{S}^{2}\rangle \in \vec{\gamma}(\vec{\gamma}(r^{(39)}))$$

$= \forall\sigma \in \llbracket\tau^{3}\rrbracket \,.\, \exists\overline{\sigma} \in \mathcal{S}^{2} \,.\, \langle\sigma,\, \overline{\sigma}\rangle \in \vec{\gamma}(r^{(39)})$ $\wr$definition (37) of $\vec{\gamma}\wr$

$= \mathsf{let}\ \rho(c, \overline{c}) \triangleq \exists v, x, y, i, \overline{v}, \overline{y}, \overline{i} \,.\, c = \langle \boldsymbol{\lambda}\, t \bullet \langle v(t),\, x(t),\, y(t)\rangle,\, i\rangle \wedge \overline{c} = \langle \boldsymbol{\lambda}\, t \bullet \langle \overline{v}(t),\, \overline{y}(t)\rangle,\, i\rangle \wedge i \cap \overline{i} \neq \emptyset \wedge \forall t \in i\ \cap i \,.\, v(t) = \overline{v}(t) \wedge y(t) = \overline{y}(t)\ \mathsf{in}$
$\forall\sigma \in \llbracket\tau^{3}\rrbracket \,.\, \exists\overline{\sigma} \in \mathcal{S}^{2} \,.\, \forall j < |\sigma| \,.\, (\mathsf{e}(\sigma_j) \leqslant \rrbracket\overline{\sigma}\rrbracket) \Longrightarrow (\exists k < |\overline{\sigma}| \,.\, \rho(\sigma_j, \overline{\sigma}_k) \wedge$
$\forall k < |\overline{\sigma}| \,.\, (\mathsf{e}(\overline{\sigma}_k) \leqslant \rrbracket\sigma\rrbracket) \Longrightarrow (\exists j < |\sigma| \,.\, \rho(\sigma_j, \overline{\sigma}_k)$

$\wr$by (41)$\wr$

$= \mathsf{let}\ \rho(c, \overline{c}) \triangleq \exists v, x, y, i, \overline{v}, \overline{y}, \overline{i} \,.\, c = \langle \boldsymbol{\lambda}\, t \bullet \langle v(t),\, x(t),\, y(t)\rangle,\, i\rangle \wedge \overline{c} = \langle \boldsymbol{\lambda}\, t \bullet \langle \overline{v}(t),\, \overline{y}(t)\rangle,\, i\rangle \wedge i \cap \overline{i} \neq \emptyset \wedge \forall t \in i\ \cap i \,.\, v(t) = \overline{v}(t) \wedge y(t) = \overline{y}(t)\ \mathsf{in}$
$\forall\sigma \in \llbracket\tau^{3}\rrbracket \,.\, \exists\overline{\sigma} \,.\, |\overline{\sigma}_0| = \infty \wedge P(\overline{\sigma}_0) \wedge \forall j \in \mathbb{N} \,.\, \rho(\sigma_j, \overline{\sigma}_0)$

$\wr$definition (21) of $\mathcal{S}^{2}$ so that $k = 0$, $|\overline{\sigma}| = \mathsf{e}(\overline{\sigma}_0) = \rrbracket\sigma\rrbracket = \rrbracket\overline{\sigma}\rrbracket = \infty$,definition (25) of $\tau^{3}$ so that $|\sigma| = \infty$, and choosing $j = 0$ in $\exists j \in \mathbb{N} \,.\, \rho(\sigma_j, \overline{\sigma}_0)\wr$

$\Longrightarrow \forall\sigma \in \llbracket\tau^{3}\rrbracket \,.\, \exists\overline{\sigma} \,.\, P(\overline{\sigma}) \wedge \forall t \geqslant 0 \,.\, \sigma(t).y = \overline{\sigma}(t).y \wedge \sigma(t).v = \overline{\sigma}(t).v$

$\wr$by convention (12) and knowing than intervals $i$ and $\overline{i}$ are infinite$\wr$ □

# D Proofs for Section 7 (Transition-based hybrid trajectory semantics abstraction)

*Proof of theorem 1.* Let $\langle\sigma_i,\ i \in [0, |\sigma|[\rangle \in \llbracket\tau\rrbracket$. Then $\forall i \in [0, |\sigma| - 1[\ .\ \langle\sigma_i,\, \sigma_{i+1}\rangle \in \tau$ so, by (47), $\langle h(\sigma_i),\, h(\sigma_{i+1})\rangle \in \alpha_h(\tau)$. Because $h$ does not change the timings, $\alpha_h(\tau)$ satisfies (22). It follows that $\langle h(\sigma_i),\, i \in [0, |\sigma|[\rangle \in \alpha_h(\tau)$ proving $\alpha_h(\llbracket\tau\rrbracket) \subseteq \llbracket\alpha_h(\tau)\rrbracket$.

Conversely, if $\langle\overline{\sigma}_i,\ i \in [0, |\overline{\sigma}|[\rangle \in \llbracket\alpha_h(\tau)\rrbracket$ then $\forall i \in [0, |\overline{\sigma} - 1|[\ .\ \langle\overline{\sigma}_i,\, \overline{\sigma}_{i+1}\rangle \in \alpha_h(\tau)$ and so, by def. (47) of $\alpha_h(\tau)$, there exist concrete configurations $\sigma_i$ and

$\sigma_{i+1}$ such that $h(\sigma_i) = \overline{\sigma}_i$ and $h(\sigma_{i+1}) = \overline{\sigma}_{i+1}$ with the same timings and so $\langle \sigma_i, \sigma_{i+1}\rangle \in \tau$. It follows that $\langle \sigma_i,\ i \in [0, |\overline{\sigma}|[\rangle = \langle \sigma_i,\ i \in [0, |\sigma|[\rangle \in [\![\tau]\!]$, proving that $[\![\alpha_h(\tau)]\!] \subseteq \alpha_h([\![\tau]\!])$, and therefore $\alpha_h([\![\tau]\!]) = [\![\alpha_h(\tau)]\!]$ by antisymmetry. □

*Proof of theorem 2.* By $\alpha_h(\tau) \subseteq \overline{\tau}$ and (26), lemma 1 implies that $[\![\alpha_h(\tau)]\!] \subseteq [\![\overline{\tau}]\!]$. By hypothesis and transitivity, $[\![\alpha_h(\tau)]\!] \subseteq P$. By (48), $\alpha_h([\![\tau]\!]) \subseteq P$, and therefore, by the Galois connection (46), we get the conclusion $[\![\tau]\!] \subseteq \gamma_h(\overline{P})$ of (49). □

*Proof of theorem 3.* Given a trajectory $\sigma$, we have the trajectory commutation property

$$
\begin{aligned}
&- \ \alpha_\delta(\alpha_h(\sigma))\\
&= \ \alpha_\delta(\alpha_h(\langle\langle f_j,\ i_j\rangle,\ j \in [0,|\sigma|[\rangle))\\
&\qquad \wr\sigma = \langle\langle \sigma_j,\ i_j\rangle,\ j \in [0,|\sigma|[\rangle \text{ where } \sigma_j = \langle f_j,\ i_j\rangle \text{ by def.(15) of trajectories and (10) of configurations}\wr\\
&= \ \alpha_\delta(\langle \alpha_h(\langle f_j,\ i_j\rangle),\ j \in [0,|\sigma|[\rangle) && \wr\text{def.(44) of } \alpha_h\wr\\
&= \ \alpha_\delta(\langle\langle h \circ f_j,\ i_j\rangle,\ j \in [0,|\sigma|[\rangle) && \wr\text{def.(50) of } \alpha_h\wr\\
&= \ \langle(\langle\langle h \circ f_j,\ i_j\rangle,\ j \in [0,|\sigma|[\rangle)_{n\delta},\ n \in \mathbb{N} \wedge n\delta < ]\!\!]\sigma]\!\!]\rangle && \wr\text{def.(27) of } \alpha_\delta\ \wr\\
&= \ \langle h \circ f_{n\delta},\ n \in \mathbb{N} \wedge n\delta < ]\!\!]\sigma]\!\!]\rangle) && \wr\text{function application (17)}\wr\\
&= \ \langle \alpha_h(f_{n\delta}),\ n \in \mathbb{N} \wedge n\delta < ]\!\!]\sigma]\!\!]\rangle) && \wr\text{def.of } \alpha_h \text{ (similar to (43) for traces)}\wr\\
&= \ \langle \alpha_h(\sigma_{n\delta}),\ n \in \mathbb{N} \wedge n\delta < ]\!\!]\sigma]\!\!]\rangle) && \wr\text{def.configurations } \sigma_{n\delta} = \langle f_{n\delta},\ i_{n\delta}\rangle\wr\\
&= \ \alpha_h(\langle \sigma_{n\delta},\ n \in \mathbb{N} \wedge n\delta < ]\!\!]\sigma]\!\!]\rangle) && \wr\text{def.}\alpha_h \text{ similar to (44) for traces}\wr\\
&= \ \alpha_h(\alpha_\delta(\langle \sigma_j,\ j \in [0,|\sigma|[\rangle)) && \wr\text{def.(27) of } \alpha_\delta\wr
\end{aligned}
$$

$$
\begin{aligned}
&- \ \alpha_\delta(\alpha_h(T))\\
&= \ \{\alpha_\delta(\sigma) \mid \sigma \in \alpha_h(T)\} && \wr\text{def.(27) of } \alpha_\delta\wr\\
&= \ \{\alpha_\delta(\sigma) \mid \sigma \in \{\alpha_h(\sigma') \mid \sigma' \in T\}\} && \wr\text{def.(45) of } \alpha_h\wr\\
&= \ \{\alpha_\delta(\alpha_h(\sigma')) \mid \sigma' \in T\} && \wr\text{def.}\in\wr\\
&= \ \{\alpha_h(\alpha_\delta(\sigma')) \mid \sigma' \in T\} && \wr\text{trajectory commutation property}\wr\\
&= \ \{\alpha_h(\varsigma) \mid \varsigma \in \{\alpha_\delta(\sigma') \mid \sigma' \in T\}\} && \wr\text{def.}\in\wr\\
&= \ \{\alpha_h(\varsigma) \mid \varsigma \in \alpha_\delta(T)\} && \wr\text{def.(27) of } \alpha_\delta\wr\\
&= \ \alpha_h(\alpha_\delta(T)) && \wr\text{def.(27) of } \alpha_\delta\wr \quad \square
\end{aligned}
$$

*Proof of (52).*

$$
\begin{aligned}
&\forall c, \overline{c}, c' \,.\, \exists \overline{c}' \,.\, (\langle c,\ \overline{c}\rangle \in R \wedge (\langle c,\ c'\rangle \in \tau \vee c' = \varepsilon)) \Longrightarrow\\
&(((\langle \overline{c},\ \overline{c}'\rangle \in \overline{\tau} \vee \overline{c}' = \varepsilon) \wedge \langle c \fatsemi c'(\!|\min(\mathsf{b}(c'), \mathsf{b}(\overline{c}')), \min(\mathsf{e}(c'), \mathsf{e}(\overline{c}'))|\!),\ \overline{c} \fatsemi\\
&\overline{c}'(\!|\min(\mathsf{b}(c'), \mathsf{b}(\overline{c}')), \min(\mathsf{e}(c'), \mathsf{e}(\overline{c}'))|\!)\rangle \in R) && \wr(51)\wr\\
\Longleftrightarrow\ &\forall c, \overline{c}, c' \,.\, \exists \overline{c}' \,.\, (\langle c,\ \overline{c}\rangle \in R \wedge (\langle c,\ c'\rangle \in \tau)) \Longrightarrow\\
&(((\langle \overline{c},\ \overline{c}'\rangle \in \overline{\tau} \vee \overline{c}' = \varepsilon) \wedge \langle c \fatsemi c'(\!|\min(\mathsf{b}(c'), \mathsf{b}(\overline{c}')), \min(\mathsf{e}(c'), \mathsf{e}(\overline{c}'))|\!),\ \overline{c} \fatsemi\\
&\overline{c}'(\!|\min(\mathsf{b}(c'), \mathsf{b}(\overline{c}')), \min(\mathsf{e}(c'), \mathsf{e}(\overline{c}'))|\!)\rangle \in R)
\end{aligned}
$$

$$\begin{aligned}
&\quad \text{\textlbrackdbl absence of concrete blocking configurations (but a several concrete configurations may abstract to the same abstract configuration)\textrbrackdbl}\\
\Longleftrightarrow\ & \forall c, \overline{c}, c' \,.\, \exists \overline{c}' \,.\, (\langle c,\ \overline{c}\rangle \in R \wedge (\langle c,\ c'\rangle \in \tau)) \Longrightarrow\\
& (((\langle \overline{c},\ \overline{c}'\rangle \in \overline{\tau} \vee \overline{c}' = \varepsilon) \wedge \langle c \mathbin{⨟} c'(\!|\mathsf{b}(c'), \mathsf{e}(c')|\!),\ \overline{c} \mathbin{⨟} \overline{c}'(\!|\mathsf{b}(c'), \mathsf{e}(c')|\!)\rangle \in R)\\
&\quad \text{(well-nested configurations)}\\
\Longleftrightarrow\ & \forall c, \overline{c}, c' \,.\, \exists \overline{c}' \,.\, (\langle c,\ \overline{c}\rangle \in R \wedge (\langle c,\ c'\rangle \in \tau)) \Longrightarrow\\
& (((\langle \overline{c},\ \overline{c}'\rangle \in \overline{\tau} \vee \overline{c}' = \varepsilon) \wedge \langle c',\ \overline{c}'(\!|\mathsf{b}(c'), \mathsf{e}(c')|\!)\rangle \in R) \quad \text{(definition (13) of ⨟)} \qquad \square
\end{aligned}$$

*Details of example 6.* Let $\epsilon > \zeta$.

$$\begin{aligned}
\mathsf{S} &\triangleq \{\mathit{open}, \mathit{shut}, \mathit{on}, \mathit{off}\} \times \mathbb{R} \times \mathbb{R}\\
\mathsf{C}^{\mathit{shut}} &\triangleq \{\langle f,\ [t_1, t_2[\rangle \mid \exists x, y \,.\, \forall t \in [t_1, t_2] \,.\, f(t) = \langle \mathit{shut},\ x(t),\ y(t)\rangle \wedge\\
&\qquad (t = t_1 \Longrightarrow y(t) = 0) \wedge x(t) \leqslant 3 - \epsilon \wedge\\
&\qquad (x(t) = 3 - \epsilon \Longrightarrow t = t_2) \wedge \dot{x}(t) = 1 \wedge \dot{y}(t) = 3/(3 - 2\epsilon)\}\\
\mathsf{C}^{\mathit{open}} &\triangleq \{\langle f,\ [t_1, t_2[\rangle \mid \exists x, y \,.\, \forall t \in [t_1, t_2] \,.\, f(t) = \langle \mathit{open},\ x(t),\ y(t)\rangle \wedge\\
&\qquad (t = t_1 \Longrightarrow x(t) = 0) \wedge y(t) \geqslant 0 \wedge (y(t) = 0 \Longrightarrow t = t_2)\\
&\qquad \wedge\, \dot{x}(t) = 1 \wedge \dot{y}(t) = -2\}\\
\mathsf{C}^{\mathit{off}} &\triangleq \{\langle f,\ [t_1, t_1 + \epsilon[\rangle \mid \exists x, y \,.\, \forall t \in [t_1, t_1 + \epsilon] \,.\, f(t) = \langle \mathit{off},\ x(t),\ 0\rangle \wedge\\
&\qquad (x(t) = x(t_1) + \epsilon \Longrightarrow t = t_1 + \epsilon) \wedge \dot{x}(t) = 1\}\\
\mathsf{C}^{\mathit{on}} &\triangleq \{\langle f,\ [t_1, t_1 + \epsilon[\rangle \mid \exists x, y \,.\, \forall t \in [t_1, t_1 + \epsilon] \,.\, f(t) = \langle \mathit{on},\ x(t),\ y(t_1)\rangle \wedge\\
&\qquad (x(t) = 3 \Longrightarrow t = t_1 + \epsilon) \wedge \dot{x}(t) = 1\}\\
\mathsf{C} &\triangleq \mathsf{C}^{\mathit{shut}} \cup \mathsf{C}^{\mathit{open}} \cup \mathsf{C}^{\mathit{on}} \cup \mathsf{C}^{\mathit{off}}\\
\mathsf{C}^0 &\triangleq \{\langle\langle \mathit{off},\ x,\ 0\rangle,\ [0, \epsilon[\rangle \in \mathsf{C}^{\mathit{off}} \mid \forall t \in [0, \epsilon] \,.\, \dot{x}(t) = 1\}\\
\tau^{6} &\triangleq (\mathsf{C}^{\mathit{off}} \times \mathsf{C}^{\mathit{shut}}) \cup (\mathsf{C}^{\mathit{shut}} \times \mathsf{C}^{\mathit{on}}) \cup (\mathsf{C}^{\mathit{on}} \times \mathsf{C}^{\mathit{open}}) \cup (\mathsf{C}^{\mathit{open}} \times \mathsf{C}^{\mathit{off}}) \qquad (77)\\
&\quad \text{as restricted by (22)} \qquad \square
\end{aligned}$$

*Proof of (55).*

$$\begin{aligned}
& r^{(53)}(t) = \alpha(R^{(53)})(t)\\
=\ & \{\langle f(t),\ \overline{f}(t)\rangle \mid \exists i, \overline{i} \,.\, t \in i \cap \overline{i} \wedge \langle\langle f,\ i\rangle,\ \langle \overline{f},\ \overline{i}\rangle\rangle \in R^{(53)}\} \quad \text{(definition (31) of } \alpha\text{)}\\
=\ & \{\langle f(t),\ \overline{f}(t)\rangle \mid \exists i, \overline{i} \,.\, t \in i \cap \overline{i} \wedge \langle\langle f,\ i\rangle,\ \langle \overline{f},\ \overline{i}\rangle\rangle \in \big\{\langle\langle \boldsymbol{\lambda}\, t \bullet \langle m_t,\ x_t,\ y_t\rangle,\ [t_1, t_2[\rangle,\\
& \langle \boldsymbol{\lambda}\, t \bullet \langle \overline{m}_t,\ \overline{x}_t,\ \overline{y}_t\rangle,\ [\overline{t}_1, \overline{t}_2[\rangle\rangle \bigm| P^{(53)}(m, x, y, t_1, t_2, \overline{m}, \overline{x}, \overline{y}, \overline{t}_1, \overline{t}_2)\big\}\}\\
&\quad \text{(definition (53) of } R^{(53)}\text{)}\\
=\ & \{\langle\langle m_t,\ x_t,\ y_t\rangle,\ \langle \overline{m}_t,\ \overline{x}_t,\ \overline{y}_t\rangle\rangle \mid \exists [t_1, t_2[, [\overline{t}_1, \overline{t}_2[ \,.\, t \in [t_1, t_2[ \cap [\overline{t}_1, \overline{t}_2[ \wedge\\
& P^{(53)}(m, x, y, t_1, t_2, \overline{m}, \overline{x}, \overline{y}, \overline{t}_1, \overline{t}_2)\} \quad \text{(definition of } \in\text{)}\\
=\ & \big\{\langle\langle m_t,\ x_t,\ y_t\rangle,\ \langle \overline{m}_t,\ \overline{x}_t,\ \overline{y}_t\rangle\rangle \bigm| \exists [t_1, t_2[ \subseteq [\overline{t}_1, \overline{t}_2[ \,.\, t \in [t_1, t_2[ \wedge\\
& P^{(53)}(m_t, x_t, y_t, t_1, t_2, \overline{m}_t, \overline{x}_t, \overline{y}_t, \overline{t}_1, \overline{t}_2)\\
&\quad \text{(definition (54) of } P^{(53)} \text{ implying well-nesting, that is } [t_1, t_2[ \subseteq [\overline{t}_1, \overline{t}_2[\text{)}\\
=\ & r^{(53)} \quad \text{(definition (55) of } r^{(53)}\text{)} \qquad \square
\end{aligned}$$

*Proof of theorem 4.* Given $\sigma \in \llbracket\tau\rrbracket$, we construct $\overline{\sigma} \in \llbracket\overline{\tau}\rrbracket$ satisfying (34), that is $\langle\sigma,\ \overline{\sigma}\rangle \in \vec{\gamma}(r)$. We conclude, by (37), that $\langle\llbracket\tau\rrbracket,\ \llbracket\overline{\tau}\rrbracket\rangle \in \vec{\gamma}(\vec{\gamma}(r))$, as required.

So let be given $\sigma \in \llbracket\tau\rrbracket$. By def.(23) of $\llbracket\tau\rrbracket$, we have $\sigma_0 \in \mathsf{C}^0$ with $\mathsf{b}(c) = 0$. So by the initialization hypothesis (56), there exists a related abstract initial condition $\overline{\sigma}_0 \in \overline{\mathsf{C}}^0$ such that $\langle\sigma_0,\ \overline{\sigma}_0\rangle \in \gamma(r)$. Moreover, by (22), $\mathsf{b}(\overline{\sigma}_0) = 0$. By def.(30) of $\gamma(r)$, and considering the prefix traces $\sigma_0$ of duration $\rrbracket\sigma_0\llbracket = \mathsf{e}(\sigma_0)$ and $\overline{\sigma}_0$ of duration $\rrbracket\overline{\sigma}_0\llbracket = \mathsf{e}(\overline{\sigma}_0)$, we have $\forall t \in [0, \min(\rrbracket\sigma_0\llbracket, \rrbracket\overline{\sigma}_0\llbracket)[ \cap \mathsf{dom}(r)\ .\ \langle(\sigma_0)_t,$ $(\overline{\sigma}_0)_t\rangle \in r(t)$, that is $\langle\sigma_0,\ \overline{\sigma}_0\rangle \in \vec{\gamma}(r)$ by (34).

Assume by induction hypothesis, that we have constructed an abstract trajectory $\overline{\sigma}_0 \ldots \overline{\sigma}_k$ from a prefix trajectory $\sigma_0 \ldots \sigma_j$ of $\sigma$ such that $\langle\sigma_0 \ldots \sigma_j,$ $\overline{\sigma}_0 \ldots \overline{\sigma}_k\rangle \in \vec{\gamma}(r)$, the intersection of the time intervals of $\sigma_j$ and $\overline{\sigma}_k$ is not empty (meaning that $\max(\mathsf{b}(\sigma_j), \mathsf{b}(\overline{\sigma}_k)) \leqslant \min(\mathsf{e}(\sigma_j), \mathsf{e}(\overline{\sigma}_k)))$ and $\langle\sigma_j,\ \overline{\sigma}_k\rangle \in \gamma(r)$. We have shown that to hold for the basis $j = k = 0$.

For the induction step there are two cases.

— (A) If $\sigma_j$ has no successor by $\tau$, that is $\forall c'\ .\ \langle\sigma_j, c'\rangle \notin \tau$, then $j = |\sigma|$ and by (23), $\sigma = \sigma_0 \ldots \sigma_j \in \llbracket\tau\rrbracket$. Consider any finite or infinite maximal extension $\overline{\sigma} \in \llbracket\overline{\tau}\rrbracket$ of the prefix $\overline{\sigma}_0 \ldots \overline{\sigma}_k$ by transitions $\llbracket\overline{\tau}\rrbracket$. By induction hypothesis, $\langle\sigma_0 \ldots \sigma_j,$ $\overline{\sigma}_0 \ldots \overline{\sigma}_k\rangle \in \vec{\gamma}(r)$, $\max(\mathsf{b}(\sigma_j), \mathsf{b}(\overline{\sigma}_k)) \leqslant \min(\mathsf{e}(\sigma_j), \mathsf{e}(\overline{\sigma}_k))$, and $\langle\sigma_j,\ \overline{\sigma}_k\rangle \in \gamma(r)$. There are two three cases, as shown in figure 12.

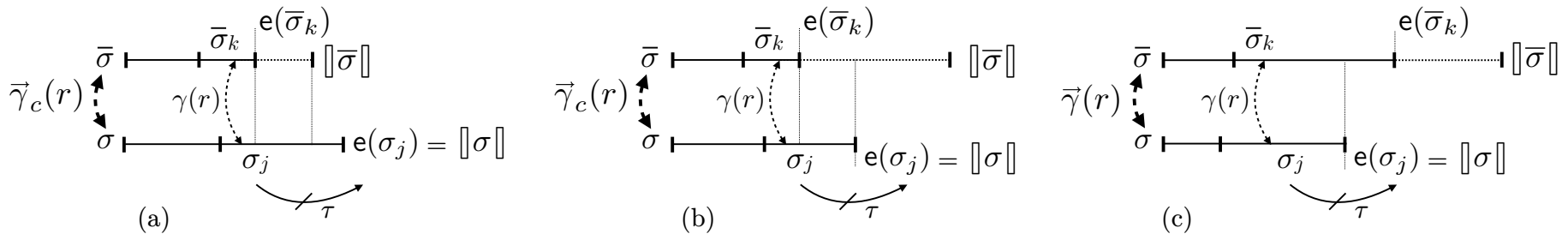

**Fig. 12.** Termination of concrete versus abstract trajectories

(a) If $\mathsf{e}(\overline{\sigma}_k) \leqslant \rrbracket\overline{\sigma}\llbracket \leqslant \mathsf{e}(\sigma_j) = \rrbracket\sigma\llbracket$ then, by induction hypothesis $\langle\sigma_0 \ldots \sigma_j,$ $\overline{\sigma}_0 \ldots \overline{\sigma}_k\rangle \in \vec{\gamma}(r)$, we have $\langle\sigma_t,\ \overline{\sigma}_t\rangle \in r(t)$ for $t \in [0, \mathsf{e}(\overline{\sigma}_k)[$ but maybe not for $t \in [\mathsf{e}(\overline{\sigma}_k), \rrbracket\overline{\sigma}\llbracket]$. Moreover, by induction hypothesis, $\langle\sigma_j,\ \overline{\sigma}_k\rangle \in \gamma(r)$ so that, by (35.a), $\langle\sigma,\ \overline{\sigma}\rangle \in \vec{\gamma}_c(r)$, Q.E.D.

(b) Else, if $\mathsf{e}(\overline{\sigma}_k) \leqslant \mathsf{e}(\sigma_j) = \rrbracket\sigma\llbracket \leqslant \rrbracket\overline{\sigma}\llbracket$ then, by induction hypothesis $\langle\sigma_0 \ldots \sigma_j,$ $\overline{\sigma}_0 \ldots \overline{\sigma}_k\rangle \in \vec{\gamma}(r)$, we have $\langle\sigma_t,\ \overline{\sigma}_t\rangle \in r(t)$ for $t \in [0, \mathsf{e}(\overline{\sigma}_k)[$ but maybe not for $t \in [\mathsf{e}(\overline{\sigma}_k), \mathsf{e}(\sigma_j)] = [\mathsf{e}(\overline{\sigma}_k), \rrbracket\sigma\llbracket]$. Again by induction hypothesis, $\langle\sigma_j,$ $\overline{\sigma}_k\rangle \in \gamma(r)$ so that, by (35.a), $\langle\sigma,\ \overline{\sigma}\rangle \in \vec{\gamma}_c(r)$, Q.E.D.

(c) Otherwise $\mathsf{e}(\sigma_j) = \rrbracket\sigma\llbracket \leqslant \mathsf{e}(\overline{\sigma}_k) \leqslant \rrbracket\overline{\sigma}\llbracket$ then, by induction hypothesis $\langle\sigma_0 \ldots \sigma_j,$ $\overline{\sigma}_0 \ldots \overline{\sigma}_k\rangle \in \vec{\gamma}(r)$, we have $\langle\sigma_t, \overline{\sigma}_t\rangle \in r(t)$ for $t \in [0, \mathsf{e}(\sigma_j)[ = [0, \min(\rrbracket\sigma\llbracket, \rrbracket\overline{\sigma}\llbracket)[$ proving, by (34), that $\langle\sigma,\ \overline{\sigma}\rangle \in \vec{\gamma}(r)$, which implies $\langle\sigma,\ \overline{\sigma}\rangle \in \vec{\gamma}_c(r)$, Q.E.D.

If, in addition, the blocking condition (57) *does* hold then in cases (a) and (b), the last configuration $\overline{\sigma}_k$ has no successor by $\overline{\tau}$ so that, by (23), $\overline{\sigma} = \overline{\sigma}_0 \ldots \overline{\sigma}_k \in \llbracket\overline{\tau}\rrbracket$ is maximal. Then, by induction hypothesis, $\langle\sigma, \overline{\sigma}\rangle = \langle\sigma_0 \ldots \sigma_j, \overline{\sigma}_0 \ldots \overline{\sigma}_k\rangle \in \vec{\gamma}(r)$, Q.E.D.

— (B) Otherwise, $\sigma \in [\![\tau]\!]$ and (23) imply that $\langle \sigma_j,\ \sigma_{j+1}\rangle \in \tau$ with $\mathsf{e}(\sigma_j) = \mathsf{b}(\sigma_{j+1})$ by (22). By $\langle \sigma_j,\ \overline{\sigma}_k\rangle \in \gamma(r)$ and the simulation hypothesis (51) there exists $\overline{\sigma}_{k+1}$ such that $\langle \overline{\sigma}_k,\ \overline{\sigma}_{k+1}\rangle \in \overline{\tau}$ so that $\mathsf{e}(\overline{\sigma}_k) = \mathsf{b}(\overline{\sigma}_{k+1})$ and

$$\langle \sigma_j \mathbin{⨟} \sigma_{j+1}(\!|\min(\mathsf{b}(\sigma_{j+1}), \mathsf{b}(\overline{\sigma}_{k+1})), \min(\mathsf{e}(\sigma_{j+1}), \mathsf{e}(\overline{\sigma}_{k+1}))|\!),\\ \overline{\sigma}_k \mathbin{⨟} \overline{\sigma}_{k+1}(\!|\min(\mathsf{b}(\sigma_{j+1}), \mathsf{b}(\overline{\sigma}_{k+1})), \min(\mathsf{e}(\sigma_{j+1}), \mathsf{e}(\overline{\sigma}_{k+1}))|\!)\rangle \in \gamma(r)$$

By induction hypothesis, the intersection of the time intervals of $\sigma_j$ and $\overline{\sigma}_k$ is not empty and so $\mathsf{b}(\sigma_j) \leqslant \mathsf{e}(\overline{\sigma}_k)$. Depending on how the time intervals of configurations $\sigma_j$, $\sigma_{j+1}$, $\overline{\sigma}_k$, and $\overline{\sigma}_{k+1}$ overlay, we have six cases, illustated in figure 13.

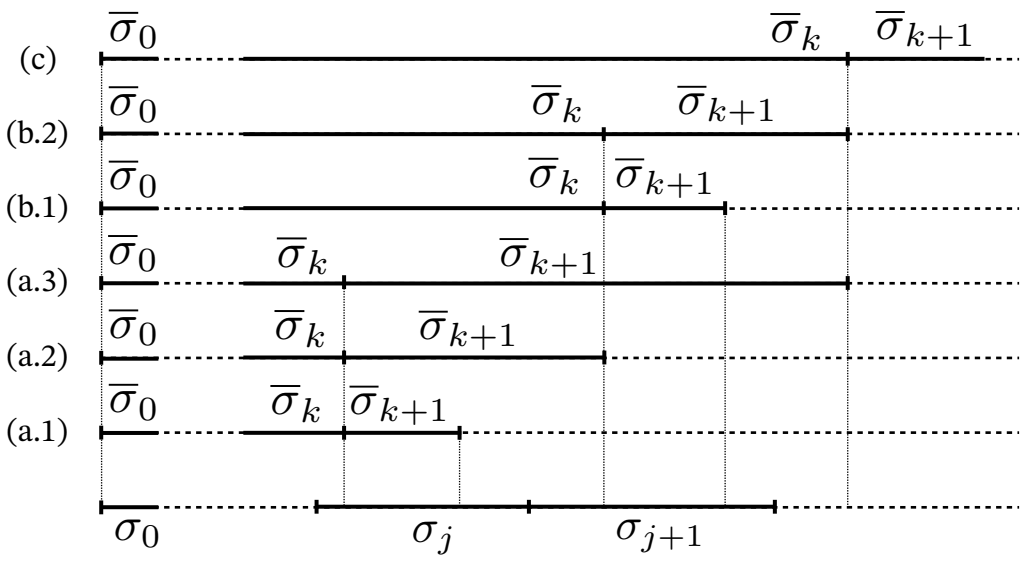

**Fig. 13.** Possible transition steps

(a.1) In this case $\mathsf{e}(\overline{\sigma}_k) \leqslant \mathsf{e}(\sigma_j)$ and $\mathsf{e}(\overline{\sigma}_{k+1}) \leqslant \mathsf{e}(\sigma_j)$, so we have to prove that $\langle \sigma_0 \ldots \sigma_j,\ \overline{\sigma}_0 \ldots \overline{\sigma}_k\overline{\sigma}_{k+1}\rangle \in \vec{\gamma}(r)$.
We have $[\!]\sigma_0 \ldots \sigma_j[\!] = \mathsf{e}(\sigma_j)$ and $[\!]\overline{\sigma}_0 \ldots \overline{\sigma}_k[\!] = \mathsf{e}(\overline{\sigma}_k)$ so that $\min([\!]\sigma_0 \ldots \sigma_j[\!], [\!]\overline{\sigma}_0 \ldots \overline{\sigma}_k[\!]) = \mathsf{e}(\overline{\sigma}_k)$.
By induction hypothesis, $\langle \sigma_0 \ldots \sigma_j,\ \overline{\sigma}_0 \ldots \overline{\sigma}_k\rangle \in \vec{\gamma}(r)$ so that, by def.(34) of $\vec{\gamma}(r)$, we have $\forall t \in [0, \mathsf{e}(\overline{\sigma}_k)[ \cap \mathsf{dom}(r)\ .\ \langle (\sigma_0 \ldots \sigma_j)_t,\ (\overline{\sigma}_0 \ldots \overline{\sigma}_k)_t\rangle \in r(t)$, that is, $(\alpha)$ $\forall t \in [0, \mathsf{e}(\overline{\sigma}_k)[ \cap \mathsf{dom}(r)\ .\ \langle (\sigma_0 \ldots \sigma_j)_t,\ (\overline{\sigma}_0 \ldots \overline{\sigma}_k\overline{\sigma}_{k+1})_t\rangle \in r(t)$ since $\mathsf{e}(\overline{\sigma}_k) = \mathsf{b}(\overline{\sigma}_{k+1})$.
By def.(51) of the simulation, we have $\langle \sigma_j \mathbin{⨟} \sigma_{j+1}(\!|\min(\mathsf{b}(\sigma_{j+1}), \mathsf{b}(\overline{\sigma}_{k+1})), \min(\mathsf{e}(\sigma_{j+1}), \mathsf{e}(\overline{\sigma}_{k+1}))|\!), \overline{\sigma}_k \mathbin{⨟} \overline{\sigma}_{k+1}(\!|\min(\mathsf{b}(\sigma_{j+1}), \mathsf{b}(\overline{\sigma}_{k+1})), \min(\mathsf{e}(\sigma_{j+1}), \mathsf{e}(\overline{\sigma}_{k+1}))|\!)\rangle \in \vec{\gamma}(r)$, that is, in case (a.1), $\langle \sigma_j \mathbin{⨟} \sigma_{j+1}(\!|\mathsf{b}(\overline{\sigma}_{k+1}), \mathsf{e}(\overline{\sigma}_{k+1})|\!),\ \overline{\sigma}_k \mathbin{⨟} \overline{\sigma}_{k+1}(\!|\mathsf{b}(\overline{\sigma}_{k+1}), \mathsf{e}(\overline{\sigma}_{k+1})|\!)\rangle \in \vec{\gamma}(r)$. Since $\mathsf{b}(\overline{\sigma}_{k+1}) = \mathsf{e}(\overline{\sigma}_k)$ and $\mathsf{e}(\overline{\sigma}_{k+1}) \leqslant \mathsf{e}(\sigma_j)$, we get $\langle \sigma_j(\!|\mathsf{e}(\overline{\sigma}_k), \mathsf{e}(\overline{\sigma}_{k+1})|\!),\ \overline{\sigma}_k \mathbin{⨟} \overline{\sigma}_{k+1}(\!|\mathsf{e}(\overline{\sigma}_k), \mathsf{e}(\overline{\sigma}_{k+1})|\!)\rangle \in \vec{\gamma}(r)$. By def.(34) of $\vec{\gamma}(r)$ and (13) of $⨟$, we get $\forall t \in [\mathsf{e}(\overline{\sigma}_k), \mathsf{e}(\overline{\sigma}_{k+1})[ \cap \mathsf{dom}(r)\ .\ \langle (\sigma_j)_t,\ (\overline{\sigma}_k\overline{\sigma}_{k+1})_t\rangle \in r(t)$, which implies $(\beta)$ $\forall t \in [\mathsf{e}(\overline{\sigma}_k), \mathsf{e}(\overline{\sigma}_{k+1})[ \cap \mathsf{dom}(r)\ .\ \langle (\sigma_0 \ldots \sigma_j)_t,\ (\overline{\sigma}_0 \ldots \overline{\sigma}_k\overline{\sigma}_{k+1})_t\rangle \in r(t)$.
Combining $(\alpha)$ and $(\beta)$, we get $\forall t \in [0, \mathsf{e}(\overline{\sigma}_{k+1})[ \cap \mathsf{dom}(r)\ .\ \langle (\sigma_0 \ldots \sigma_j)_t,\ (\overline{\sigma}_0 \ldots \overline{\sigma}_k\overline{\sigma}_{k+1})_t\rangle \in r(t)$, that is $\langle \sigma_0 \ldots \sigma_j,\ \overline{\sigma}_0 \ldots \overline{\sigma}_k\overline{\sigma}_{k+1}\rangle \in \vec{\gamma}(r)$, Q.E.D.

The other proofs are similar and we just provide sketches.

(a.2) In this case $\mathsf{e}(\overline{\sigma}_k) \leqslant \mathsf{e}(\sigma_j)$ and $\mathsf{e}(\sigma_j) \leqslant \mathsf{e}(\overline{\sigma}_{k+1}) \leqslant \mathsf{e}(\sigma_{j+1})$, so we have $\forall t \in [0, \mathsf{b}(\overline{\sigma}_{k+1})[ \cap \mathsf{dom}(r) \,.\, \langle(\sigma_0 \ldots \sigma_j\sigma_{j+1})_t,\ (\overline{\sigma}_0 \ldots \overline{\sigma}_k\overline{\sigma}_{k+1})_t\rangle \in r(t)$ by induction hypothesis and $\forall t \in [\mathsf{b}(\overline{\sigma}_{k+1}), \mathsf{e}(\overline{\sigma}_{k+1})[ \cap \mathsf{dom}(r) \,.\, \langle(\sigma_0 \ldots \sigma_j\sigma_{j+1})_t,$ $(\overline{\sigma}_0 \ldots \overline{\sigma}_k\overline{\sigma}_{k+1})_t\rangle \in r(t)$ by the simulation. By composition, we conclude that $\langle\sigma_0 \ldots \sigma_j\sigma_{j+1},\ \overline{\sigma}_0 \ldots \overline{\sigma}_k\overline{\sigma}_{k+1}\rangle \in \vec{\gamma}(r)$.

(a.3) In this case $\mathsf{e}(\overline{\sigma}_k) \leqslant \mathsf{e}(\sigma_j)$ and $\mathsf{e}(\sigma_{j+1}) \leqslant \mathsf{e}(\overline{\sigma}_{k+1})$, so we have $\forall t \in [0, \mathsf{b}(\overline{\sigma}_{k+1})[ \cap \mathsf{dom}(r) \,.\, \langle(\sigma_0 \ldots \sigma_j\sigma_{j+1})_t,\ (\overline{\sigma}_0 \ldots \overline{\sigma}_k\overline{\sigma}_{k+1})_t\rangle \in r(t)$ by induction hypothesis and $\forall t \in [\mathsf{b}(\overline{\sigma}_{k+1}), \mathsf{e}(\sigma_{j+1})[ \cap \mathsf{dom}(r) \,.\, \langle(\sigma_0 \ldots \sigma_j\sigma_{j+1})_t,\ (\overline{\sigma}_0 \ldots \overline{\sigma}_k$ $\overline{\sigma}_{k+1})_t\rangle \in r(t)$ by the simulation. By composition, we conclude that $\langle\sigma_0 \ldots \sigma_j$ $\sigma_{j+1},\ \overline{\sigma}_0 \ldots \overline{\sigma}_k\overline{\sigma}_{k+1}\rangle \in \vec{\gamma}(r)$.

(b.1) In this case $\mathsf{e}(\overline{\sigma}_k) > \mathsf{e}(\sigma_j) = \mathsf{b}(\sigma_{j+1})$ and $\mathsf{e}(\overline{\sigma}_{k+1}) \leqslant \mathsf{e}(\sigma_{j+1})$, so we have $\forall t \in [0, \mathsf{b}(\overline{\sigma}_{k+1})[ \cap \mathsf{dom}(r) \,.\, \langle(\sigma_0 \ldots \sigma_j\sigma_{j+1})_t,\ (\overline{\sigma}_0 \ldots \overline{\sigma}_k\overline{\sigma}_{k+1})_t\rangle \in r(t)$ by induction hypothesis and $\forall t \in [\mathsf{b}(\overline{\sigma}_{k+1}), \mathsf{e}(\overline{\sigma}_{k+1})[ \cap \mathsf{dom}(r) \,.\, \langle(\sigma_0 \ldots \sigma_j\sigma_{j+1})_t,$ $(\overline{\sigma}_0 \ldots \overline{\sigma}_k\overline{\sigma}_{k+1})_t\rangle \in r(t)$ by the simulation. By composition, we conclude that $\langle\sigma_0 \ldots \sigma_j\sigma_{j+1},\ \overline{\sigma}_0 \ldots \overline{\sigma}_k\overline{\sigma}_{k+1}\rangle \in \vec{\gamma}(r)$.

(b.2) In this case $\mathsf{e}(\overline{\sigma}_k) > \mathsf{e}(\sigma_j) = \mathsf{b}(\sigma_{j+1})$ and $\mathsf{e}(\overline{\sigma}_{k+1}) > \mathsf{e}(\sigma_{j+1})$, so we have $\forall t \in [0, \mathsf{b}(\overline{\sigma}_{k+1})[ \cap \mathsf{dom}(r) \,.\, \langle(\sigma_0 \ldots \sigma_j\sigma_{j+1})_t,\ (\overline{\sigma}_0 \ldots \overline{\sigma}_k\overline{\sigma}_{k+1})_t\rangle \in r(t)$ by induction hypothesis and $\forall t \in [\mathsf{b}(\overline{\sigma}_{k+1}), \mathsf{e}(\sigma_{j+1})[ \cap \mathsf{dom}(r) \,.\, \langle(\sigma_0 \ldots \sigma_j\sigma_{j+1})_t,$ $(\overline{\sigma}_0 \ldots \overline{\sigma}_k\overline{\sigma}_{k+1})_t\rangle \in r(t)$ by the simulation. By composition, we conclude that $\langle\sigma_0 \ldots \sigma_j\sigma_{j+1},\ \overline{\sigma}_0 \ldots \overline{\sigma}_k\overline{\sigma}_{k+1}\rangle \in \vec{\gamma}(r)$.

(c) In this case $\mathsf{e}(\overline{\sigma}_k) > \mathsf{e}(\sigma_{j+1})$, so we have $\forall t \in [0, \mathsf{e}(\sigma_j)[ \cap \mathsf{dom}(r) \,.$ $\langle(\sigma_0 \ldots \sigma_j\sigma_{j+1})_t,\ (\overline{\sigma}_0 \ldots \overline{\sigma}_k)_t\rangle \in r(t)$ by induction hypothesis and for all $t \in [\mathsf{b}(\sigma_{j+1}), \mathsf{e}(\sigma_{j+1})[ \cap \mathsf{dom}(r) \,.\, \langle(\sigma_0 \ldots \sigma_j\sigma_{j+1})_t,\ (\overline{\sigma}_0 \ldots \overline{\sigma}_k)_t\rangle \in r(t)$ by the simulation. By composition, we conclude that $\langle\sigma_0 \ldots \sigma_j\sigma_{j+1},\ \overline{\sigma}_0 \ldots \overline{\sigma}_k\rangle \in \vec{\gamma}(r)$.

In all cases, after this induction step at least one of the concrete and abstract prefix trajectories has one more configuration. Either this extension will terminate for $\sigma$ as in case (A) and we are done. Otherwise, this extension never terminate for $\sigma$, and so, by the blocking hypothesis (57), never terminate either for $\overline{\sigma}$. By the nonzeno condition we will get two infinite trajectories such that $\langle\sigma,\ \overline{\sigma}\rangle \in \vec{\gamma}(r)$. Otherwise there would be a smallest time $t$ such that $\langle\sigma_t,\ \overline{\sigma}_t\rangle \notin r$ which would be in contradiction with the induction hypothesis for configurations with sufficiently long duration. □

*Proof of example 7.* To prove (51), we have, by def.(77) of $\tau \in (\mathsf{C}^{off} \times \mathsf{C}^{shut}) \cup (\mathsf{C}^{shut} \times \mathsf{C}^{on}) \cup (\mathsf{C}^{on} \times \mathsf{C}^{open}) \cup (\mathsf{C}^{open} \times \mathsf{C}^{off})$, def.(77) of $\overline{\tau}$, and def.(53) of $r$, to consider only the concrete $\tau$ and abstract $\overline{\tau}$ transitions defined by (25) and (77), as illustrated in figure 14 by the configurations $c$ of the form $c_1$, $c_2$, $c_3$, or $c_4$. In particular, the cases $c' = \varepsilon$ is impossible in (51) since these concrete transitions are shorter that the abstract ones.

1. In case $\langle c_1,\ \overline{c}_1\rangle \in \gamma(r) \wedge \langle c_2,\ c_1\rangle \in \tau \in (\mathsf{C}^{off} \times \mathsf{C}^{shut})$, there exists $\overline{c}' = \varepsilon$ such that $\langle c_1 \fatsemi c_2 (\!| \min(\mathsf{b}(c_2), \mathsf{b}(\overline{c}')), \min(\mathsf{e}(c_2), \mathsf{e}(\overline{c}')) |\!),\ \overline{c}_1 \fatsemi \overline{c}' (\!| \min(\mathsf{b}(c_2), \mathsf{b}(\overline{c}')),$ $\min(\mathsf{e}(c_2), \mathsf{e}(\overline{c}')) |\!)\rangle = \langle c_1 \fatsemi c_2 (\!| \mathsf{b}(c_2), \mathsf{e}(c_2) |\!),\ \overline{c}_1 (\!| \mathsf{b}(c_2), \mathsf{e}(c_2) |\!)\rangle = \langle c_2,\ \overline{c}_1 (\!| \mathsf{b}(c_2),$

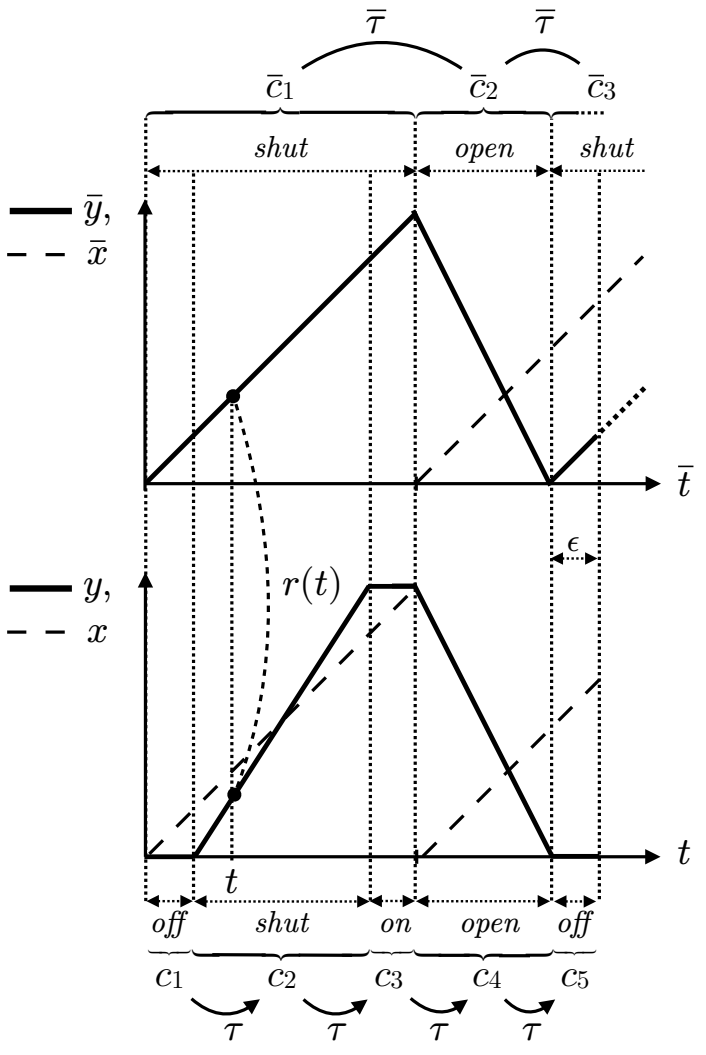

**Fig. 14.** Concrete and abstract transitions for the tank

$\mathsf{e}(c_2)\rrparenthesis\rangle \in \gamma(r)$ by def.(25) of $\tau$, def.(53) of $r$ since $x = \overline{x}$ and $y$ starts $\epsilon$-time after $\overline{y}$ and terminates with the final value of $y$ $\epsilon$-time before $\overline{y}$ does, and def.(30) of $\gamma(r)$, proving (51).

2. In case $\langle c_2,\ \overline{c}_1\rangle \in \gamma(r) \wedge \langle c_2,\ c_3\rangle \in \tau \in (\mathsf{C}^{shut} \times \mathsf{C}^{on})$, there exists $\overline{c}' = \varepsilon$ such that $\langle c_2 \fatsemi c_3 \llparenthesis \min(\mathsf{b}(c_3), \mathsf{b}(\overline{c}')), \min(\mathsf{e}(c_3), \mathsf{e}(\overline{c}'))\rrparenthesis, \overline{c}_1 \fatsemi \overline{c}' \llparenthesis \min(\mathsf{b}(c_3), \mathsf{b}(\overline{c}')), \min(\mathsf{e}(c_3), \mathsf{e}(\overline{c}'))\rrparenthesis\rangle = \langle c_2 \fatsemi c_3 \llparenthesis \mathsf{b}(c_3), \mathsf{e}(c_3)\rrparenthesis, \overline{c}_1 \fatsemi \overline{c}' \llparenthesis \mathsf{b}(c_3), \mathsf{e}(c_3)\rrparenthesis\rangle = \langle c_3, \overline{c}_1 \llparenthesis \mathsf{b}(c_3), \mathsf{e}(c_3)\rrparenthesis\rangle \in \gamma(r)$ by def.(25) of $\tau$, def.(53) of $r$ since $x = \overline{x}$ and $y$ is constant equal to the final value of $\overline{y}$, and def.(30) of $\gamma(r)$, proving (51).
3. In case $\langle c_3, \overline{c}_1\rangle \in \gamma(r) \wedge \langle c_3, c_4\rangle \in \tau \in (\mathsf{C}^{on} \times \mathsf{C}^{open})$, there exists $\overline{c}_2 \,.\, \langle \overline{c}_1, \overline{c}_2\rangle \in \overline{\tau}$ such that $\langle c_3 \fatsemi c_4 \llparenthesis \min(\mathsf{b}(c_4), \mathsf{b}(\overline{c}_2)), \min(\mathsf{e}(c_4), \mathsf{e}(\overline{c}_2))\rrparenthesis, \overline{c}_1 \fatsemi \overline{c}_2 \llparenthesis \min(\mathsf{b}(c_4), \mathsf{b}(\overline{c}_2)), \min(\mathsf{e}(c_4), \mathsf{e}(\overline{c}_2))\rrparenthesis\rangle = \langle c_4,\ \overline{c}_2\rangle \in \gamma(r)$ by def.(25) of $\tau$, def.(22) of $\overline{\tau}$, def.(53) of $r$ since $x = \overline{x}$ and $y = \overline{y}$, and def.(30) of $\gamma(r)$, proving (51).
4. In case $\langle c_4, \overline{c}_2\rangle \in \gamma(r) \wedge \langle c_4, c_5\rangle \in \tau \in (\mathsf{C}^{open} \times \mathsf{C}^{off})$, there exists $\overline{c}_3 \,.\, \langle \overline{c}_2, \overline{c}_3\rangle \in \overline{\tau}$ such that $\langle c_4 \fatsemi c_5 \llparenthesis \min(\mathsf{b}(c_5), \mathsf{b}(\overline{c}_3)), \min(\mathsf{e}(c_5), \mathsf{e}(\overline{c}_3))\rrparenthesis, \overline{c}_2 \fatsemi \overline{c}_3 \llparenthesis \min(\mathsf{b}(c_5), \mathsf{b}(\overline{c}_3)), \min(\mathsf{e}(c_5), \mathsf{e}(\overline{c}_3))\rrparenthesis\rangle = \langle c_5,\ \overline{c}_3\rangle \in \gamma(r)$ by def.(25) of $\tau$, def.(22) of $\overline{\tau}$, def.(53) of $r$ since $x = \overline{x}$, $\overline{y}$ is a strictly increasing function of the time spent in the configuration, and $y = 0$ for $\epsilon$ time, and def.(30) of $\gamma(r)$, proving (51). ☐

*Proof of theorem 5.* Let us first calculate

$$\langle T,\ \overline{T}\rangle \in \vec{\gamma}(\vec{\gamma}_c(r))$$

$$= \quad \forall \sigma \in T \,.\, \exists \overline{\sigma} \in \overline{T} \,.\, \langle \sigma,\ \overline{\sigma}\rangle \in \vec{\gamma}_c(r) \qquad \text{(definition (37) of } \vec{\gamma}\text{)}$$

$$= \quad \forall \sigma \in T \,.\, \exists \overline{\sigma} \in \overline{T} \,.\, \forall j < |\sigma| \,.\, (\mathsf{e}(\sigma_j) \leqslant \llbracket \overline{\sigma} \rrbracket) \Longrightarrow (\exists k < |\overline{\sigma}| \,.\, \langle \sigma_j,\ \overline{\sigma}_k\rangle \in \gamma(r))$$

$$\text{(definition (35) of } \vec{\gamma}_c\text{)}$$

$$= \quad \forall \langle \sigma_j,\ j \in [0, \ell[\rangle \in T \,.\, \exists \langle \overline{\sigma}_k,\ k \in [0, \overline{\ell}[\rangle \in \overline{T} \,.\, \forall j < \ell \,.\, (\mathsf{e}(\sigma_j) < \llbracket \langle \overline{\sigma}_k,\ k \in [0, \overline{\ell}[\rangle \rrbracket) \Longrightarrow (\exists k < \overline{\ell} \,.\, \langle \sigma_j,\ \overline{\sigma}_k\rangle \in \gamma(r))$$

$\wr$trajectory notation $\sigma = \langle \sigma_i,\ i \in [0, \ell[ \rangle$, letting $\ell = |\sigma|$ and $\overline{\ell} = |\overline{\sigma}|$, $|\langle \sigma_i,\ i \in [0, \ell[\rangle| = \ell$, $\langle \sigma_i,\ i \in [0, \ell[\rangle_j = \sigma_j$ when $j < \ell$ $\wr$

$= \forall \langle \langle f_j,\ i_j \rangle,\ j \in [0, \ell[ \rangle \in T \ .\ \exists \langle \langle \overline{f}_k,\ \overline{i}_k \rangle,\ k \in [0, \overline{\ell}[ \rangle \in \overline{T} \ .\ \forall j < \ell \ .\ (\mathsf{e}(\langle f_j,\ i_j \rangle) < \llbracket \langle \langle \overline{f}_k,\ \overline{i}_k \rangle,\ k \in [0, \overline{\ell}[ \rangle \rrbracket) \Longrightarrow (\exists k < \overline{\ell} \ .\ \langle \langle f_j,\ i_j \rangle,\ \langle \overline{f}_k,\ \overline{i}_k \rangle \rangle \in \gamma(r))$

$\wr$letting $\langle f_j,\ i_j \rangle = \langle f_j,\ i_j \rangle$ and $\overline{\sigma}_k = \langle \overline{f}_k,\ \overline{i}_k \rangle$ $\wr$

$= \forall \langle \langle f_j,\ i_j \rangle,\ j \in [0, \ell[ \rangle \in T \ .\ \exists \langle \langle \overline{f}_k,\ \overline{i}_k \rangle,\ k \in [0, \overline{\ell}[ \rangle \in \overline{T} \ .\ \forall j < \ell \ .\ (\mathsf{e}(i_j) < \llbracket \langle \langle \overline{f}_k,\ \overline{i}_k \rangle,\ k \in [0, \overline{\ell}[ \rangle \rrbracket) \Longrightarrow (\exists k < \overline{\ell} \ .\ \langle \langle f_j,\ i_j \rangle,\ \langle \overline{f}_k,\ \overline{i}_k \rangle \rangle \in \gamma(r))$

$\wr \mathsf{e}(\langle f,\ i \rangle) = \mathsf{e}(i) \wr$

$= \forall \langle \langle f_j,\ i_j \rangle,\ j \in [0, \ell[ \rangle \in T \ .\ \exists \langle \langle \overline{f}_k,\ \overline{i}_k \rangle,\ k \in [0, \overline{\ell}[ \rangle \in \overline{T} \ .\ \forall j < \ell \ .\ (\mathsf{e}(i_j) < \llbracket \langle \langle \overline{f}_k,\ \overline{i}_k \rangle,\ k \in [0, \overline{\ell}[ \rangle \rrbracket) \Longrightarrow (\exists k < \overline{\ell} \ .\ i_j \cap \overline{i}_k \neq \emptyset \wedge \forall t \in i_j \cap \overline{i}_k \ .\ \langle f_j(t),\ \overline{f}_k(t) \rangle \in r(t))$ $\wr$definition (30) of $\gamma(r)$ $\wr$

So if $\langle T,\ \overline{T} \rangle \in \vec{\gamma}(\vec{\gamma}_c(r_1))$ and $\langle \overline{T},\ \overline{\overline{T}} \rangle \in \vec{\gamma}(\vec{\gamma}_c(r_2))$ then we have

$\forall \langle \langle f_j,\ i_j \rangle,\ j \in [0, \ell[ \rangle \in T \ .\ \exists \langle \langle \overline{f}_k,\ \overline{i}_k \rangle,\ k \in [0, \overline{\ell}[ \rangle \in \overline{T} \ .\ \forall j < \ell \ .\ (\mathsf{e}(i_j) < \llbracket \langle \langle \overline{f}_k,\ \overline{i}_k \rangle,\ k \in [0, \overline{\ell}[ \rangle \rrbracket) \Longrightarrow (\exists k < \overline{\ell} \ .\ i_j \cap \overline{i}_k \neq \emptyset \wedge \forall t \in i_j \cap \overline{i}_k \ .\ \langle f_j(t),\ \overline{f}_k(t) \rangle \in r_1(t)) \wedge$

$\forall \langle \langle \overline{f}_{\overline{j}},\ \overline{i}_{\overline{j}} \rangle,\ \overline{j} \in [0, \overline{\ell}[ \rangle \in \overline{T} \ .\ \exists \langle \langle \overline{\overline{f}}_{\overline{k}},\ \overline{\overline{i}}_{\overline{k}} \rangle,\ \overline{k} \in [0, \overline{\overline{\ell}}[ \rangle \in \overline{\overline{T}} \ .\ \forall \overline{j} < \overline{\ell} \ .\ (\mathsf{e}(\overline{i}_{\overline{j}}) < \llbracket \langle \langle \overline{\overline{f}}_{\overline{k}},\ \overline{\overline{i}}_{\overline{k}} \rangle,\ \overline{k} \in [0, \overline{\overline{\ell}}[ \rangle \rrbracket) \Longrightarrow (\exists \overline{k} < \overline{\overline{\ell}} \ .\ \overline{i}_{\overline{j}} \cap \overline{\overline{i}}_{\overline{k}} \neq \emptyset \wedge \forall t \in \overline{i}_{\overline{j}} \cap \overline{\overline{i}}_{\overline{k}} \ .\ \langle \overline{f}_{\overline{j}}(t),\ \overline{\overline{f}}_{\overline{k}}(t) \rangle \in r_2(t))$

Observe that the situation which is illustrated in figure 6 will not allow for the composition of $r_1 \circ r_2$ since this relation is restricted to the duration of the intermediate trajectory $\llbracket \langle \langle \overline{f}_k,\ \overline{i}_k \rangle,\ k \in [0, \overline{\ell}[ \rangle \rrbracket$, which cannot be recovered from the duration of $\langle \langle f_j,\ i_j \rangle,\ j \in [0, \ell[ \rangle$ and $\langle \langle \overline{\overline{f}}_{\overline{k}},\ \overline{\overline{i}}_{\overline{k}} \rangle,\ \overline{k} \in [0, \overline{\overline{\ell}}[ \rangle$. But for the case of trajectories, which duration is infinite, we have a simplification into

$\forall \langle \langle f_j,\ i_j \rangle,\ j \in [0, \infty[ \rangle \in T \ .\ \exists \langle \langle \overline{f}_k,\ \overline{i}_k \rangle,\ k \in [0, \infty[ \rangle \in \overline{T} \ .\ \forall j \in \mathbb{N} \ .\ (\exists k \in \mathbb{N} \ .\ i_j \cap \overline{i}_k \neq \emptyset \wedge \forall t \in i_j \cap \overline{i}_k \ .\ \langle f_j(t),\ \overline{f}_k(t) \rangle \in r_1(t)) \wedge$

$\forall \langle \langle \overline{f}_{\overline{j}},\ \overline{i}_{\overline{j}} \rangle,\ \overline{j} \in [0, \infty[ \rangle \in \overline{T} \ .\ \exists \langle \langle \overline{\overline{f}}_{\overline{k}},\ \overline{\overline{i}}_{\overline{k}} \rangle,\ \overline{k} \in [0, \infty[ \rangle \in \overline{\overline{T}} \ .\ \forall \overline{j} \in \mathbb{N} \ .\ (\exists \overline{k} \in \mathbb{N} \ .\ \overline{i}_{\overline{j}} \cap \overline{\overline{i}}_{\overline{k}} \neq \emptyset \wedge \forall t \in \overline{i}_{\overline{j}} \cap \overline{\overline{i}}_{\overline{k}} \ .\ \langle \overline{f}_{\overline{j}}(t),\ \overline{\overline{f}}_{\overline{k}}(t) \rangle \in r_2(t))$

$\Longrightarrow \forall \langle \langle f_j,\ i_j \rangle,\ j \in \mathbb{N} \rangle \in T \ .\ \exists \langle \langle \overline{f}_k,\ \overline{i}_k \rangle,\ k \in \mathbb{N} \rangle \in \overline{T} \ .\ \forall j \in \mathbb{N} \ .\ (\exists k \in \mathbb{N} \ .\ i_j \cap \overline{i}_k \neq \emptyset \wedge \forall t \in i_j \cap \overline{i}_k \ .\ \langle f_j(t),\ \overline{f}_k(t) \rangle \in r_1(t)) \wedge \exists \langle \langle \overline{\overline{f}}_{\overline{k}},\ \overline{\overline{i}}_{\overline{k}} \rangle,\ \overline{k} \in \mathbb{N} \rangle \in \overline{\overline{T}} \ .\ (\exists \overline{k} \in \mathbb{N} \ .\ \overline{i}_k \cap \overline{\overline{i}}_{\overline{k}} \neq \emptyset \wedge \forall t \in \overline{i}_k \cap \overline{\overline{i}}_{\overline{k}} \ .\ \langle \overline{f}_k(t),\ \overline{\overline{f}}_{\overline{k}}(t) \rangle \in r_2(t))$

$\wr$in case $\langle \langle \overline{f}_{\overline{j}},\ \overline{i}_{\overline{j}} \rangle,\ \overline{j} \in \mathbb{N} \rangle \in \overline{T}$ is $\langle \langle \overline{f}_k,\ \overline{i}_k \rangle,\ k \in \mathbb{N} \rangle \in \overline{T}$ with $\overline{j} = k$ $\wr$

$\Longrightarrow \forall \langle \langle f_j,\ i_j \rangle,\ j \in \mathbb{N} \rangle \in T \ .\ \exists \langle \langle \overline{f}_k,\ \overline{i}_k \rangle,\ k \in \mathbb{N} \rangle \in \overline{T} \ .\ \forall j \in \mathbb{N} \ .\ (\exists k \in \mathbb{N} \ .\ i_j \subseteq \overline{i}_k \wedge \forall t \in i_j \cap \overline{i}_k \ .\ \langle f_j(t),\ \overline{f}_k(t) \rangle \in r_1(t)) \wedge \exists \langle \langle \overline{\overline{f}}_{\overline{k}},\ \overline{\overline{i}}_{\overline{k}} \rangle,\ \overline{k} \in \mathbb{N} \rangle \in \overline{\overline{T}} \ .\ (\exists \overline{k} \in \mathbb{N} \ .\ \overline{i}_k \subseteq \overline{\overline{i}}_{\overline{k}} \wedge \forall t \in \overline{i}_k \cap \overline{\overline{i}}_{\overline{k}} \ .\ \langle \overline{f}_k(t),\ \overline{\overline{f}}_{\overline{k}}(t) \rangle \in r_2(t))$

$\wr$by the well-nested hypothesis (59) so $i_j \cap \overline{i}_k \neq \emptyset$ implies $i_j \subseteq \overline{i}_k$ and conversely since time intervals in the trajectories are non-empty $\wr$

$\Longrightarrow \forall \langle \langle f_j,\ i_j \rangle,\ j \in \mathbb{N} \rangle \in T \ .\ \exists \langle \langle \overline{\overline{f}}_{\overline{k}},\ \overline{\overline{i}}_{\overline{k}} \rangle,\ \overline{k} \in \mathbb{N} \rangle \in \overline{\overline{T}} \ .\ (\exists \overline{k} \in \mathbb{N} \ .\ \overline{i}_j \subseteq \overline{\overline{i}}_{\overline{k}} \wedge \forall t \in \overline{i}_j \cap \overline{\overline{i}}_{\overline{k}} \ .\ \langle \overline{f}_j(t),\ \overline{\overline{f}}_{\overline{k}}(t) \rangle \in r_1 \circ r_2(t))$

$\wr$Assume that $t \in \mathbb{R}_{\geqslant 0}$, then $T \subseteq \mathsf{T}^{\infty}_{\mathsf{C}}$ and (15), implies that $t \in i_j$ for some $j \in \mathbb{N}$. So there exist $k$ and $\overline{k}$ such that $i_j \subseteq \overline{i}_k \subseteq \overline{\overline{i}}_{\overline{k}}$ with $\langle f_j(t),\, \overline{f}_k(t)\rangle \in r_1(t)$ and $\langle \overline{f}_k(t),\, \overline{\overline{f}}_{\overline{k}}(t)\rangle \in r_2(t)$ so that $\langle f_j(t),\, \overline{\overline{f}}_{\overline{k}}(t)\rangle \in r_1(t) \circ r_2(t) = (r_1(t) \circ r_2)(t)$.$\wr$

$\Longrightarrow \forall \langle\langle f_j,\, i_j\rangle,\, j \in [0,\ell[\rangle \in T \,.\, \exists \langle\langle \overline{f}_k,\, \overline{i}_k\rangle,\, k \in [0,\overline{\ell}[\rangle \in \overline{\overline{T}} \,.\, \forall j < \ell \,.\, (\mathsf{e}(i_j) < \llbracket\langle\langle \overline{f}_k,\, \overline{i}_k\rangle,\, k \in [0,\overline{\ell}[\rangle\rrbracket) \Longrightarrow (\exists k < \overline{\ell} \,.\, i_j \cap \overline{i}_k \neq \emptyset \wedge \forall t \in i_j \cap \overline{i}_k \,.\, \langle f_j(t),\, \overline{f}_k(t)\rangle \in (r_1 \circ r_2)(t))$

$\wr$by $\ell = \overline{\ell} = \infty$ so $\llbracket\langle\langle \overline{f}_k,\, \overline{i}_k\rangle,\, k \in [0,\overline{\ell}[\rangle\rrbracket = \infty$, renaming, $i_j \subseteq \overline{i}_k$ and non-empty intervals (9) implies $i_j \cap \overline{i}_k \neq \emptyset$ $\wr$

$= \quad \langle T,\, \overline{\overline{T}}\rangle \in \vec{\gamma}(\vec{\gamma}_c(r_1 \circ r_2))$ $\quad\wr$as shown at the beginning of this proof$\wr$ $\square$

*Proof of* (60).

$$r^{(53)} \circ r^{(39)}$$

$= \{\langle c,\, \overline{\overline{c}}\rangle \mid \exists \overline{c} \,.\, \langle c,\, \overline{c}\rangle \in r^{(53)} \wedge \langle \overline{c},\, \overline{\overline{c}}\rangle \in r^{(39)}\}$ $\quad\wr$definition of the composition $\circ\wr$

$= \{\langle\langle m_t,\, x_t,\, y_t\rangle,\, \overline{\overline{c}}\rangle \mid \exists \langle \overline{m}_t,\, \overline{x}_t,\, \overline{y}_t\rangle \,.\, \exists [t_1,t_2[ \,\subseteq\, [\overline{t}_1,\overline{t}_2[ \,.\, t \in [t_1,t_2[ \wedge P^{(53)}(m_t,x_t,y_t,t_1,t_2,\overline{m}_t,\overline{x}_t,\overline{y}_t,\overline{t}_1,\overline{t}_2) \wedge \langle\langle \overline{m}_t,\, \overline{x}_t,\, \overline{y}_t\rangle,\, \overline{\overline{c}}\rangle \in r^{(39)}\}$ $\quad\wr$by (55)$\wr$

$= \{\langle\langle m_t,\, x_t,\, y_t\rangle,\, \langle \overline{m}_t,\, \overline{y}_t\rangle\rangle \mid \exists \overline{x}_t \,.\, \exists [t_1,t_2[ \,\subseteq\, [\overline{t}_1,\overline{t}_2[ \,.\, t \in [t_1,t_2[ \wedge P^{(53)}(m_t,x_t,y_t,t_1,t_2,\overline{m}_t,\overline{x}_t,\overline{y}_t,\overline{t}_1,\overline{t}_2)\}$ $\quad\wr$by (39)$\wr$

$= \{\langle\langle m_t,\, x_t,\, y_t\rangle,\, \langle \overline{m}_t,\, \overline{y}_t\rangle\rangle \mid \exists [t_1,t_2[ \,\subseteq\, [\overline{t}_1,\overline{t}_2[ \,.\, t \in [t_1,t_2[ \wedge P^{(53)}(m_t,x_t,y_t,t_1,t_2,\overline{m}_t,x_t,\overline{y}_t,\overline{t}_1,\overline{t}_2)\}$

$\wr$by definition (54) of $P^{(53)}$ enforcing $\overline{x}_t = x_t\wr$ $\square$

*Proof of* (62). By theorem 4, the premise $R \subseteq F^s_{\tau,\overline{\tau}}(R) \wedge (56) \wedge (57)$ implies that $\langle\llbracket\tau\rrbracket,\, \llbracket\overline{\tau}\rrbracket\rangle \in \vec{\gamma}(\vec{\gamma}(r))$. Since $\vec{\gamma}(\vec{\gamma}(r)) \in \wp(\mathsf{T}^{+\infty}_{\mathsf{C}}) \otimes \wp(\mathsf{T}^{+\infty}_{\mathsf{C}})$ by (38), we get by reflexivity, and $\llbracket\overline{\tau}\rrbracket \subseteq \overline{P}$, and (7.a) that $\langle\llbracket\tau\rrbracket,\, \overline{P}\rangle \in \vec{\gamma}(\vec{\gamma}(r))$. $\square$

*Proof of theorem 6.* — We first show that $\alpha_\delta(\llbracket\tau\rrbracket) \subseteq \llbracket\alpha_\delta(\tau)\rrbracket$. Let $\sigma = \langle \sigma_i,\, i \in [0,|\sigma|[\rangle$ be a concrete trajectory of $\tau$ and its discretization $\alpha_\delta(\sigma) \triangleq \langle\langle \sigma_{n\delta},\, n\rangle,\, n \in \mathbb{N} \wedge n\delta < \llbracket\sigma\rrbracket\rangle$ in (64). We must prove that $\alpha_\delta(\sigma)$ is generated by $\alpha_\delta(\tau)$ in (66).

We have $\sigma_0 \in \mathsf{C}^0$ by (23) so $\langle(\sigma_0)_0,\, 0\rangle \in \alpha_\delta(\mathsf{C}^0)$ by (66.e). So we have constructed an abstract trajectory of the transition system $\langle \mathsf{S} \times \mathbb{N}, \alpha_\delta(\mathsf{C}^0), \alpha_\delta(\tau)\rangle$ up to time $t = 0 = 0\delta$.

Assume that we have constructed the abstract trajectory up to time $n\delta$ (initially $n = 0$). The last state in the abstract trajectory is $\langle \sigma_{n\delta},\, n\rangle$. $n\delta$ belongs to a unique time interval of the configurations of $\sigma$ says $\sigma_k$ so the last state in the abstract trajectory is $\langle(\sigma_k)_{n\delta},\, n\rangle$. There are four cases.

– If $n\delta$ is the last discretized step in the configuration $\sigma_k$, that is, $n\delta = \mathsf{e}(\sigma_k)$, then there are two cases.
  - If $\sigma_k$ has no successor by $\tau$ then, by (66.b), $\langle \sigma_{n\delta},\, n\rangle = \langle(\sigma_k)_{\mathsf{e}(\sigma_k)},\, n\rangle$ has no successor by $\alpha_\delta(\tau)$ so that we have constructed a finite trace of $\llbracket\alpha_\delta(\tau)\rrbracket$;

- Otherwise, $\langle \sigma_k,\ \sigma_{k+1} \rangle \in \tau$, so that the next discretized state is, by (66.b), $\langle \sigma_{(n+1)\delta},\ n+1 \rangle = \langle (\sigma_{k+1})_{(n+1)\delta},\ (n+1) \rangle$. So we have extended the discrete trace by one more step.
- Otherwise, $n\delta$ is not the last discretized step in the configuration $\sigma_k$, that is $(n+1)\delta \leqslant \mathsf{e}(\sigma_k)$, so that, by (66), there are two cases.
  - If $k = 0$, for the initial configuration $\sigma_0$, we can, by (66.a), extend the discretized trajectory by next state $\langle (\sigma_0)_{(n+1)\delta},\ (n+1) \rangle$;
  - Otherwise $k > 0$ and by (66.b), we can extend the discretized trajectory by the transition $\langle \langle (\sigma_k)_{n\delta},\ n \rangle,\ \langle (\sigma_k)_{(n+1)\delta},\ n+1 \rangle \rangle \in [\![\alpha_\delta(\tau)]\!]$.

Repeating this construction we get a maximal finite trajectory of $[\![\alpha_\delta(\tau)]\!]$ ending by a blocking state or, because $\delta > 0$ an infinite, hence maximal, trajectory of $[\![\alpha_\delta(\tau)]\!]$. Q.E.D.

— Conversely, we show that $[\![\alpha_\delta(\tau)]\!] \subseteq \alpha_\delta([\![\tau]\!])$. Given a trace $\langle \varsigma_k,\ k \in [0, |\varsigma|[ \rangle$ with $\varsigma_k = \langle s_k,\ k \rangle$ of $\langle \mathsf{S} \times \mathbb{N},\ \alpha_\delta(\mathsf{C}^0),\ \alpha_\delta(\tau) \rangle$, we can rebuilt, by aggregation of successive states corresponding to the same configuration in (66.e), a trajectory of $\tau$.

For the basis, we have $\varsigma_0 \in \alpha_\delta(\mathsf{C}^0)$ so that by (66), there exists a configuration $\sigma_0 \in \mathsf{C}^0$ such that $\varsigma_0 = \langle (\sigma_0)_0,\ 0 \rangle = \langle (\sigma_0)_{0\delta},\ 0 \rangle$.

Assume by induction hypothesis that we have constructed a prefix trajectory $\sigma_0 \ldots \sigma_k$ (with an initial configuration $\sigma_0$ and transitions in $\tau$ between consecutive configurations) such that $\varsigma_j = \langle (\sigma_k)_{j\delta},\ j \rangle$ (which is the base case for $k = j = 0$).

By hypothesis that the durations of configurations are multiples of $\delta$ and def.(65) of $\alpha_\delta(\sigma)$ there exists $m \geqslant$ such that $(j+m)\delta = \mathsf{e}(\sigma_k)$. At this point there are two cases.

- If $\langle (\sigma_k)_{(j+m)\delta},\ j+m \rangle$ has no successor by $\alpha_\delta(\tau)$ then, by (66.b), $\sigma_k$ has no successor by $\tau$ so that we have built a maximal trajectory $\sigma_0 \ldots \sigma_k \in [\![\tau]\!]$ such that $\alpha_\delta(\sigma_0 \ldots \sigma_k) = \varsigma \in [\![\alpha_\delta(\tau)]\!]$.
- Otherwise $\langle (\sigma_k)_{(j+m)\delta},\ j+m \rangle$ has a successor $\varsigma_{j+m+1}$ by $\alpha_\delta(\tau)$, and so, by (66.b), $\exists \sigma_{k+1} \,.\, \langle \sigma_k,\ \sigma_{k+1} \rangle \in \tau$. Therefore we have constructed a prefix longer trajectory $\sigma_0 \ldots \sigma_k \sigma_{k+1}$ (with an initial configuration $\sigma_0$ and transitions in $\tau$ between consecutive configurations) such that $\varsigma'_j = \langle (\sigma_{k+1})_{j'\delta},\ j' \rangle$ with $j' = j+m+1$. So that the induction hypothesis is satisfied.

Repeating this construction we eventually reach a final configuration as in the first case or we can extend the trajectory for ever, in both case gearing a maximal trajectory $\sigma$ in $[\![\tau]\!]$ such that $\alpha_\delta(\sigma) = \varsigma \in [\![\alpha_\delta(\tau)]\!]$, proving that $[\![\alpha_\delta(\tau)]\!] \subseteq \alpha_\delta([\![\tau]\!])$. Q.E.D.

— By antisymmetry, we conclude that $\alpha_\delta([\![\tau]\!]) = [\![\alpha_\delta(\tau)]\!]$, that is (67), holds. □

*Proof of theorem 7.* Assume, by hypothesis, that $\vec{\gamma}(r)$ is a hybrid simulation (51) between $\tau$ and $\overline{\tau}$ (or equivalently $\vec{\gamma}(r) \subseteq F^s_{\tau,\overline{\tau}}(\vec{\gamma}(r))$ by (61)), both with configuration durations which are positive multiples of $\delta > 0$ (no configuration can have an empty interval).

Let $\alpha_\delta(\tau)$ and $\alpha_\delta(\overline{\tau})$ be the discretization of these transition systems, as defined by (66).

Assume, by hypothesis, that $\langle\langle\overline{s}, \overline{n}\rangle, \langle s', n'\rangle\rangle \in \alpha_\delta(r)^{-1} \circ \alpha_\delta(\tau)$ so that there exists a state $\langle s, n\rangle$ such that $\langle\langle s, n\rangle, \langle\overline{s}, \overline{n}\rangle\rangle \in \alpha_\delta(r)$ and $\langle\langle s, n\rangle, \langle s', n'\rangle\rangle \in \alpha_\delta(\tau)$.

By def.(66) of $\alpha_\delta$ for transitions, $\langle\langle s, n\rangle, \langle s', n'\rangle\rangle \in \alpha_\delta(\tau)$ implies that $n' = n + 1$.

By def.(65) of $\alpha_\delta$ for timed relations, $\langle\langle s, n\rangle, \langle\overline{s}, \overline{n}\rangle\rangle \in \alpha_\delta(r)$ implies that $\overline{n} = n$, $n\delta \in \mathsf{dom}(r)$, and $\langle s, \overline{s}\rangle \in r(n\delta)$.

Again, by def.(65) of $\alpha_\delta$ for timed relations, $\langle\langle s', n'\rangle, \langle\overline{s}', \overline{n}'\rangle\rangle \in \alpha_\delta(r)$, that is $\langle\langle s', n+1\rangle, \langle\overline{s}', \overline{n}'\rangle\rangle \in \alpha_\delta(r)$ is equivalent to $\overline{n}' = n + 1$, $(n + 1)\delta \in \mathsf{dom}(r)$ (which follows from hypothesis (68)) and $\langle s', \overline{s}'\rangle \in r((n + 1)\delta)$.

By definition of Milner's simulation, given the hypothesis $\langle\langle s, n\rangle, \langle\overline{s}, \overline{n}\rangle\rangle \in \alpha_\delta(r)$ and $\langle\langle s, n\rangle, \langle s', n'\rangle\rangle \in \alpha_\delta(\tau)$, we must prove that $\langle\langle\overline{s}, \overline{n}\rangle, \langle s', n'\rangle\rangle \in \alpha_\delta(\overline{\tau}) \circ \alpha_\delta(r)^{-1}$, that is, find an abstract successor state $\langle\overline{s}', \overline{n}'\rangle$ such that $\langle\langle\overline{s}, \overline{n}\rangle, \langle\overline{s}', \overline{n}'\rangle\rangle \in \alpha_\delta(\overline{\tau})$ and $\langle\langle s', n'\rangle, \langle\overline{s}', \overline{n}'\rangle\rangle \in \alpha_\delta(r)$.

Equivalently, by the previous remarks, assuming that $n\delta, (n + 1)\delta \in \mathsf{dom}(r)$, $\langle s, \overline{s}\rangle \in r(n\delta)$, $\langle\langle s, n\rangle, \langle s', n + 1\rangle\rangle \in \alpha_\delta(\tau)$, we must find $\overline{s}'$ such that $\langle\langle\overline{s}, n\rangle, \langle\overline{s}', n + 1\rangle\rangle \in \alpha_\delta(\overline{\tau})$ and $\langle s', \overline{s}'\rangle \in r((n + 1)\delta)$.

By (66), $\langle\langle s, n\rangle, \langle s', n + 1\rangle\rangle \in \alpha_\delta(\tau)$ can only follow from three cases.

1. In case (66.a), $\exists c \;.\; (c \in \mathsf{C}^0 \vee \exists c' \;.\; \langle c', c\rangle \in \tau) \wedge \mathsf{b}(c) \leqslant n\delta < (n + 1)\delta < \mathsf{e}(c) \wedge s = c_{n\delta} \wedge s' = c_{(n+1)\delta}$. By $\mathsf{b}(c) \leqslant n\delta < \mathsf{e}(c)$, we have $n\delta \in \mathsf{dom}(c)$. By (68), we have $n\delta \in \mathsf{dom}(r)$. Therefore, by (69) and $\langle s, \overline{s}\rangle = \langle c_{n\delta}, \overline{s}\rangle \in r(n\delta)$, we have $\exists\overline{c} \in \overline{\mathsf{C}} \;.\; (\overline{c} \in \overline{\mathsf{C}}^0 \vee \exists\overline{c}' \;.\; \langle\overline{c}', \overline{c}\rangle \in \overline{\tau}) \wedge n\delta \in \mathsf{dom}(\overline{c}) \wedge \overline{c}_{n\delta} = \overline{s}$.
   There are three subcases.
   (a) A first case is $n\delta < (n + 1)\delta < \mathsf{e}(\overline{c})$.

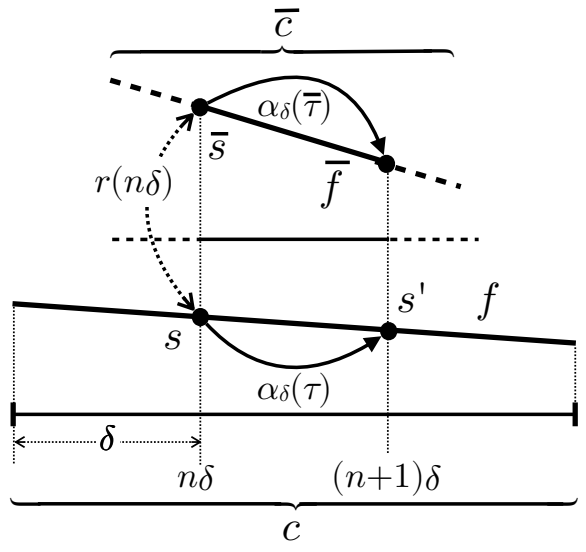

   By (66.a), we have $\langle\overline{c}_{n\delta}, \overline{c}_{(n+1)\delta}\rangle \in \alpha_\delta(\overline{\tau})$ so that choosing $\overline{s}' = \overline{c}_{(n+1)\delta}$, we have $\langle\langle\overline{s}, n\rangle, \langle\overline{s}', n + 1\rangle\rangle \in \alpha_\delta(\overline{\tau})$.
   By hypothesis $\langle s, \overline{s}\rangle \in r(n\delta)$, that is $\langle c_{n\delta}, \overline{c}_{n\delta}\rangle \in r(n\delta)$ and (71.a), we have $\langle c_{(n+1)\delta}, \overline{c}_{(n+1)\delta}\rangle \in r(n\delta)$, that is $\langle s', \overline{s}'\rangle \in r(n\delta)$. By (65), $\langle\langle s', n + 1\rangle, \langle\overline{s}', n + 1\rangle\rangle \in \alpha_\delta(r)$, Q.E.D.

   (b) A second case is $n\delta < (n + 1)\delta = \mathsf{e}(\overline{c})$, with two subcases, depending on whether
      i. $\overline{c}$ has a successor $\overline{c}'$ by $\overline{\tau}$.

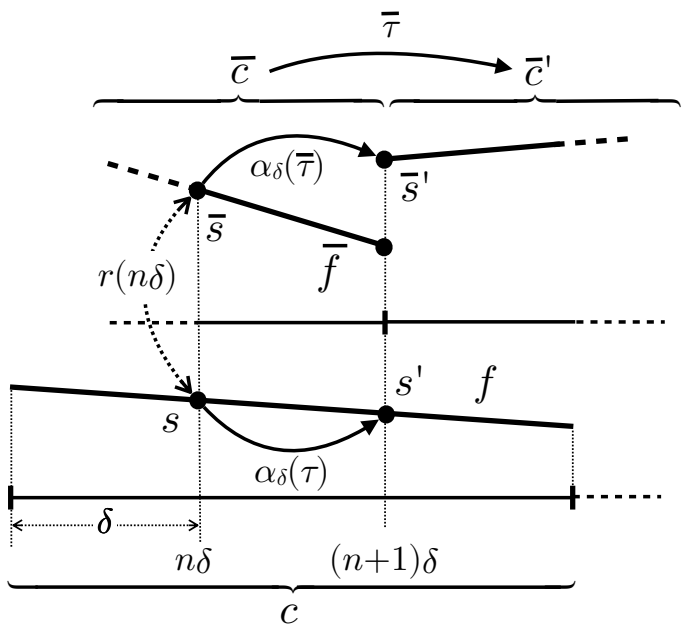

By (66.b), we have $\langle \overline{c}_{n\delta},\, \overline{c}'_{(n+1)\delta}\rangle \in \alpha_\delta(\overline{\tau})$ so that choosing $\overline{s}' = \overline{c}'_{(n+1)\delta}$, we have $\langle\langle \overline{s},\, n\rangle,\, \langle \overline{s}',\, n+1\rangle\rangle \in \alpha_\delta(\overline{\tau})$.

By hypothesis $\langle s,\, \overline{s}\rangle \in r(n\delta)$, that is $\langle c_{n\delta},\, \overline{c}_{n\delta}\rangle \in r(n\delta)$ and so, by (71.c.1) for $n+1$, we have $(n+1)\delta = \mathsf{e}(\overline{c}) \in \mathsf{dom}(c)$ that implies $\langle c_{(n+1)\delta},\, \overline{c}'_{(n+1)\delta}\rangle \in r((n+1)\delta)$, that is $\langle s',\, \overline{s}'\rangle \in r((n+1)\delta)$. By (65), $\langle\langle s',\, n+1\rangle,\, \langle \overline{s}',\, n+1\rangle\rangle \in \alpha_\delta(r)$, Q.E.D.

ii. $\overline{c}$ has no successor by $\overline{\tau}$.

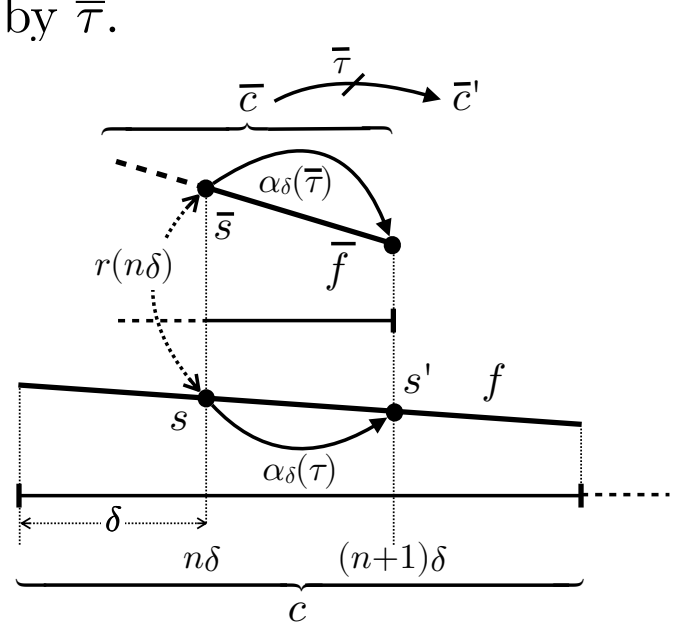

By (66.c), we have $\langle\langle \overline{c}_{n\delta}, n\rangle, \langle \overline{c}_{(n+1)\delta},\, \rangle\rangle n+1 \in \alpha_\delta(\overline{\tau})$ so that choosing $\overline{s}' = \overline{c}_{(n+1)\delta}$, we have $\langle\langle \overline{s},\, n\rangle,\, \langle \overline{s}',\, n+1\rangle\rangle \in \alpha_\delta(\overline{\tau})$.

By hypothesis $\langle s,\, \overline{s}\rangle \in r(n\delta)$, that is $\langle c_{n\delta},\, \overline{c}_{n\delta}\rangle \in r(n\delta)$ and so, by (71.c.3) for $n+1$, we have $(n+1)\delta = \mathsf{e}(\overline{c}) \in \mathsf{dom}(c)$ that implies $\langle c_{(n+1)\delta},\, \overline{c}_{(n+1)\delta}\rangle \in r((n+1)\delta)$. that is $\langle s',\, \overline{s}'\rangle \in r((n+1)\delta)$. By (65), $\langle\langle s',\, n+1\rangle,\, \langle \overline{s}',\, n+1\rangle\rangle \in \alpha_\delta(r)$, Q.E.D.

(c) The third and last case is $n\delta = \mathsf{e}(\overline{c})$, with two subcases, depending on whether

i. $\overline{c}$ has a successor $\overline{c}'$ by $\overline{\tau}$, that is $\langle \overline{c},\, \overline{c}'\rangle \in \overline{\tau}$, so that we are in case (71.c.1) or (71.c.2) of figure 10. In both cases, we have already considered the situation of a transition inside $c$ and $\overline{c}'$ in case 1.(a), Q.E.D.

ii. $\overline{c}$ has no successor by $\overline{\tau}$ of figure 10, we are in case (71.c.3) and obviously we don't have a Milner simulation. But this case is prohibited by the non-blocking hypothesis (70), Q.E.D.

2. The second case is (66.b), $\exists\langle c, c'\rangle \in \tau \,.\, (n+1)\delta = \mathsf{e}(c) \wedge s = c_{n\delta} \wedge s' = c'_{(n+1)\delta}$. Recall that assuming $\langle s,\, \overline{s}\rangle \in r(n\delta)$, $\langle\langle s,\, n\rangle,\, \langle s',\, n+1\rangle\rangle \in \alpha_\delta(\tau)$, we must find $\overline{s}'$ such that $\langle\langle \overline{s},\, n\rangle,\, \langle \overline{s}',\, n+1\rangle\rangle \in \alpha_\delta(\overline{\tau})$ and $\langle s',\, \overline{s}'\rangle \in r((n+1)\delta)$.

Because the duration of configuration $c$ is at least $\delta$, we have $\mathsf{b}(c) \leqslant n\delta < (n+1)\delta = \mathsf{e}(c)$. By (68), we have $n\delta, (n+1)\delta \in \mathsf{dom}(r)$. Therefore, by (69) and $\langle s,\ \overline{s}\rangle = \langle c_{n\delta},\ \overline{s}\rangle \in r(n\delta)$, we have $\exists \overline{c} \in \overline{\mathsf{C}}\ .\ (\overline{c} \in \overline{\mathsf{C}}^0 \vee \exists \overline{c}'\ .\ \langle \overline{c}',\ \overline{c}\rangle \in \overline{\tau}) \wedge n\delta \in \mathsf{dom}(\overline{c}) \wedge \overline{c}_{n\delta} = \overline{s}$.

There are three subcases.

(a) A first case is $n\delta < (n+1)\delta < \mathsf{e}(\overline{c})$.

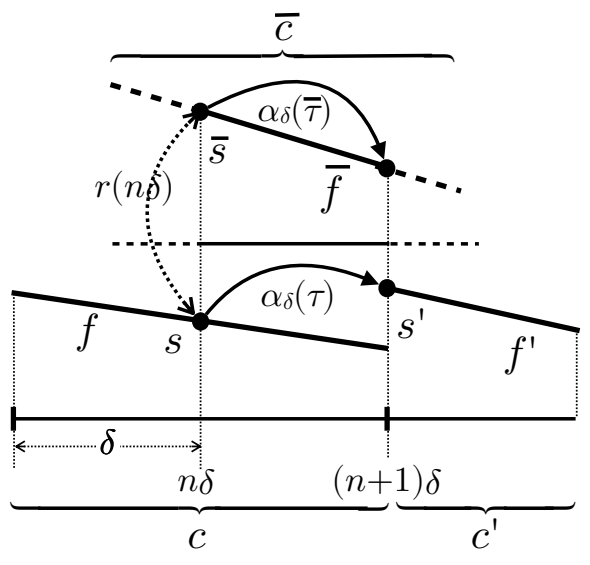

We choose $\overline{s}' = \overline{c}_{(n+1)\delta}$. Then, by (66.a), we have $\langle\langle \overline{s},\ n\rangle,\ \langle \overline{s}',\ n+1\rangle\rangle = \langle\langle \overline{c}_{n\delta},\ n\rangle,\ \langle \overline{c}_{(n+1)\delta},\ n+1\rangle\rangle \in \alpha_\delta(\overline{\tau})$.

Moreover, by (71.b.1) for $n+1$, we have $\langle c_t, \overline{c}_t\rangle = \langle s, \overline{s}\rangle \in r(t)$ for $t = n\delta$, $(n+1)\delta \neq \mathsf{e}(\overline{c})$, and $\langle c, c'\rangle \in \tau$ imply that $\langle c'_{(n+1)\delta}, \overline{c}_{(n+1)\delta}\rangle \in r((n+1)\delta)$. By (65), $\langle\langle s', n+1\rangle, \langle \overline{s}', n+1\rangle\rangle = \langle\langle c_{(n+1)\delta}, n+1\rangle, \langle \overline{c}_{(n+1)\delta}, n+1\rangle\rangle \in \alpha_\delta(r)$, Q.E.D.

(b) A second case is $n\delta < (n+1)\delta = \mathsf{e}(\overline{c})$, with two subcases.

i. $\overline{c}$ has a successor $\overline{c}'$ by $\overline{\tau}$

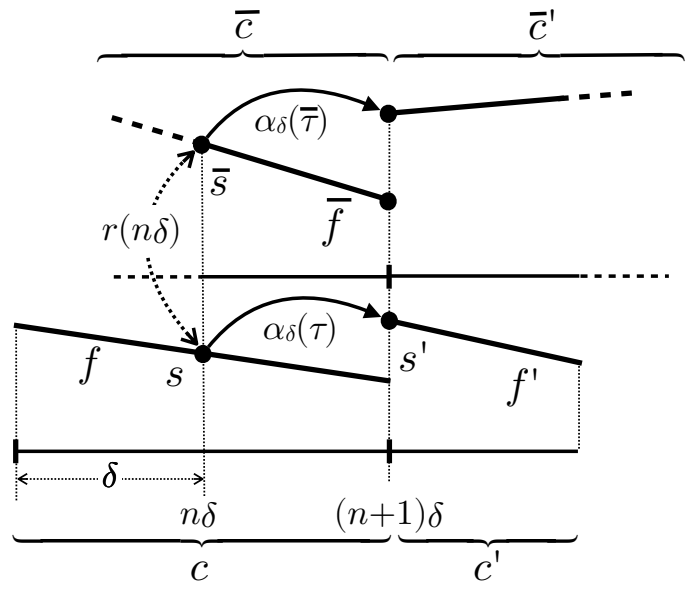

By (66.b), we have $\langle\langle \overline{c}_{n\delta},\ n\rangle,\ \langle \overline{c}'_{(n+1)\delta},\ n+1\rangle\rangle \in \alpha_\delta(\overline{\tau})$ so we choose $\overline{s}' = \langle \overline{c}'_{(n+1)\delta},\ n+1\rangle$ and therefore $\langle\langle \overline{s},\ n\rangle,\ \langle \overline{s}',\ n+1\rangle\rangle \in \alpha_\delta(\overline{\tau})$. Then by (71.b.1) or (71.c.1) for $n+1$, we have $\langle c'_{(n+1)\delta},\ \overline{c}'_{(n+1)\delta}\rangle \in r((n+1)\delta)$. By (65), $\langle\langle s',\ n+1\rangle,\ \langle \overline{s}',\ n+1\rangle\rangle = \langle\langle c'_{(n+1)\delta},\ n+1\rangle,\ \langle \overline{c}'_{(n+1)\delta},\ n+1\rangle\rangle \in \alpha_\delta(r)$, Q.E.D.

ii. $\overline{c}$ has no successor by $\overline{\tau}$

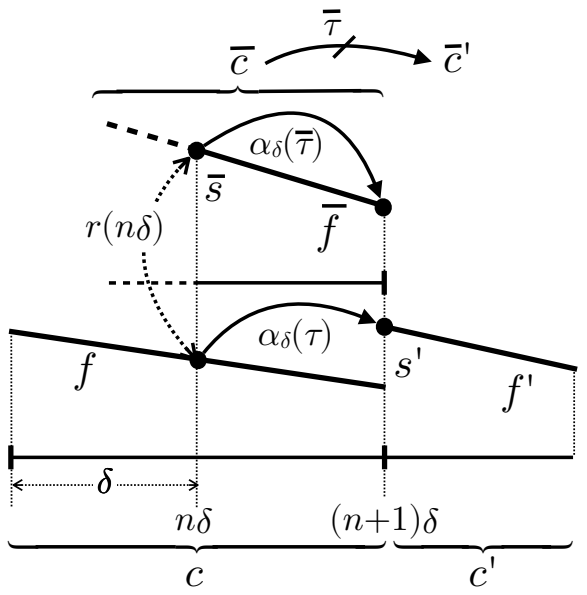

We are in case $s = c_{n\delta}$, $s' = c'_{(n+1)\delta}$, $\langle c_{n\delta},\, \overline{c}_{n\delta}\rangle \in r(n\delta)$, $\langle\langle c_{n\delta},\, n\rangle,$ $\langle c'_{(n+1)\delta},\, n+1\rangle\rangle \in \alpha_\delta(\tau)$, and $\forall \overline{c}'\ .\ \langle \overline{c},\, \overline{c}'\rangle \notin \overline{\tau}$. By (66.c) for $\overline{\tau}$, we have $\langle\langle \overline{c}_{n\delta}, n\rangle, \langle \overline{c}_{(n+1)\delta}, n{+}1\rangle\rangle \in \alpha_\delta(\overline{\tau})$ so we choose $\overline{s}' = \overline{c}_{(n+1)\delta}$. By (71.b.1) for $n+1$, we have $\langle c'_{(n+1)\delta},\, \overline{c}_{(n+1)\delta}\rangle \in r((n+1)\delta)$. By (65), $\langle\langle s', n{+}1\rangle, \langle \overline{s}', n{+}1\rangle\rangle = \langle\langle c'_{(n+1)\delta}, n{+}1\rangle, \langle \overline{c}_{(n+1)\delta}, n{+}1\rangle\rangle \in \alpha_\delta(r)$, Q.E.D.

(c) The third and last case is $n\delta = \mathsf{e}(\overline{c})$, with two subcases.

i. $\overline{c}$ has a successor $\overline{c}'$ by $\overline{\tau}$.

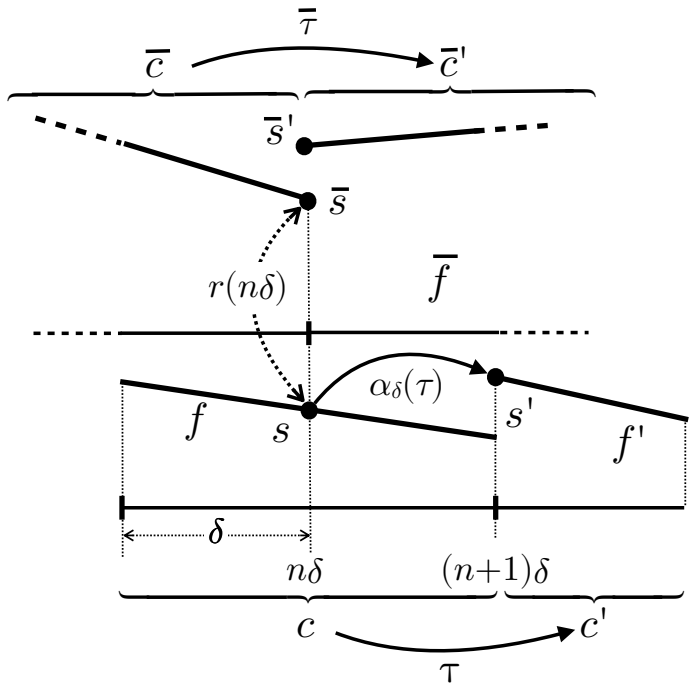

We are in case $s = c_{n\delta}$, $s' = c'_{(n+1)\delta}$ and by (71.c.1), we have $\langle c_{n\delta},$ $\overline{c}'_{n\delta}\rangle \in r_{n\delta}$. Because $n\delta = \mathsf{b}(\overline{c}'_{n\delta})$ and the duration of $\overline{c}'$ is at least $\delta$, there are two subcases.

A. If the duration of $\overline{c}'$ is strictly greater than $\delta$, then we have $(n{+}1)\delta \in \mathsf{dom}(\overline{c}')$, and so, by (66.a), $\langle\langle \overline{c}'_{n\delta}, n\rangle, \langle \overline{c}'_{(n+1)\delta}, n{+}1\rangle\rangle \in \alpha_\delta(\overline{\tau})$. By (61), we have $\langle c\,{}_{9}^{\circ}\,c'(\!|\min(\mathsf{b}(c'), \mathsf{b}(\overline{c}')), \min(\mathsf{e}(c'), \mathsf{e}(\overline{c}'))|\!),$ $\overline{c}\,{}_{9}^{\circ}\,\overline{c}'(\!|\min(\mathsf{b}(c'), \mathsf{b}(\overline{c}')), \min(\mathsf{e}(c'), \mathsf{e}(\overline{c}'))|\!)\rangle \in \gamma(r)$ that is $\langle c\,{}_{9}^{\circ}\,c'(\!|n\delta,$ $\min(\mathsf{e}(c'), \mathsf{e}(\overline{c}'))|\!), \overline{c}'(\!|n\delta, \min(\mathsf{e}(c'), \mathsf{e}(\overline{c}'))|\!)\rangle \in \gamma(r)$ which, by definition (30) of $\gamma(r)$ implies that $\langle c'_{(n+1)\delta}, \overline{c}'_{(n+1)\delta}\rangle \in r((n+1)\delta)$, Q.E.D.

B. Otherwise, the duration of $\overline{c}'$ is exactly $\delta$. So there are two subcases.

– If $\overline{c}'$ has a successor configuration $\overline{c}''$ by $\overline{\tau}$. We are in the case $\langle c_{n\delta},\, \overline{c}'_{n\delta}\rangle \in r_{n\delta}$ and $\langle\langle c_{n\delta},\, n\rangle,\, \langle c'_{(n+1)\delta},\, n+1\rangle\rangle \in \alpha_\delta(\tau)$. By (71.c.1), we have $\langle c'_{(n+1)\delta}, \overline{c}''_{(n+1)\delta}\rangle \in r_{(n+1)\delta}$. By (61), we have $\langle c \mathbin{;} c'(\!|\min(\mathsf{b}(c'), \mathsf{b}(\overline{c}'')), \min(\mathsf{e}(c'), \mathsf{e}(\overline{c}''))|\!), \overline{c}' \mathbin{;} \overline{c}''(\!|\min(\mathsf{b}(c'), \mathsf{b}(\overline{c}'')), \min(\mathsf{e}(c'), \mathsf{e}(\overline{c}''))|\!)\rangle = \langle c \mathbin{;} c'(\!|(n+1)\delta, \min(\mathsf{e}(c'), \mathsf{e}(\overline{c}''))|\!), \overline{c}' \mathbin{;} \overline{c}''(\!|(n+1)\delta, \min(\mathsf{e}(c'), \mathsf{e}(\overline{c}''))|\!)\rangle \in \gamma(r)$ which implies, by def.(30) of $\gamma(r)$, that $\langle c'_{(n+1)\delta}, \overline{c}''_{(n+1)\delta}\rangle \in r((n+1)\delta)$, Q.E.D.

– Otherwise, $\overline{c}'$ has duration $\delta$ and no successor configuration by $\overline{\tau}$. Again, we are in the case $\langle c_{n\delta}, \overline{c}'_{n\delta}\rangle \in r_{n\delta}$ and $\langle\langle c_{n\delta}, n\rangle, \langle c'_{(n+1)\delta}, n+1\rangle\rangle \in \alpha_\delta(\tau)$. By (71.c.3), we have $\langle c'_{(n+1)\delta}, \overline{c}'_{(n+1)\delta}\rangle \in r_{(n+1)\delta}$. By (66.a) we have $\langle \overline{c}'_{n\delta}, \overline{c}'_{(n+1)\delta}\rangle \in \alpha_\delta(\overline{\tau})$. Now we have the situation that $\langle c'_{(n+1)\delta}, \overline{c}'_{(n+1)\delta}\rangle \in r_{(n+1)\delta}$ and $\langle\langle c'_{(n+1)\delta}, n+1\rangle, \langle c'_{(n+2)\delta}, n+2\rangle\rangle \in \alpha_\delta(\tau)$ but forbidden by $\overline{c}'$ has no successor by $\overline{\tau}$, a situation which is forbidden by the non-blocking condition of (70). Q.E.D.

ii. $\overline{c}$ has no successor by $\overline{\tau}$.

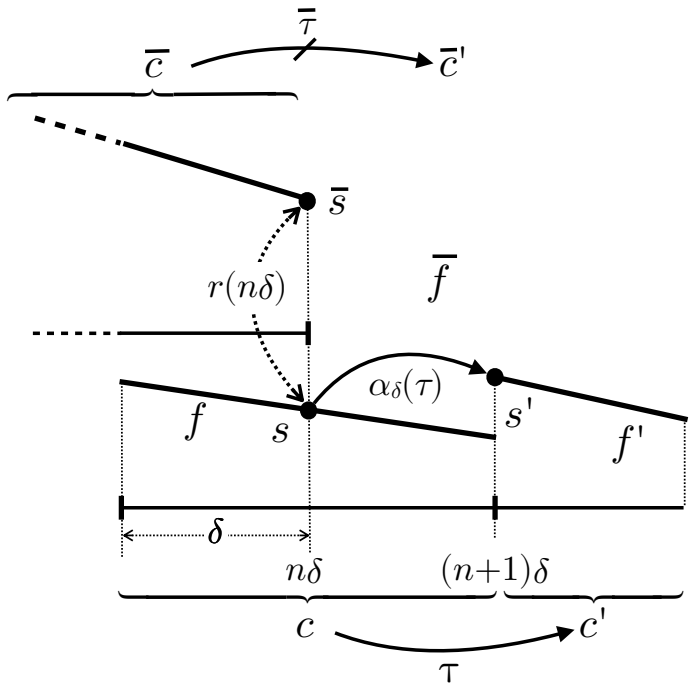

We are in the case of $\langle c_{n\delta}, \overline{c}_{n\delta}\rangle \in r_{n\delta}$ and $\langle\langle c_{n\delta}, n\rangle, \langle c_{(n+1)\delta}, n+1\rangle\rangle \in \alpha_\delta(\tau)$ and $\overline{c}$ has no successor by $\overline{\tau}$, a situation which is excluded by the non-blocking condition of (70). Q.E.D.

3. In case (66.c), $\exists c \in \mathsf{C} \,.\, \forall c' \,.\, \langle c, c'\rangle \notin \tau \wedge (n+1)\delta = \mathsf{e}(c) \wedge s = c_{n\delta} \wedge s' = c_{(n+1)\delta}$. Again, there are three cases

   (a) A first case is $n\delta < (n+1)\delta < \mathsf{e}(\overline{c})$.

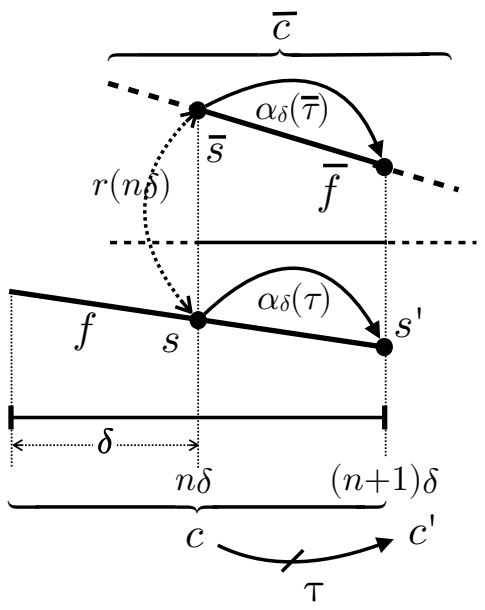

We are in case $s = c_{n\delta}$, $s' = c'_{(n+1)\delta}$, $\langle c_{n\delta}, \overline{c}_{n\delta}\rangle \in r(n\delta)$, $\langle\langle c_{n\delta}, n\rangle, \langle c'_{(n+1)\delta}, n+1\rangle\rangle \in \alpha_\delta(\tau)$, and $\forall c' \,.\, \langle c, c'\rangle \notin \tau$. By (66.a) for $\overline{\tau}$, we have $\langle\langle \overline{c}_{n\delta}, n\rangle, \langle \overline{c}_{(n+1)\delta}, n+1\rangle\rangle \in \alpha_\delta(\overline{\tau})$ so we choose $\overline{s}' = \overline{c}_{(n+1)\delta}$. By (71.b.1) for $n+1$, we have $\langle c'_{(n+1)\delta},\ \overline{c}_{(n+1)\delta}\rangle \in r((n+1)\delta)$. By (65), $\langle\langle s',\ n+1\rangle,\ \langle \overline{s}',\ n+1\rangle\rangle = \langle\langle c'_{(n+1)\delta},\ n+1\rangle,\ \langle \overline{c}_{(n+1)\delta},\ n+1\rangle\rangle \in \alpha_\delta(r)$, Q.E.D.

(b) A second case is $n\delta < (n+1)\delta = \mathsf{e}(\overline{c})$, with two subcases, depending on whether $\overline{c}$ has a successor $\overline{c}'$ by $\overline{\tau}$ or not.

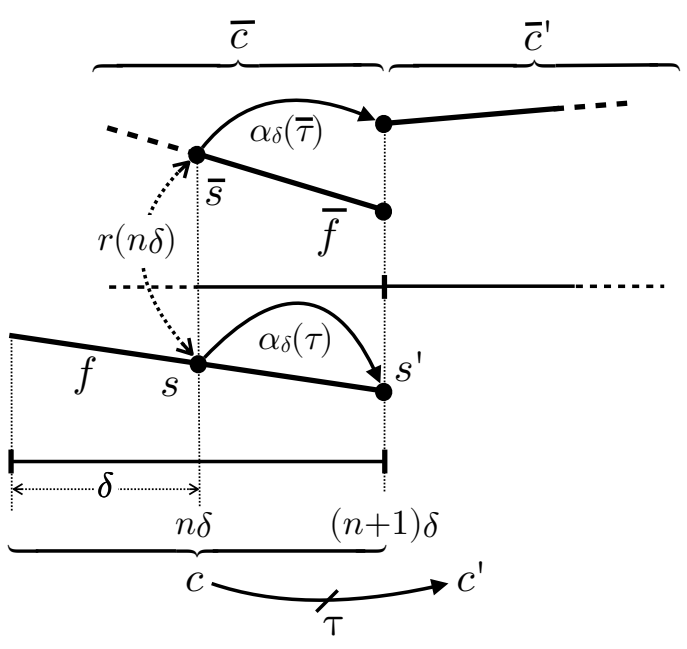

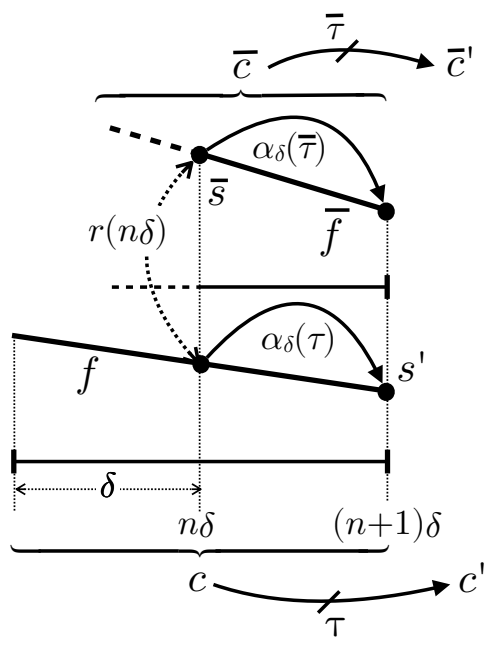

In both cases, $s' = c_{(n+1)\delta}$ has no successor by $\alpha_\delta(\tau)$. Therefore for all concrete states $s''$, we have $\langle s',\ s''\rangle \notin \alpha_\delta(r)^{-1} \circ \alpha_\delta(\tau)$. It follows that (72) holds vacuously. Q.E.D.

(c) The third and last case is $n\delta = \mathsf{e}(\overline{c})$, with two subcases, depending on whether

i. $\overline{c}$ has a successor $\overline{c}'$ by $\overline{\tau}$

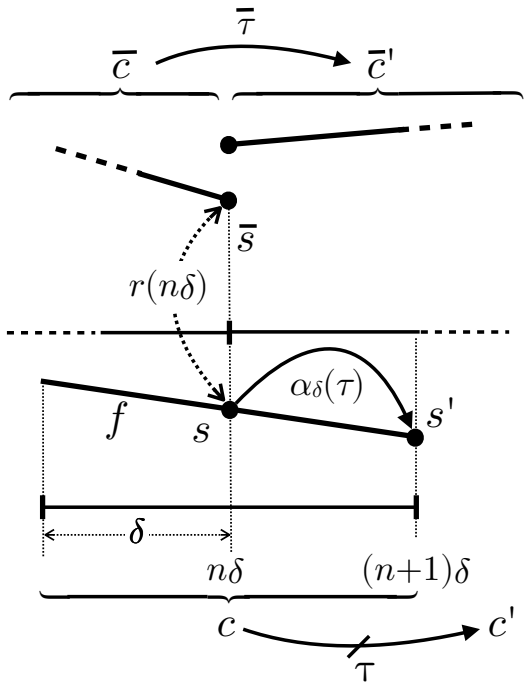

We are in case $s = c_{n\delta}$, $s' = c'_{(n+1)\delta}$ with $\langle\langle c_{n\delta},\ n\rangle,\ \langle c_{(n+1)\delta},\ n+1\rangle\rangle \in \alpha_\delta(\tau)$. By (71.c.1), we have $\langle c_{n\delta},\ \overline{c}'_{n\delta}\rangle \in r_{n\delta}$. The duration of $\overline{c}'$ is at least $\delta$ and so, by ((66).a), we have $\langle\langle \overline{c}'_{n\delta},\ n\rangle,\ \langle \overline{c}'_{(n+1)\delta},\ n+1\rangle\rangle \in \alpha_\delta(\overline{c}')$.
By (51) where $c' = \varepsilon$, we have $\langle c(\!|\min(\mathsf{b}(c'),\mathsf{b}(\overline{c}')),\min(\mathsf{e}(c'),\mathsf{e}(\overline{c}'))|\!),\ \overline{c}\,⨾\,\overline{c}'(\!|\min(\mathsf{b}(c'),\mathsf{b}(\overline{c}')),\min(\mathsf{e}(c'),\mathsf{e}(\overline{c}'))|\!)\rangle = \langle c(\!|\min(+\infty,\mathsf{b}(\overline{c}')),\min(+\infty,\mathsf{e}(\overline{c}'))|\!),\ \overline{c}\,⨾\,\overline{c}'(\!|\min(+\infty,\mathsf{b}(\overline{c}')),\min(+\infty,\mathsf{e}(\overline{c}'))|\!)\rangle = \langle c(\!|n\delta,\mathsf{e}(\overline{c}')|\!),\ \overline{c}\,⨾\,\overline{c}'(\!|n\delta,\mathsf{e}(\overline{c}')|\!)\rangle \in \gamma(r)$ with $\mathsf{e}(\overline{c}') \geqslant (n+1)\delta$ which implies, by def.(30) of $\gamma(r)$, that $\langle c'_{(n+1)\delta},\ \overline{c}'_{(n+1)\delta}\rangle \in r((n+1)\delta)$, Q.E.D.

ii. $\overline{c}$ has no successor by $\overline{\tau}$

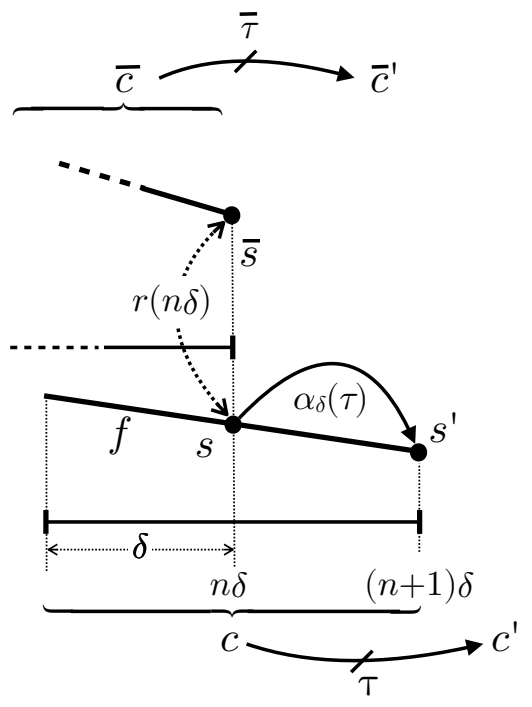

We have $t = n\delta$ such that $t \in \mathsf{dom}(c) \cap \mathsf{dom}(\overline{c}) \cap \mathsf{dom}(r)$ . $\langle c_t, \overline{c}_t\rangle \in r(t)$ but thehypothesis (71) is not satisfied and so (72) holds vacuously. Q.E.D. □

# E Proofs for Section A (Transition-based hybrid trajectory semantics abstraction (continued))

*Proof of theorem 8.* Progress (76) and preservation (74) implies simulation (51) so that, by theorem 4, the initialization (56) implies that $\langle [\![\tau]\!], [\![\overline{\tau}]\!]\rangle \in \vec{\gamma}(\vec{\gamma}_c(r))$. Moreover, having a simulation, the blocking condition (57) implies, again by theorem 4, that $\langle [\![\tau]\!], [\![\overline{\tau}]\!]\rangle \in \vec{\gamma}(\vec{\gamma}(r))$. □